\documentclass{aa}   

\usepackage{euclid}

\usepackage{natbib}
\bibpunct{(}{)}{;}{a}{}{,} 

\usepackage{soul}
\usepackage{float,lscape}
\usepackage{graphicx}
\usepackage{subcaption}
\usepackage{txfonts}
\usepackage{amsmath}
\usepackage{physics}    
\usepackage{color}              
\usepackage{verbatim}   
\usepackage{threeparttable} 
\usepackage[colorlinks,linkcolor=blue,urlcolor=blue,citecolor=blue]{hyperref} 

\usepackage{geometry}
\usepackage{pdflscape}

\graphicspath{{./}{figures/}}

\begin{document}


\definecolor{blue}{rgb}{0,0.08,0.9}
\definecolor{red}{rgb}{0.85,0.08,0.05}
\definecolor{green}{rgb}{0.35,0.45,0.25}
\definecolor{orange}{rgb}{1.0,0.5,0.15}
\definecolor{yellow}{rgb}{1.,1.,0.}
\definecolor{brown}{rgb}{0.7,0.25,0.0}
\definecolor{cyan}{rgb}{0.1,0.66,0.66}
\definecolor{pink}{rgb}{0.858, 0.688, 0.688}

\newcommand{\shuntov}[1]{{\color{red}{SHUNTOV: \textbf\small{#1}}}}
\newcommand{\hjmcc}[1]{{\color{green}{hjmcc: \textbf\small{#1}}}}
\newcommand{\gavazzi}[1]{{\color{blue}{GAVAZZI: \textbf\small{#1}}}}
\newcommand{\ilbert}[1]{{\color{brown}{ILBERT: \textbf\small{#1}}}}
\newcommand{\yo}[1]{{\color{magenta}{DUBOIS: \textbf\small{#1}}}}
\newcommand{\weaver}[1]{{\color{orange}{WEAVER: \textbf\small{#1}}}}
\newcommand{\moneti}[1]{{\color{cyan}{MONETI: \textbf\small{#1}}}}
\newcommand{\davidzon}[1]{{\color{purple}{DAVIDZON: \textbf\small{#1}}}}
\newcommand{\toft}[1]{{\color{yellow}{TOFT: \textbf\small{#1}}}}
\newcommand{\milvangjensen}[1]{{\color{cyan}{MILVANG-JENSEN: \textbf\small{#1}}}}
\newcommand{\mobasher}[1]{{\color{magenta}{MOBASHER: \textbf\small{#1}}}}
\newcommand{\laigle}[1]{{\color{pink}{ \textbf\small{#1}}}}

\newcommand{\mum}{\,\si{\micron}\xspace}
\newcommand{\vect}[1]{\boldsymbol{#1}}
\newcommand{\zphot}{z_{\rm phot}}
\newcommand{\sigmaLogMs}{\sigma_{\text{Log}M_*}}


\title{COSMOS2020: Cosmic evolution of the stellar-to-halo mass relation for central and satellite galaxies up to $z\sim5$}
\subtitle{}  

\author{
M.~Shuntov\inst{1}
\and H.~J.~McCracken\inst{1} 
\and R.~Gavazzi \inst{1,2}
\and C.~Laigle\inst{1}
\and J.~R. Weaver \inst{3,4}    
\and I.~Davidzon \inst{3,4}   
\and O.~Ilbert\inst{2}
\and O.~B.~Kauffmann\inst{2}
\and A.~Faisst\inst{5}
\and Y.~Dubois\inst{1}
\and A.~M.~Koekemoer\inst{6}
\and A.~Moneti\inst{1}
\and B.~Milvang-Jensen\inst{3,4}
\and B.~Mobasher\inst{7}
\and D.~B.~Sanders\inst{8}
\and S.~Toft\inst{3,4}
}

\institute{
Institut d'Astrophysique de Paris, UMR 7095, CNRS, and Sorbonne Universit\'e, 98 bis boulevard Arago, 75014 Paris, France
\and Aix Marseille Univ, CNRS, LAM, Laboratoire d'Astrophysique de Marseille, Marseille, France 
\and Cosmic Dawn Center (DAWN)                   
\and Niels Bohr Institute, University of Copenhagen, Jagtvej 128, 2200 Copenhagen, Denmark      
\and Infrared Processing and Analysis Center, California Institute of Technology, 1200 East California Boulevard, Pasadena, CA 91125, USA 
\and Space Telescope Science Institute, 3700 San Martin Dr., Baltimore, MD 21218, USA
\and Department of Physics and Astronomy, University of California, Riverside, 900 University Avenue, Riverside, CA 92521, USA
\and Institute for Astronomy, University of Hawaii, 2680 Woodlawn Drive, Honolulu, HI 96822, USA
}
\date{Released on TBD / Accepted date: TBD}

\abstract
{

We used the COSMOS2020 catalog to measure the stellar-to-halo mass relation (SHMR) divided by central and satellite galaxies from $z=0.2$ to $z = 5.5$.
Starting from accurate photometric redshifts, we measured the near-infrared selected two-point angular correlation and stellar mass functions in ten redshift bins. We used a phenomenological model that parametrizes the stellar-to-halo mass relation for central galaxies and the number of galaxies inside each halo to describe our observations. This model qualitatively reproduces our measurements and their dependence on the stellar mass threshold.  Surprisingly, the mean halo occupation distribution only shows  a mild evolution with redshift suggesting that galaxies occupy halos similarly throughout cosmic time.
At each redshift, we measured the ratio of stellar mass to halo mass, $M_*/M_h$, which shows the characteristic strong dependence of halo mass with a peak at $M_h^{\rm peak} \sim 2 \times 10^{12} \, M_{\odot}$. 
For the first time, using a joint modeling of clustering and abundances, we measured the evolution of $M_h^{\rm peak}$ from $z=0.2$ to $z=5.5$. $M_h^{\rm peak}$ increases gradually with redshift from log$\,M_h^{\rm peak}/M_{\odot} \sim 12.1$ at $z\sim0.3$ to log$\,M_h^{\rm peak}/M_{\odot} \sim 12.3$ at $z\sim2,$ and up to log$\,M_h^{\rm peak}/M_{\odot} \sim 12.9 $ at $z\sim5$. 
Similarly, the stellar mass peak $M_*^{\rm peak}$ increases with redshift from log$\,M_*^{\rm peak}/M_{\odot} \sim 10.5$ at $z\sim0.3$ to log$\,M_*^{\rm peak}/M_{\odot} \sim 10.9$ at $z\sim3$. The SHMR ratio at the peak halo mass remains almost constant with redshift. These  results are in accordance with the scenario in which the peak of star-formation efficiency moves toward more massive halos at higher redshifts.
We also measured the fraction of satellites as a function of stellar mass and redshift. For all stellar mass thresholds, the satellite fraction decreases at higher redshifts. At a given redshift, there is a higher fraction of low-mass satellites and this fraction reaches a plateau at $\sim 25 \%$ at $z\sim1$.  The satellite contribution to the total stellar mass budget in halos becomes more important than that of the central at halo masses of about $M_h > 10^{13} \, M_{\odot}$ and always stays below the peak, indicating that quenching mechanisms are present in massive halos that keep the star-formation efficiency low.
 Finally, we compared our results with three hydrodynamical simulations: {\sc Horizon-AGN}, {\sc TNG100} of the {\sc IllustrisTNG} project, and {\sc EAGLE}. We find that the most significant discrepancy is at the high-mass end, where the simulations generally show that satellites have a higher contribution to the total stellar mass budget than the observations. This, together with the finding that the fraction of satellites is higher in the simulations, indicates that the feedback mechanisms acting in both group- and cluster-scale halos appear to be less efficient in quenching the mass assembly of satellites -- and that quenching occurs much later in the simulations. 
}


\titlerunning{SHMR up to $z\sim5$ in COSMOS2020}
\authorrunning{Shuntov et al.}

\maketitle


\section{Introduction} 
\label{sec:intro}


Within the current paradigm of structure and galaxy formation, galaxies form and evolve within dark matter halos \citep{1978MNRAS.183..341W}. Their properties are inextricably connected in what is known as the galaxy-halo connection \citep[see e.g.,][for a review]{wechsler_connection_2018}. One facet of the galaxy-halo connection is the relationship between the mass of the dark matter halo and the stellar mass of the galaxy it hosts, referred to as the stellar-to-halo mass relation (SHMR). The SHMR expresses the efficiency of the stellar mass assembly of a galaxy integrated over the halo's lifetime and it is, at first-order, a function of the halo mass and is shaped by the physical mechanisms of galaxy formation \citep[e.g.,][for a review]{somerville_physical_2015}. 


Galaxy formation is an inefficient process: the ratio of stellar mass to halo mass, $M_*/M_h$, is quite low \citep{Shankar2006, 2006MNRAS.368..715M, zheng_galaxy_2007, conroy_connecting_2009, behroozi_comprehensive_2010}.  This quantity is a strong function of halo mass and rises to a peak at a characteristic peak halo mass, suggesting that at most only $20\%$ of all the available baryons in the halo have turned into stars. At lower and higher halo masses, the $M_*/M_h$ ratio decreases rapidly, which is seen as a signature of different feedback processes that suppress star formation and act at different halo mass scales: stellar feedback in low-mass halos and active galactic nuclei (AGN) in high-mass halos \citep[e.g.,][for a review]{silk_current_2012}.

Central and satellite galaxies contribute to the total stellar mass content of halos. As a consequence, the total SHMR can be decomposed in the contributions from both. In lower mass halos, the central galaxy makes up most of the stellar mass content, but its growth is regulated by stellar feedback mechanisms such as supernovae (SNe), stellar winds, radiation pressure, and photoheating. All of these mechanisms are important in explaining the observed amount of stellar mass in lower mass galaxies; otherwise, the galaxy masses end up overpredicted \citep{hopkins_stellar_2012}. On the other hand, in cluster-scale halos, the satellite galaxies dominate the stellar mass budget, mostly due to their high number \citep{leauthaud_new_2012, coupon_galaxy-halo_2015}. Within large halos, the stellar mass assembly in satellites is described by interplay between the hierarchical merger tree, where smaller halos accrete into larger ones and in situ star formation facilitated by cold gas inflows and the various environmental quenching mechanisms that act to suppress further growth \citep{peng_mass_2010, gabor_hot_2015}. These include the so called "hot-halo," where the infalling gas is heated by virial shock heating \citep{birnboim_virial_2003}, as well as mechanisms such as strangulation \citep{1980ApJ...237..692L, 2000ApJ...540..113B}, ram-pressure stripping \citep{1972ApJ...176....1G}, and harassment \citep{moore_galaxy_1996}. In such massive halos, AGN feedback keeps the gas hot through the so-called "radio" mode and is necessary for explaining the break in the local galaxy luminosity/stellar mass function at the bright/massive end \citep{croton_many_2006, Bower2006}. All of these factors shape the total (central $+$ satellite) SHMR, and its evolution with redshift can indicate the relative importance of these processes in shaping the star-formation efficiency as a function of the halo mass. 


Some of the most predictive models that give direct insight into the physical processes that shape the galaxy-halo connection come from hydrodynamical simulations \citep[e.g.,][]{vogelsberger_introducing_2014,dubois_dancing_2014,schaye_eagle_2015,pillepich_first_2018}.
However, the known physics implemented in the simulations imposes a strong prior on the galaxy evolution and cannot provide information about physical processes that had not been previously expected to contribute to the galaxy-halo connection \citep{behroozi_most_2018}. Additionally, an important caveat of simulations is that they cannot simulate the full physics at all scales, so they rely on various parametrizations below the resolution scale, namely, subgrid physics. Subgrid physics varies from one simulation to another, and so do its conclusions on the galaxy-halo connection. Even though hydrodynamical simulations are the most predictive, they are computationally expensive, which  makes it difficult for them to be constrained against observations using techniques such as Markov chain Monte Carlo (MCMC). 

Empirical models offer the most flexibility \citep{wechsler_connection_2018}. Many methods have been described in the literature to compute the SHMR from observational data sets or numerical simulations. One of the most widely employed techniques is abundance matching (AM), where the abundance (i.e., number density in a comoving volume) of galaxies above a given mass is matched to the abundance of dark matter (sub)halos which then gives the halo mass \citep{marinoni_mass--light_2002,kravtsov_dark_2004,2004MNRAS.353..189V, 2004ApJ...614..533T}.
Another method based on a statistical description of the way galaxies populate halos is the halo occupation distribution (HOD) model. The HOD can model one- and two-point statistical observables such as the galaxy stellar mass function (GSMF) and galaxy clustering as measured by the correlation function (2PCF) \citep[e.g.,][]{peacock_halo_2000,seljak_analytic_2000,2001ApJ...546...20S,berlind_halo_2002}. 
Within the HOD formalism, we can also model observables that directly probe dark matter halos, such as galaxy-galaxy lensing \citep{leauthaud_theoretical_2011}. In galaxy-galaxy lensing, one measures the subtle coherent distortions of the shapes of background galaxies induced by the foreground matter distribution to probe the latter. However, it relies on accurate measurements of galaxy ellipticities, which becomes increasingly difficult at $z>1$ and poses a limitation with regard to its implementation in probing a redshift evolution over a vast interval.

The phenomenological approaches of AM and HOD are based on  statistical measures (e.g., GSMF and 2PCF) that indirectly probe the dark matter halos and are agnostic with regard to the physical processes that shape their relation with the galaxies. An additional drawback of these phenomenological models is that they rely on accurate knowledge of the halo mass function, which has to be calibrated using numerical simulations; in addition, they are sensitive to various definitions of the halo profile, radius, mass, concentration, and bias. However, this type of modeling is not constrained by strong priors from the physics of galaxy evolution -- it is almost entirely constrained by observations and, as such, they can reveal signatures of new physical processes that shape galaxy properties. 

This paper aims to constrain the redshift evolution of the SHMR up to $z\sim5$ by applying an HOD-based analysis consistently on a homogeneous data set: the COSMOS2020 photometric catalog. COSMOS2020 \citep{weaver_cosmos2020_2021} is the latest iteration of the photometric catalog of the COSMOS \citep{scoville_cosmic_2007} survey that includes the latest data-releases of deep imaging, covering wavelengths from the ultraviolet to the near-infrared. The deep multi-band photometry allows for the estimation of accurate photometric redshifts and stellar masses, along with a selection of complete samples up to high redshift. 
By adopting an HOD-based model to jointly fit for galaxy abundance and clustering, our analysis is aimed at constraining the satellite contribution to the total stellar mass budget in halos across a vast redshift range. This allows us to infer a coherent picture of how the stellar mass assembles as a function of the halo mass throughout cosmic history.

The novel aspect of our work is comprised by the use of a single dataset to perform all the measurements and probe the SHMR to $z\sim 5$ for both central and satellite galaxies. Most of the investigations in the literature have relied on observables from heterogeneous data sets to constrain their models \citep[e.g.,][]{behroozi_average_2013, moster_emerge_2018, behroozi_universemachine_2019}. Different data sets can have different selection functions and methods of estimating galaxies' physical parameters that have the capacity to propagate various systematic biases, which may muddle the interpretations \citep{behroozi_comprehensive_2010}. Therefore, our work is free from such "inter-observational" systematics.

Our work builds up on the literature in several ways. \cite{legrand_cosmos-ultravista_2019} is the only work that has measured the SHMR using a single data set \citep[COSMOS2015 of][]{laigle_cosmos2015_2016} up to $z\sim 5$ using sub-halo abundance matching. A shortcoming of this approach is the satellites are treated as centrals in their own sub-halo so that they only predict the SHMR of central galaxies. Our approach allows for centrals and satellites to be decoupled in order to compute the contribution of both to the total mass content of halos. Previous works that have measured both central and satellite SHMR are limited only to $z<1$ (e.g., \citealp{leauthaud_new_2012, coupon_galaxy-halo_2015}), or a single $z$-bin measurement at $2<z<3$ as in \cite{cowley_stellar--halo_2019}. Therefore, this paper presents the only measurement of the SHMR for both centrals and satellites up to $z\sim5$ using a homogeneous dataset: COSMOS2020.


The organization of this paper is as follows. In Section \ref{sec:data}, we describe the COSMOS dataset  applied here and the mass-selected samples in the ten redshift bins comprised by our analysis. In Section \ref{sec:measurements}, we present the methods we employ to perform our measurements of galaxy abundance and clustering. In Section \ref{sec:modeling}, we lay out the HOD-based modeling of our observables with a parametrization of the SHMR as a starting point. In Section \ref{sec:results}, we present our measurements of the observables and the results on the redshift evolution of the HOD and SHMR. In Section \ref{sec:discussion}, we discuss the physical mechanisms that may regulate the growth of central and satellite galaxies in dark matter halos. We also compare our results with hydrodynamical simulations and discuss the possible origins of the discrepancies that we find. Finally, we summarize our finding in Section \ref{sec:summary-conclusion}

Throughout this paper we adopt a standard $\Lambda$CDM cosmology with $H_0=70$\,km\,s$^{-1}$\,Mpc$^{-1}$, $\Omega_{\rm m,0}=0.3$, where $\Omega_{\rm b,0}=0.04$, $\Omega_{\Lambda,0}=0.7$, $\sigma_8 = 0.82$ and $n_s = 0.97$. Galaxy stellar masses, when derived from spectral energy distribution (SED) fitting, scale as the square of the luminosity distance (i.e., $D_L^2$), therefore, as $h^{-2}$; dark matter halo masses, usually derived from dynamics in numerical simulations, scale as $h^{-1}$. The $h$ scaling factors are retained implicitly for all relevant measurements, unless explicitly noted otherwise \citep[see][for an overview of $h$ and best practices]{croton_damn_2013}. When making comparisons to the literature, we rescale all the measurements to the cosmology adopted for this paper. All magnitudes are expressed in the AB system \citep{1983ApJ...266..713O}. Stellar masses are obtained assuming  \cite{chabrier_galactic_2003} initial mass functions (IMF) and when comparing to the literature, stellar masses are rescaled to match the IMF adopted in this paper.

\section{Data} \label{sec:data}
\subsection{COSMOS2020 catalog}

This work makes use of the COSMOS2020 catalog \citep{weaver_cosmos2020_2021}. This deep multi-wavelength near-infrared selected catalog uses deep observations over the  $2\, \rm deg^2$ COSMOS field in 35 photometric bands from ultraviolet (UV) to near-infrared (NIR). This unique combination of depth, area, and wavelength coverage allows an  accurate estimation of photometry, photometric redshifts, and stellar masses for around a million sources up to $z\sim10$. 

Briefly, COSMOS2020 comprises two photometric catalogs extracted on a "chi-squared' combination \citep{szalay_simultaneous_1999} of deep near-IR images in $izYJHK_s$ (AB mag $3\sigma$ depth in $2\arcsec$ apertures of $27.6,\, 27.2,\, 25.3,\, 25.2,\, 24.9,\, 25.3$). One catalog uses the traditional approach of measuring fluxes in fixed  apertures using \texttt{SExtractor} \citep{bertin_sextractor_1996}, while the second uses a profile-fitting technique using \texttt{The Farmer} (Weaver et al. in prep) built around \texttt{The Tractor}. Photometric redshifts (photo-$z$s), stellar masses, and other physical parameters are estimated for these two photometric catalogs using two spectral energy distribution (SED) fitting codes \texttt{LePhare} \citep{arnouts_measuring_2002, ilbert_accurate_2006} and \texttt{EAZY} \citep{brammer_eazy_2008}. 

In this analysis, we use the \textsc{Classic} catalog with photometric redshifts estimated using \texttt{LePhare}. This is principally because photometric measurements carried out with \texttt{The Farmer}, which is based on a model fitting technique, can fail to converge in certain cases, especially in crowded regions or near bright sources. This spatially variable completeness is problematic for measurements of the angular correlation function; we have measured relative correlation function differences of up to about $30 \%$ on scales of $1 \arcmin$ between the two catalogs. The \textsc{Classic} catalog, on the other hand, contains photometric measurements for almost all sources within the survey masks. There is a caveat, however, since aperture photometry is not very reliable in crowder regions and around bright sources neither. So whereas \textsc{Farmer} is a pure catalog, since it photometers all the reliable sources, \textsc{Classic} is a more complete catalog, photometrying almost all sources.

Compared to COSMOS2015, we know that COSMOS2020 reaches similar photometric redshift  precision almost one magnitude fainter -- this is shown in Fig. 17 of \cite{weaver_cosmos2020_2021}, where the $1\sigma$ uncertainty of the photo-$z$s is plotted as a function of redshift and magnitude bin. The normalized median absolute deviation ($\sigma_{\text{NMAD}}$\footnote{$\sigma_{\text{NMAD}} = 1.48 \times \text{ median}[(|\Delta z - \text{median}(\Delta z)|)/(1+z_{\rm spec})]$;  $\Delta z = \zphot - z_{\rm spec}$}) at $i < 22.5$ is below $0.01\,(1+z)$ and stays below $0.05\,(1+z)$ to $25 < i < 27$. The outlier fraction\footnote{defined as $|\Delta z| > 0.15 (1+z_{\rm spec})$} is below $1\%$ and $20\%$ for the corresponding magnitude bins. The bias\footnote{defined as median$(\Delta z)$} ranges from $-0.003$ to $-0.014$ in the bright and faint magnitude bins, respectively. 

Accurate and complete stellar mass estimates over a wide redshift range are necessary for obtaining an accurate measurement of the SHMR. In COSMOS2020, this is enabled by the inclusion of the deep near-IR data from the UltraVISTA survey \citep{mccracken_ultravista_2012}, DR4 in $YJHK_s$, and mid-IR data from the Cosmic Dawn Survey \citep{moneti_euclid_2021}, as well as \textit{Spitzer}/IRAC observations in channels 1-2 ($3.6\, \mu \rm m,\; 4.5\, \mu \rm m$). Stellar masses are estimated with \texttt{LePhare} using SED templates produced from stellar population synthesis models by \cite{bruzual_stellar_2003} and initial mass functions by \cite{chabrier_galactic_2003}. The SEDs are fixed at $z=\zphot$ then and fitted to the multi-wavelength photometry (for more details, see Weaver et al. and \citealt{laigle_cosmos2015_2016}). The point estimate of the stellar mass is the median of the resulting PDF marginalized over all other parameters, with the $16^{\rm th}$ and $84^{\rm th}$ percentiles of the PDF giving the $1\sigma$ confidence interval. The improved depth (e.g., $K_s = 25.3$, $[3.6] = 26.4$ at 3$\sigma$) translates to higher stellar mass completeness compared to the previous versions of the catalog. This enables a selection of samples based on stellar mass that are complete down to log $M_*/M_{\odot} \sim 8.2$ at $z\sim0.3$ and log $M_*/M_{\odot} \sim 9.3$ at $z\sim4$.

Throughout this work, we used $2 \arcsecond$ aperture magnitudes and apply aperture-to-total and Milky Way extinction corrections using the \cite{schlafly_measuring_2011} dust map. We applied masks to remove sources near bright stars and in regions near artefacts. This leaves an effective area of $1.27\, \rm deg^2$ that corresponds to the footprint of the UltraVISTA survey\footnote{In the catalog this is selected using the keyword \texttt{FLAG\_COMBINED}}.

Finally, we use two star-galaxy classifications to remove uncorrelated stars from the catalog. One uses morphological information from \textit{HST}/ACS and Subaru/HSC images, where half-light radii and magnitudes classify as stars all point-like sources at $i<23$ and $i<21.5$ in ACS and HSC images, respectively. This criterion is also satisfied by point-like AGN sources. The second, which is an SED-based criterion, classifies as stars those sources with the $\chi^2$ of the best-fit stellar template lower than the $\chi^2$ of the best-fit galaxy template (for more details see Weaver et al). We performed tests by measuring the correlation function of sources classified as stars, while further removing point-like AGN sources based on their $\chi^2$. The correlation function of our stellar sample is zero, indicating a clean separation. 

\subsection{Sample selection} \label{sec:sample-selection}
Measuring galaxy clustering and abundance requires complete stellar mass-selected and volume-limited samples. To select stellar mass-complete samples we use the stellar mass completeness limit, which is computed following the method prescribed by \cite{pozzetti_zcosmos_2010} and is described in catalog paper of \cite{weaver_cosmos2020_2021}. 
To ensure complete samples, the channel 1 limiting magnitude $[3.6]_{\rm lim} = 26$ is computed with the help of the deeper CANDELS-COSMOS catalog \citep{nayyeri_candels_2017} which is used for completeness check. All samples throughout this work are selected to be brighter than $[3.6]_{\rm lim} = 26$ across the full redshift range, despite the fact that a $K_s$ based selection is also suitable at low redshifts. 

\begin{figure}
    \centering
    \includegraphics[width=1\columnwidth]{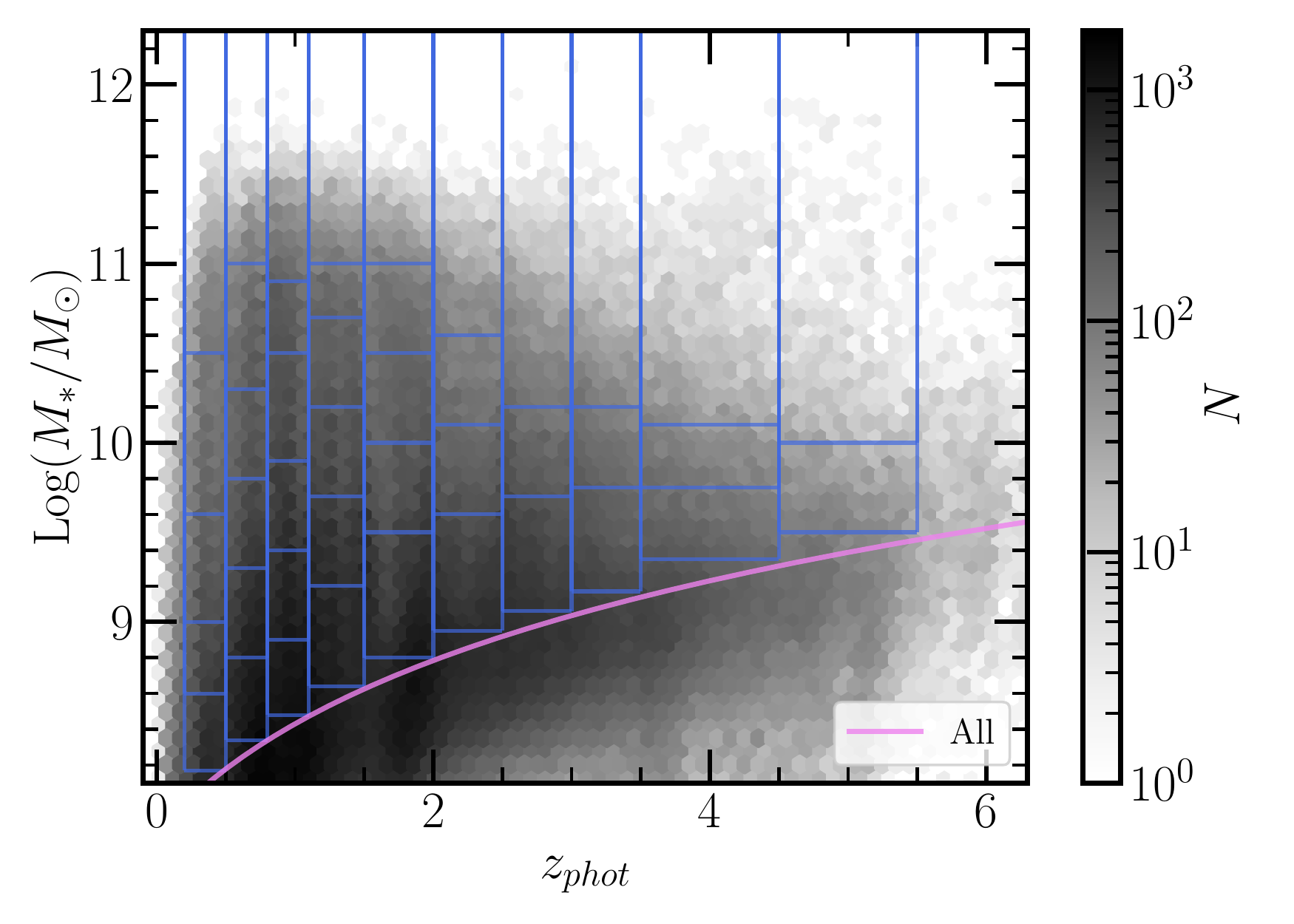}
    \caption{Sample selection in the stellar mass-redshift plane. The solid grid lines show the mass threshold for each sample. The solid violet curve indicates the stellar mass completeness limit. We note that the histogram does not correspond to the final selection since further selection criteria (e.g., based on ${\rm PDF}(z)$ width) are applied.}
    \label{fig:M-vs-z}
\end{figure}

To probe the cosmic evolution of the observables we bin the samples in ten redshift bins from $z = 0$ to $z = 5.5$ with varying widths. The widths were chosen to ensure roughly the same number of galaxies in each bin. The redshift bins are listed in Table \ref{tab:SampleSelection}. Additionally, we require that each galaxy has its lower and upper $1\sigma$ values ($z_{\rm low}$, $z_{\rm up}$) within $\pm 0.5$ of the $z$-bin limits. This criterion removes any galaxies with highly uncertain redshifts that can introduce errors. To ensure that the samples remain as complete as possible, we don't impose any other selection criteria, for example, based on $S/N$, number of bands in which a source was photometered, $\chi^2$ of the SED fit etc.

One of the ingredients of the model of the galaxy correlation function is the redshift distribution $N(z)$. We used the $z$-likelihood from \texttt{LePhare} to build $N(z)$. Formally, for each galaxy, there is a likelihood of the observed photometry (denoted by the vector of fluxes $\vect{o}$) given the redshift $\mathcal{L}(\vect{o}|z)$. $N(z)$ is then constructed by simply stacking the individual $\mathcal{L}(\vect{o}|z)$ in each $z$-bin:
\begin{equation}
    N(z) = \displaystyle\sum_{i}^{N_{\rm sample}} \mathcal{L}_i(\vect{o}|z),
\end{equation}
where the sum is over the number of objects of the redshift and stellar mass threshold-selected sample $N_{\rm sample}$.

\begin{figure}
    \centering
    \includegraphics[width=1\columnwidth]{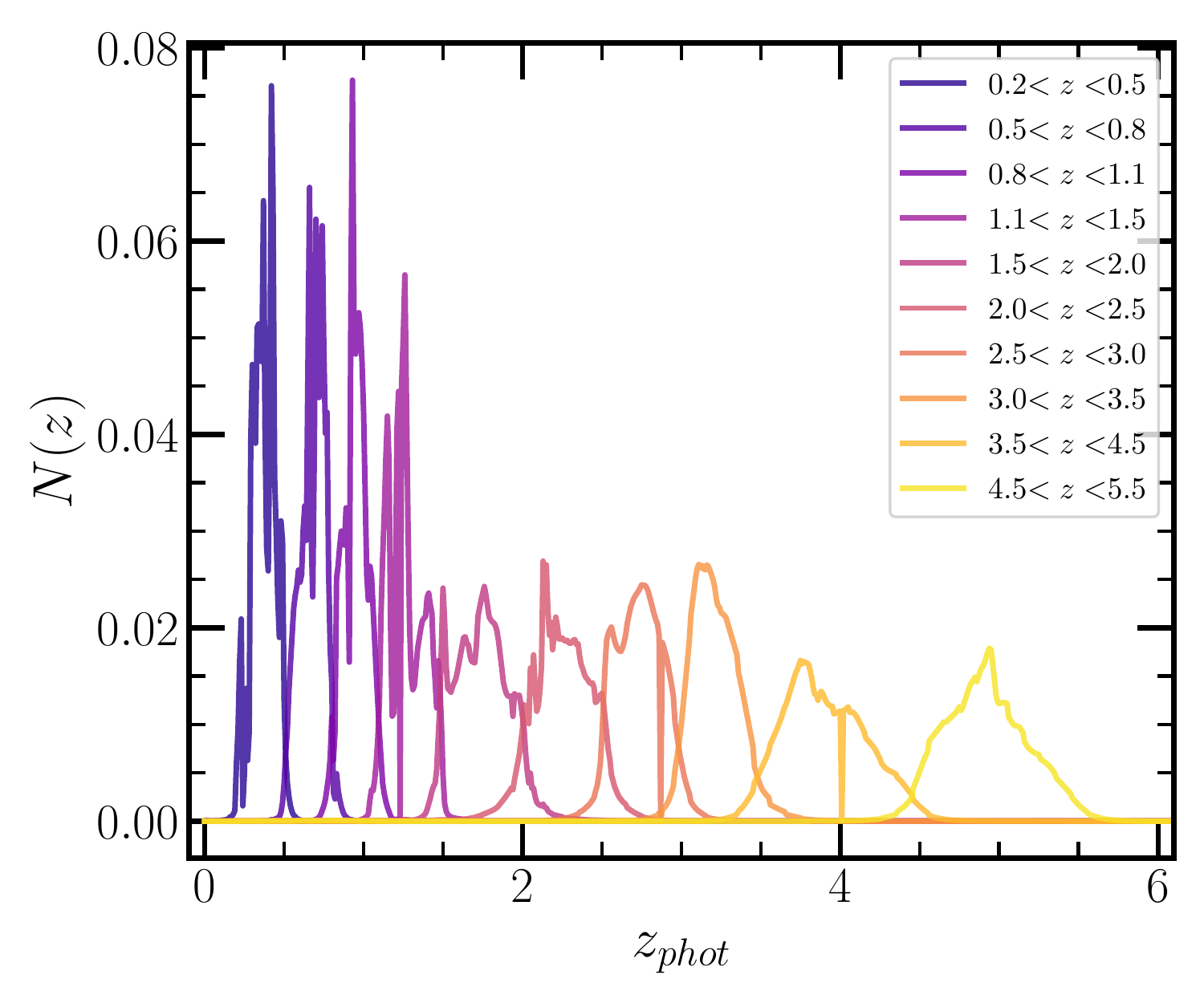}
    \caption{Redshift distribution of the ten galaxy samples used for the clustering and abundance measurements. The redshift distribution is obtained by stacking the posterior photo-$z$ distributions for all the sources in a given bin as described in Section \ref{sec:sample-selection}. 
    }
    \label{fig:nofz}
\end{figure}

Constructing the $N(z)$ in this way directly accounts for the uncertainty in the photometric redshift when selecting galaxies in a bin, namely, the fact that galaxies can still have their true $z$ outside the bin limits. However, it has been shown by \cite{ilbert_euclid_2021} that such a procedure can lead to biases in the mean redshift that can be inferred from the $N(z)$. They have quantified the notion that these biases can reach up to $\sim 0.01 \, (1+z)$. By using the model of $w(\theta)$ (described in Section \ref{sec:woth-model}) we tested the effects of bias in the mean redshift, as well as different estimates of $N(z)$. Our conclusion is that biases on an order of magnitude of $\sim 0.01 \, (1+z)$ result in relative differences in the value of $w(\theta)$ of less than about $3\%$, which is considerably smaller than the typical relative error of the measurement (about $10\%)$. On the other hand, the shape of $N(z)$ (notably, the width of its wings) can lead to significant differences in $w(\theta)$ that can bias the inferred SHMR parameters. The reason for this is the mix of physical scales when considering larger volumes: the angular correlation of galaxies selected in a wider radial interval is inevitably lower, since they can be far apart in the radial direction but close in angular separation. For example, considering $N(z)$ to be a Gaussian distribution centered at the $z$-bin mean and with width half of the $z$-bin width, can lead to a relative difference of about $20\%$ at $z\sim 0.4$ and more than $50\%$ at $z\sim2.5$ (see appendix \ref{apdx:nofz-on-w}). \cite{ilbert_euclid_2021} have shown that the $\mathcal{L}_i(\vect{o}|z)$ as output from \texttt{LePhare} can lead to biased and miss-calibrated $N(z)$ as evidenced by comparison with the true redshift histogram and the Probability Integral Transform (PIT) statistic \citep[see Fig. 4 in][]{ilbert_euclid_2021}. The authors show that an $N(z)$ that is better representative of the true distribution can be obtained by using a posterior distribution, such as:
\begin{equation} \label{eq:nofz-posterior}
    N(z) = \displaystyle\sum_{i}^{N_{\rm sample}} \mathcal{P}_i(\vect{o}|z) = \displaystyle\sum_{i}^{N_{\rm sample}} \mathcal{L}_i(\vect{o}|z) \, \text{Pr}(z|m_0),
\end{equation}
where $\text{Pr}(z|m_0)$ is the so called `photo-$z$ prior' \citep{brodwin_canada-france_2006}. This prior can be constructed for every $z$-bin by summing the likelihoods per magnitude bins such as:
\begin{equation} \label{eq:photz-prior}
    \text{Pr}(z|m_0) = \displaystyle\sum_{i}^{N_{z-\rm bin}} \mathcal{L}_i(\vect{o}|z) \, \Theta(m_{0,i}|m_0),
\end{equation}
where $\Theta(m_{0,i}|m_0)$ is equal to $1$ if the object's magnitude $m_{0,i}$ is within the magnitude bin centred at $m_0$ and zero otherwise. The outcome of this procedure and the effects on $w(\theta)$ are presented in more detail in Appendix \ref{apdx:nofz-on-w}. We adopt the $N(z)$ obtained using Eq. \ref{eq:nofz-posterior} for our analyses. The $N(z)$ is constructed for each considered sample including the mass threshold-selected samples. The resulting distributions for all $z$-bins for galaxies above the mass-completeness limit are shown in Fig. \ref{fig:nofz}. We note that there are some narrow (of a width of $\sim0.01$) dips at several $z$-values (e.g., $z\sim 1.3,~2.9,~4.0$). These come from the individual likelihoods being close to zero at these exact values. The template fitting outputs minimal likelihoods at exactly these three $z$-points, and as they are narrow ($\sim0.01),$  they do not affect the selection nor the modeling.


To probe the clustering strength as a function of galaxy mass, we further select samples in stellar mass thresholds. In each $z$-bin we define samples selected in several stellar mass thresholds starting from the mass completeness limit; these are indicated by the horizontal solid lines in Fig. \ref{fig:M-vs-z} and listed in Table \ref{tab:SampleSelection}. We chose the thresholds rather arbitrarily to ensure a good signal-to-noise (S/N) for the clustering measurement of each mass threshold-selected sample within a $z$-bin. 

\begin{table}
\centering
\caption{Sample selection in redshift and stellar mass thresholds. The columns indicate the redshift bin, mass threshold, median mass, and number of galaxies in for each sample used to measure clustering. }
\footnotesize
\renewcommand{\arraystretch}{1.1}
\setlength{\tabcolsep}{5pt} 
\begin{threeparttable}
\begin{tabular}{cccc}
\hline\hline
$z$-bin & log$(M_*^{\rm threshold}/M_{\odot})$ & Median log$(M_*/M_{\odot})$  & $N$ \\
\hline

$0.2 < z < 0.5$ & $8.17$ & $8.86$ & $23\,346$   \\
                & $8.60$ & $9.25$ & $15\,234$   \\
                & $9.00$ & $9.64$ & $10\,000$   \\
                & $9.60$ & $10.16$ & $5\,229$   \\
                & $10.50$ & $10.76$ & $1\,495$   \\
\hline
$0.5 < z < 0.8$  & $8.34$ & $8.99$ & $43\,752$  \\
                 & $8.80$ & $9.40$ & $27\,104$  \\
                 & $9.30$ & $9.90$ & $15\,325$  \\
                 & $9.80$ & $10.32$ & $8\,588$  \\
                 & $10.30$ & $10.65$ & $4\,454$  \\
                 & $11.00$ & $11.15$ & $736$  \\
\hline
$0.8 < z < 1.1$  & $8.48$ & $9.18$ & $50\,964$  \\
                 & $8.90$ & $9.55$ & $34\,306$  \\
                 & $9.40$ & $10.02$ & $20\,211$  \\
                 & $9.90$ & $10.43$ & $11\,683$  \\
                 & $10.50$ & $10.80$ & $5\,167$  \\
                 & $10.90$ & $11.07$ & $1\,864$  \\
\hline
$1.1 < z < 1.5$  & $8.64$ & $9.33$ & $53\,285$  \\
                 & $9.20$ & $9.78$ & $30\,882$  \\
                 & $9.70$ & $10.22$ & $17\,095$  \\
                 & $10.20$ & $10.58$ & $8\,823$  \\
                 & $10.70$ & $10.91$ & $3\,239$  \\
                 & $11.00$ & $11.15$ & $1\,139$  \\
\hline
$1.5 < z < 2.0$  & $8.80$ & $9.41$ & $47\,100$  \\
                 & $9.50$ & $10.03$ & $20\,941$  \\
                 & $10.00$ & $10.43$ & $10\,910$  \\
                 & $10.50$ & $10.77$ & $4\,745$  \\
                 & $11.00$ & $11.14$ & $936$  \\
\hline
$2.0 < z < 2.5$  & $8.95$ & $9.45$ & $31\,247$  \\
                 & $9.60$ & $10.01$ & $12\,205$  \\
                 & $10.10$ & $10.47$ & $5\,209$  \\
                 & $10.60$ & $10.82$ & $1\,869$  \\
\hline
$2.5 < z < 3.0$  & $9.06$ & $9.54$ & $25\,660$  \\
                 & $9.70$ & $10.04$ & $9\,625$  \\
                 & $10.20$ & $10.48$ & $3\,475$  \\
\hline
$3.0 < z < 3.5$  & $9.17$ & $9.62$ & $15\,018$  \\
                 & $9.75$ & $10.06$ & $5\,739$  \\
                 & $10.20$ & $10.43$ & $2\,059$  \\
\hline
$3.5 < z < 4.5$  & $9.35$ & $9.70$ & $12\,249$  \\
                 & $9.75$ & $10.02$ & $5\,569$  \\
                 & $10.10$ & $10.35$ & $2\,247$  \\
\hline
$4.5 < z < 5.5$  & $9.50$ & $9.79$ & $3\,374$  \\
                 & $10.00$ & $10.24$ & $932$  \\

\hline





\hline
\end{tabular}
\end{threeparttable}
\label{tab:SampleSelection}
\end{table}

\section{Measurements} \label{sec:measurements}

\subsection{Galaxy clustering} \label{sec:clustering}
We measured the two-point angular correlation function $w(\theta)$ using the \cite{landy_bias_1993} estimator\footnote{The correlation functions are computed using the \texttt{TreeCorr} code \citep{jarvis_skewness_2004}}
\begin{equation}
    w(\theta) = \dfrac{DD - 2\,DR + RR}{RR},
\end{equation}
where $DD$ is the number of data-data pairs in a given angular separation bin $[\theta, \theta+\delta \theta]$, $RR$ is the number of pairs in the random catalog in the same bin, and $DR$ is the number of pairs between the data and the random catalog. The data and random pairs are normalized by the total number of galaxies and random objects. We constructed the random catalog with the same survey geometry mask as the data catalog and $N_{\rm random} \sim 3\times10^6$, which is more than $50$ times the number of galaxies in each considered bin.

The covariance matrix is computed using the jackknife method by subdividing the full area in $N_{\rm patch} = 22$ patches and recomputing the correlation function removing one patch at a time. The covariance matrix is then estimated as:
\begin{equation}
    C_{lk} = \dfrac{N_{\rm patch} -1 }{N_{\rm patch}} \displaystyle\sum_{i=1}^{N_{\rm patch}}\left(w_i(\theta_l) - \Bar{w}\right)^T\left(w_i(\theta_k) - \Bar{w}\right),
\end{equation}
where $\Bar{w}$ is the mean correlation function and $w_i$ is the correlation function with the $i$th patch removed. The final covariance matrix thus includes uncertainties due to sample (cosmic) variance, dominating at larger scales, and due to Poisson statistics from counting objects in bins dominating mostly at small scales. However, due to the limited survey size, this method still cannot accurately estimate the uncertainties due to cosmic variance. To capture the cosmic variance effects in the covariance matrix one can use computationally expensive simulations, which is out of the scope of this work. This, for example, is done in \cite{leauthaud_theoretical_2011}, where they show that cosmic variance has an impact on the covariance matrix on large scales (i.e., in the two-halo regime), namely, as the cosmic variance increases the correlation in the data at large scales.


Due to the limited size of the survey ($1.27\, \rm deg^2$), the clustering measurements suffer from the effects of the integral constraint  \citep[IC,][]{groth_statistical_1977}. This leads to an underestimation of $w(\theta)$ at large scales comparable to angular size of the survey by a constant factor $w_{\rm IC}$, such that the true correlation function is:
\begin{equation}
    w(\theta) = w_{\rm meas}(\theta) + w_{\rm IC}.
\end{equation}
We incorporate this correction into our model, which is described in detail later in this work.

\subsection{Galaxy abundance} \label{sec:smfs}
We measured stellar mass functions across ten redshift bins using the $1/V_{\rm max}$ technique \citep{schmidt_space_1968}. This estimator weighs each galaxy by the maximum volume in which it would be observed given the redshift range of the sample and magnitude limit of the survey. 
The comoving volume $V_{\rm max}$ is computed between $z_{\rm min}$ and $z_{\rm max}$, where  $z_{\rm min}$ is the lower redshift limit of the $z$-bin and $z_{\rm max} = \text{min}(z_{\rm bin,up}, z_{\rm lim})$, then $z_{\rm bin,up}$ is the upper redshift limit of $z$-bin, and $z_{\rm lim}$ is the maximum redshift up to which a galaxy of a given magnitude can be observed given the magnitude limit of the survey. For this purpose, we used IRAC channel-1 magnitudes and a limit of $[3.6] = 26$. 
We computed the SMFs in the mass range starting from the mass completeness limit of each $z$-bin, with a bin size of $\Delta \text{log} M = 0.25$.

Uncertainties in the SMF include contributions from Poisson noise ($\sigma_{\rm Pois}$), cosmic variance ($\sigma_{\rm cv}$), and SED fitting uncertainties ($\sigma_{\rm fit}$). We computed the uncertainties due to cosmic variance following \cite{Steinhardt2021}. The starting point is the cosmic variance "cookbook" code of \cite{moster_cosmic_2011}, that computes the stellar-mass dependent cosmic variance and performs well up to intermediate redshifts and masses but becomes increasingly underestimated at high redshift and mass. \citeauthor{Steinhardt2021} extend the recipes to the early universe ($z>3$) by using linear perturbation theory.

Additionally, due to photometric errors and degeneracies in the SED fits, there are uncertainties in the $M_*$ measurements that propagate to the SMF. To estimate SED fitting uncertainties on the SMF we use the ${\rm PDF}(M_*)$. We assign to each galaxy a weight that corresponds to its probability to be found in the given mass bin:
\begin{equation}
    w_i = \dfrac{1}{\displaystyle\int\limits_{{\rm bin}} \diff M_*\;{\rm PDF}_i(M_*)}.
\end{equation}
We then used this weight to compute the combined Poisson and SED fitting uncertainties in the following way:
\begin{equation}
    \sigma_{\rm Pois + fit} = \sqrt{\displaystyle\sum_{i}^{N_{\rm g}} \dfrac{w_i^2}{V_{\text{max},i}^2}}.
\end{equation}
The final uncertainty is then $\sigma_{\Phi}^2 = \sigma_{\rm Pois + fit}^2 + \sigma_{\rm cv}^2 $. The measurements for the clustering and SMF are presented in Fig. \ref{fig:measures-fit} and discussed in Section \ref{sec:meas-and-fits}

\section{Theory and modeling} \label{sec:modeling}
The observables of galaxy clustering and abundance can be predicted within the framework of the halo model.  The halo model of the large scale structures postulates that all matter in the universe is contained in virialized dark matter halos. Using halos as the basic unit, it provides a method to statistically describe the distribution of dark matter and analytically compute its clustering. Combined with a model that describes how galaxies populate halos -- including both central and satellite galaxies, one can model various statistics of the galaxy distribution. One of the principal advantages of this modeling is the capability to infer a range of properties about the satellite galaxies, which inferred in such a large redshift span is one of the novelties of our work. The power in the clustering of all galaxies on small scales and the number densities at the low mass end of the GSMF are both shaped by satellites. Therefore, by fitting the observed correlation function and GSMF with models that consider the satellites, one can infer the parameters that govern the statistical distribution of satellites within dark matter halos. We note that, the observations don't distinguish centrals from satellites.


\subsection{Stellar-to-halo mass relation}
The stellar mass assembled by a galaxy depends most strongly on the mass assembly of the host halo. Consequently,  a strong relationship between them is evident -- the stellar-to-halo mass relation (SHMR).
Our goal is to constrain the SHMR. \cite{leauthaud_theoretical_2011}, hereafter L11, laid the theoretical framework to model galaxy clustering and SMFs based on the HOD formalism with as a starting point a functional form of the SHMR. This function must capture the different growth rates of galaxies as a function of the halo mass that is shaped by various feedback processes that operate at different mass scales. \cite{behroozi_comprehensive_2010} presented a functional form of the SHMR, which we have adopted for our analysis:
\begin{equation} \label{eq:shmr-cent}
\begin{split}
    & \log \left( f^{-1}_{\rm SHMR}(M_*)\right) = \log (M_h) = \\
    & \log (M_1) + \beta \log \left( \dfrac{M_*}{M_{*,0}} \right) + \dfrac{\left( \dfrac{M_*}{M_{*,0}} \right)^{\delta}}{1+ \left( \dfrac{M_*}{M_{*,0}} \right)^{-\gamma}} - \frac{1}{2}.
\end{split}
\end{equation}
This relation is parameterized by a characteristic halo mass and stellar mass scales given by the parameters $M_1$ and $M_{*,0}$, respectively. Here, $M_1$ controls the normalization of $M_h$ as a function of $M_*$ and $M_{*,0}$ controls the position along the $M_*$-axis. Together, two parameters govern the transition mass scale between the low-mass and high-mass regime of the SHMR. The low-mass regime ($M_* \lesssim 10^{10.5} \, M_{\odot}$) is described by a power-law regulated by the parameter $\beta$. The high-mass regime
follows a sub-exponential law regulated by the parameter $\delta$, and the transition regime is shaped by the parameter $\gamma$.


The ratio between the stellar mass and the halo mass obtained with Eq. \ref{eq:shmr-cent}, (i.e., $M_*/M_h$) can be considered as the efficiency of the galaxy formation process that encapsulates all the processes that lead to the conversion of baryons to stars (from gas cooling and star formation to stellar and AGN feedback); this can be referred to as the baryon conversion efficiency or star formation efficiency (SFE). Since we can consider that baryonic matter content of halos is equal to the universal baryonic fraction $f_b = \Omega_{\rm b}/\Omega_{\rm m} \approx 0.16$, the $M_*/M_h$ ratio will inform us of the fraction of the baryons available in the dark matter (DM) halo that have converted into stars. At a given redshift, the $M_*/M_h$ ratio gives the baryon conversion efficiency integrated over the lifetime of the halo and therefore includes the combination of all the different physical processes that regulate star formation throughout the halo life (e.g., gas accretion, mergers, feedback). The shape of the $M_h/M_*$ ratio is a strong function of halo mass, which indicates that various feedback mechanisms operate on different halo mass scales to regulate star-formation. This is the principal quantity of interest in this work and we extensively discuss it in Section \ref{sec:results}.

As L11, we assume that this SHMR concerns only central galaxies occupying the central regions of the dark matter halos. These halos can contain smaller sub-halos that orbit the potential well and can host satellite galaxies. Satellites can undergo different stellar growth than centrals, since various distinct processes affect satellites in the dark matter halos that can regulate their growth (such as stripping and harassment). This will reflect in the SHMR, therefore a different relation for satellites is a more accurate assumption. This is one of the advantages of this methodology to infer the SHMR, as compared to the commonly implemented abundance matching (AM) technique. 
In AM, one treats satellites as centrals in their own sub-halo, which then assumes that centrals and satellites follow the same SHMR. The formalism that we employ is able to constrain the contribution of both centrals and satellites in the total stellar mass content in halos of given mass (Section \ref{sec:total-SHMR}).

Due to the effects of the various galaxy formation processes, there is a scatter of galaxy stellar mass that exists (as well as other galaxy properties) at a fixed halo mass. This stochastic nature of the SHMR can be modeled with a conditional function that describes the probability of observing a central galaxy with $M_*$ at a given $M_h$, which can be chosen to be a log-normal distribution. The dispersion of the log-normal distribution $\sigmaLogMs$ describes the scatter in stellar mass at fixed halo mass and is a free parameter that can be fitted with the data. Following previous works \citep[e.g.,][]{moster_constraints_2010, leauthaud_new_2012, tinker_evolution_2013}, we consider $\sigmaLogMs$ to be independent of the halo mass, and we leave it as a free parameter to be fitted in each $z$-bin. Hydrodynamical simulations, however, show that at $z=0$, $\sigmaLogMs$ generally decreases with $M_h$ going from $\sim 0.32$ at $M_h \sim 10^{11}$ to $\sim 0.15$ at $M_h \sim 10^{12}$ and staying constant to higher masses \citep{pillepich_first_2018}. We checked that varying $\sigmaLogMs$ over this range changes the clustering correlation function insignificantly (and within the measurements errorbars) and given the fact that this parameter is mostly constrained by the GSMF, taking $\sigmaLogMs$ independent of the halo mass is a safe assumption for our purposes.

\subsection{Central occupation distribution}

The HOD describes the statistical occupation of galaxies in dark matter halos. It assumes a probability distribution of the number of galaxies residing in halos conditioned on some criteria, usually on the mass $P(N|M_h)$. Typically, centrals are assumed to follow a Bernoulli distribution, while the number of satellites follows a Poisson distribution \citep[see e.g.,][and references therein]{zheng_theoretical_2005}. Under these assumptions, the HOD is described by the average number of galaxies with stellar masses higher than some threshold in halos of a given mass $\langle N(M_h|>M_*^{\rm th}) \rangle,$ 
\begin{equation} \label{eq:Ncent}
    \begin{split}
    & \left\langle N_{\rm cent}\left(M_h|>M_*^{\rm th}\right) \right\rangle = \\
    & \dfrac{1}{2} \left[ 1 - \text{erf} \left( 
    \dfrac{\left[\log(M_*^{\rm th}) - \log(f_{\rm SHMR} (M_h) )\right]}{\sqrt{2}\,\sigmaLogMs} \right) \right].
    \end{split}
\end{equation}
$\langle N_{\rm cent}\left(M_h|>M_*^{\rm th}\right) \rangle$ is a monotonic function increasing from $0$ to $1$. $f_{\rm SHMR}(M_h)$, whose inverse function is defined by Eq. \ref{eq:shmr-cent} and gives the stellar mass at the halo mass, while $\sigmaLogMs$ is the scatter in stellar mass at fixed halo mass. We note that all the other parameters regulating the central HOD parametrize the functional form of the SHMR. An alternative approach employed by many studies \citep[e.g.,][]{zheng_galaxy_2007, zehavi_galaxy_2011, coupon_galaxy_2012, mccracken_probing_2015, ishikawa_subaru_2020} specifies the central HOD assuming the SHMR to be a simple power law, with a parameter quantifying the minimum halo mass to host a galaxy $M_{\rm min}$. The downside of this model is the difficulty in the interpretation of $M_{\rm min}$ as the halo mass at the stellar mass threshold, that is, $M_{\rm min} = f^{-1}_{\rm SHMR}(M_*^{\rm th})$, especially at high masses where the deviation from a power-law of the SHMR is clear (see L11 for more details). 

\subsection{Satellite occupation distribution}

The occupation of halos by satellites can be modeled by a power-law at high halo masses with an exponential cut-off at low masses, given by:
\begin{equation} \label{eq:Nsat}
    \begin{split}
    & \left\langle N_{\rm sat}\left(M_h|>M_*^{\rm th}\right) \right\rangle = \\
    & \left\langle N_{\rm cent}\left(M_h|>M_*^{\rm th}\right) \right\rangle\, \left( \dfrac{M_h}{M_{\rm sat}} \right)^{\alpha_{\rm sat}}\, \exp \left( \dfrac{-M_{\rm cut}}{M_h} \right),
    \end{split}
\end{equation}
where $\alpha^{\rm sat}$ is the power-law slope, $M_{\rm sat}$ is the halo mass scale for the satellites defining the amplitude of the power law and $M_{\rm cut}$ is the cutoff scale. HOD studies have shown that the satellite mass scale is proportional to $f^{-1}_{\rm SHMR}$ at the threshold stellar mass \citep[e.g.][]{zheng_galaxy_2007, zehavi_galaxy_2011}. This allows us to parametrize $M_{\rm sat}$ and $M_{\rm cut}$ as power laws by introducing four more parameters:
\begin{equation}
    \begin{split}
    & \dfrac{M_{\rm sat}}{10^{12} M_{\odot}} = B_{\rm sat} \left( \dfrac{f^{-1}_{\rm SHMR}(M_*^{\rm th})}{10^{12} M_{\odot}} \right)^{\beta_{\rm sat}}, \\
    & \dfrac{M_{\rm cut}}{10^{12} M_{\odot}} = B_{\rm cut} \left( \dfrac{f^{-1}_{\rm SHMR}(M_*^{\rm th})}{10^{12} M_{\odot}} \right)^{\beta_{\rm cut}}.
    \end{split}
\end{equation}

The HOD is fully specified by the average occupation number of galaxies in halos, as given by Eq. \ref{eq:Ncent} \& \ref{eq:Nsat}. Finally, the total number of galaxies including centrals and satellites is simply $\langle N_{\rm tot} \rangle = \langle N_{\rm cent} \rangle + \langle N_{\rm sat} \rangle $. We can also compute the average number of galaxies in a mass bin of $M_*^{\rm th1} < M_* < M_*^{\rm th2}$ by simply taking the difference
\begin{equation}
    \begin{split}
    & \left\langle N_{\rm cent/sat}\left(M_h|M_*^{\rm th1}, M_*^{\rm th2}\right) \right\rangle = \\
    & \left\langle N_{\rm cent/sat}\left(M_h|>M_*^{\rm th1}\right) \right\rangle - \left\langle N_{\rm cent/sat}\left(M_h|>M_*^{\rm th2}\right) \right\rangle.
    \end{split}
\end{equation}

The model has a total of 11 parameters. The SHMR for the centrals has five parameters ($M_1$, $M_{*,0}$, $\beta$, $\delta$, $\gamma$) with one additional parameter that describes the scatter in stellar mass at a fixed halo mass of $\sigmaLogMs$. The occupation distribution for satellites is modeled with five parameters ($\alpha_{\rm sat}$, $B_{\rm sat}$, $\beta_{\rm sat}$, $B_{\rm cut}$, $\beta_{\rm cut}$). We did not parametrize the redshift evolution of these parameters; instead we inferred them for the redshift bins from $0.2 < z < 5.5$ and then looked for their evolution with a value of $z$ determined a posteriori.

\subsection{Total stellar content in halos} \label{sec:total-SHMR}
From the model of the conditional mass function, it is possible to compute the total stellar mass contained in halos of a given mass by performing an integration over the stellar mass. Since we do not have a model of $\Phi_s(M_*|M_h)$, we can use the occupation distributions of centrals and satellites because they are also integrals of the conditional mass function. Therefore, the contribution of centrals and satellites to the total stellar mass content in halos can be computed as:
\begin{equation} \label{eq:total-SHMR}
\begin{split}
    & M^{\rm tot}_*\left(M_h|M_*^{\rm th1},M_*^{\rm th2}\right) = \\
    & M_*^{\rm tot,cent}\left(M_h|M_*^{\rm th1},M_*^{\rm th2}\right) + M_*^{\rm tot,sat}\left(M_h|M_*^{\rm th1},M_*^{\rm th2}\right) = \\
    & \displaystyle\int_{M_*^{\rm th1}}^{M_*^{\rm th2}}  \left\langle N_{\rm cent}(M_h|>M_*) \right\rangle \, \diff M_* - \left[ \left\langle N_{\rm cent}(M_h|>M_*) \right\rangle\,M_* \right]_{M_*^{\rm th1}}^{M_*^{\rm th2}} + \\
    & \displaystyle\int_{M_*^{\rm th1}}^{M_*^{\rm th2}}  \left\langle N_{\rm sat}(M_h|>M_*) \right\rangle \, \diff M_* - \left[ \left\langle N_{\rm sat}(M_h|>M_*) \right\rangle\,M_* \right]_{M_*^{\rm th1}}^{M_*^{\rm th2}}.
\end{split}
\end{equation}
This equation computes the contribution of galaxies (centrals and satellites) in a stellar mass bin $M_*^{\rm th1} < M_* < M_*^{\rm th2}$ to the total stellar mass content in halos of $M_h$. The total SHMR then shows the overall efficiency of the galaxy formation process in halos, integrated over the halo's history, that is a combination of the in situ conversion of gas to stars and ex-situ from merging with satellites.

\subsection{Model of the galaxy stellar mass function} \label{sec:smf-model}

From the defined occupation distribution of halos (Eq. \ref{eq:Ncent} and Eq. \ref{eq:Nsat}), we can obtain the number density of galaxies in a given mass bin, $\Delta {\rm log} M_* = {\rm log} M_*^{\rm th2} - {\rm log} M_*^{\rm th1}$, by integrating over the halo mass function (HMF) $\diff n / \diff M_h:$
\begin{equation} \label{eq:smf-model}
    \begin{split}
        & \Phi\left(M_*^{\rm th1}, M_*^{\rm th2}\right) \, \Delta {\rm log} M_* = \\
        & \displaystyle\int_{0}^{\infty} \diff M_h \left\langle N_{\rm tot}\left(M_h|M_*^{\rm th1}, M_*^{\rm th2}\right)  \right\rangle \dfrac{\diff n}{\diff M_h}.
    \end{split}
\end{equation}
This allows us to also compute the GSMF of centrals and satellites by using their respective occupation distributions  (shown in Eq. \ref{eq:smf-model}). The literature abounds with prescriptions of the HMF obtained under various assumptions and methods, and for our work we applied the HMF of \cite{despali_universality_2016}. The adoption of different HMFs has an effect on the modeled GSMF and inevitably on the inferred model parameters. The HMF also depends on the choice of halo mass definition, and since the models are computed using the HMF, the final results will also depend on these definitions. For example, a different halo mass definition would result in a systematic shift in halo masses. To compute the HMF, we used the  \texttt{COLOSSUS} code \citep{Diemer2018} and we used the virial overdensity \citep{BryanNorman1998} halo mass definition for the results we present in this paper.
The model for the SMF shows a high sensitivity to the parameters describing the central SHMR, and coupled with the high signal-to-noise of the measurements has the most constraining power.

\subsection{Model of the two-point angular correlation function} \label{sec:woth-model}

The model of the two-pt angular correlation function follows closely the usual prescriptions \citep[see e.g.,][]{cooray_halo_2002}. For completeness, we detail the principal equations in Appendix \ref{apdx:woftheta-model}. For the computation of the two-pt angular correlation function, we rely on LSST Dark Energy Science Collaboration's Core Cosmology Library (\texttt{CCL})\footnote{\url{https://github.com/LSSTDESC/CCL}}. It offers a library of routines to calculate a range of cosmological observables and is still under active development. The validation of the software along with a range of benchmark tests are presented in \cite{chisari_core_2019}. 
The main ingredients that enter the modeling along with the prescriptions and assumptions we adopt here are given in Table \ref{tab:assumptions}.

\begin{table} [t] 
\centering
\caption{Adopted ingredients in the halo model}
\begin{tabular}{cc}
\hline
\hline
 Ingredient & Assumption  \\
 \hline
 HMF &  \cite{despali_universality_2016}   \\
 Halo bias $b_h(M_h)$  &  \cite{tinker_large_2010}  \\
 \shortstack{Halo mass-\\concentration relation $c(M_h)$} &  \cite{duffy_dark_2008} \\
 \shortstack{Halo and satellite\\over-density profiles} & \shortstack{NFW profile, \\ \cite{navarro_universal_1997}} \\
 Halo mass definition & Virial \\
 \hline
 \hline
\end{tabular}
\label{tab:assumptions}
\end{table}

Due to the relatively small volume probed by the COSMOS survey, the integral constraint affects $w(\theta)$ at large scales. We adjust the model to take this into account. The correction factor due to the IC can be estimated from the double integration of the true correlation function over the survey area:
\begin{equation}
    w_{\rm IC} = \dfrac{1}{\Omega^2} \displaystyle\int w_{\rm true} \, \diff \Omega_1 \diff \Omega_2.
\end{equation}
This integration can be carried out using the random-random pairs from the random catalog following \cite{roche_angular_1999}
\begin{equation}
    w_{\rm IC} = \dfrac{\sum w_{\rm true}(\theta) \, RR(\theta)}{\sum RR(\theta)},
\end{equation}
where $w_{\rm true} (\theta)$ is HOD-predicted model. Finally, the model that we fit against the data is simply $w(\theta) = w_{\rm true} (\theta) - w_{\rm IC}$.

\section{Results and analysis} \label{sec:results}

\subsection{Fitting procedure}
\label{sec:smf-model}
We fit the models of the $w(\theta)$ and the SMF to our measurements using a Markov chain Monte Carlo (MCMC) approach, minimizing $\chi^2$ as:
\begin{equation} \label{eq:likelihood}
    \begin{split}
        \chi^2 = & \displaystyle\sum_{i}^{N_{M-\rm thres}} (\vect{w}_i - \Tilde{\vect{w}}_i)^T C^{-1} (\vect{w}_i - \Tilde{\vect{w}}_i) \, + \\ 
        & \displaystyle\sum_{i}^{N_{M-\rm bins}} \left( \dfrac{\Phi(M_{*,i}) - \Tilde{\Phi}(M_{*,i})}{\sigma_{\Phi}}\right)^2,
    \end{split}
\end{equation}
where $\vect{w}$ represents the measurement vector containing $w$ at $\theta$ for all mass thresholds, while $\Tilde{w}$ and $\Tilde{\Phi}$ are the models for a given set of parameter values. The first line of Eq. \ref{eq:likelihood} corresponds to the clustering likelihood and the second line to GSMF likelihood. We use the affine-invariant ensemble sampler implemented in the \texttt{emcee} code \citep{foreman-mackey_emcee_2013}. We used $200$ walkers for our $11$ parameters and relied on the auto-correlation time $\tau$ to assess the convergence of the chain. To consider the chains converged, we require that the auto-correlation time is at least $60$ times the length of the chain and that the change in $\tau$ is less than $5\%$. We discarded the first $2\times\text{max}(\tau)$ points of the chain as the burn-in phase and thin the resulting chain by $0.5\times\text{min}(\tau)$. We imposed flat priors on all parameters; for the mass parameters, the flat priors are on the log quantities.

For the best-fit parameters values, we take the medians of the resulting posterior distribution, with the 16th and 84th percentiles giving the lower and upper uncertainty estimates.
The best-fit parameters and their uncertainties for all the 10 $z$-bin are listed in the Appendix \ref{apdx:bestfit-table} The posterior distributions for the 11 parameters in the redshift bins are shown in Appendix \ref{apdx:posteriors}.

\subsection{Measurements and best-fit models} \label{sec:meas-and-fits}

\begin{figure}
    \centering
    \includegraphics[width=0.99\columnwidth]{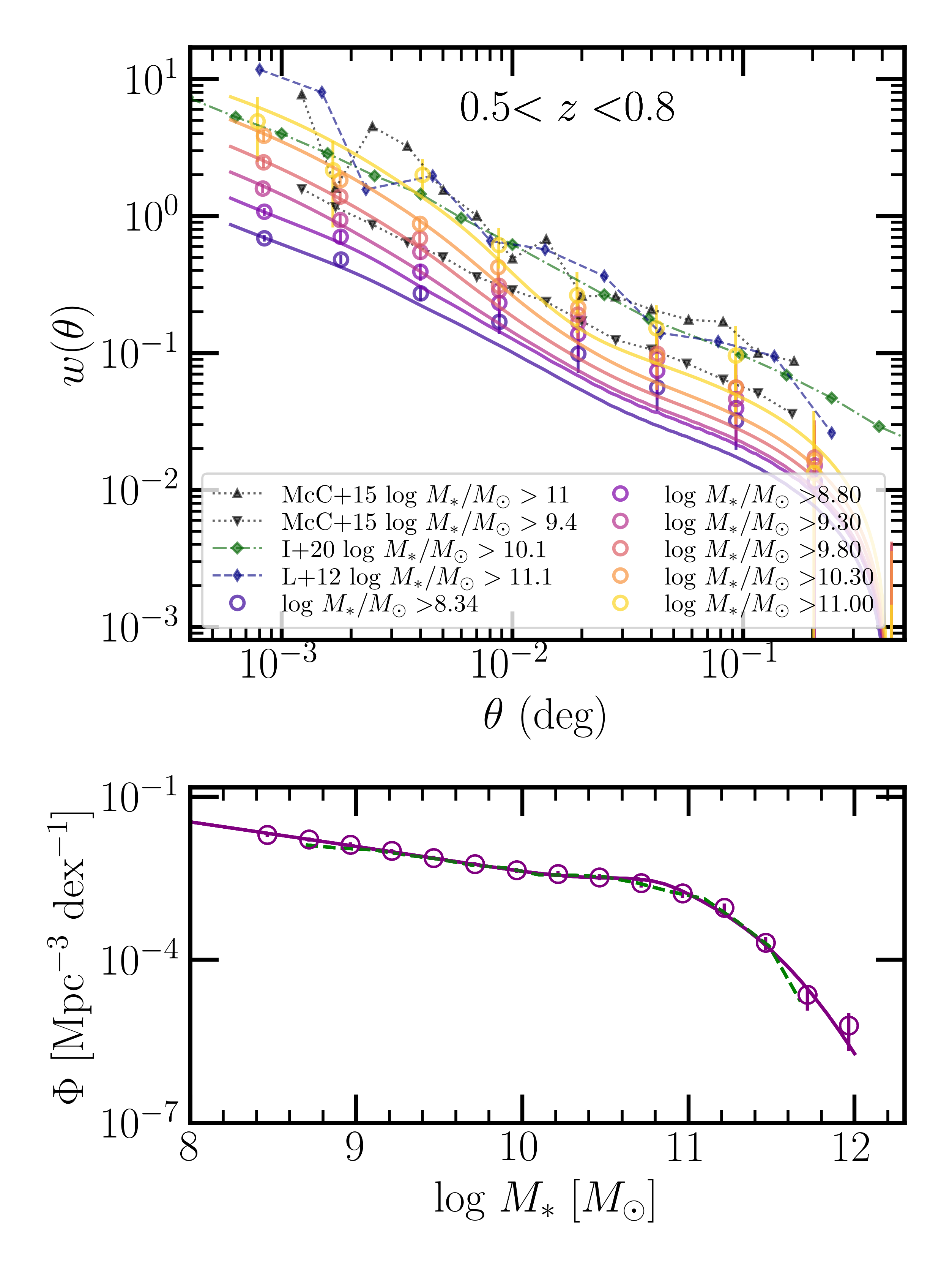}
    \caption{Best-fit models for clustering and abundance compared with measurements at $0.5 < z < 0.8$. \textit{Top}: Clustering measurements for the 6 mass threshold-selected samples (empty circles with errorbars) along with the best-fit models in solid lines in corresponding colors. \textit{Bottom}: Measurements of the stellar mass function together with the best fit model. The dashed lines show the SMF in the same redshift bin obtained by \cite{davidzon_cosmos2015_2017} for comparison.}
    \label{fig:data-fits-ex}
\end{figure}

\begin{figure*}[hp]
   \begin{subfigure}[b]{0.32\textwidth}
            \includegraphics[width=1\hsize]{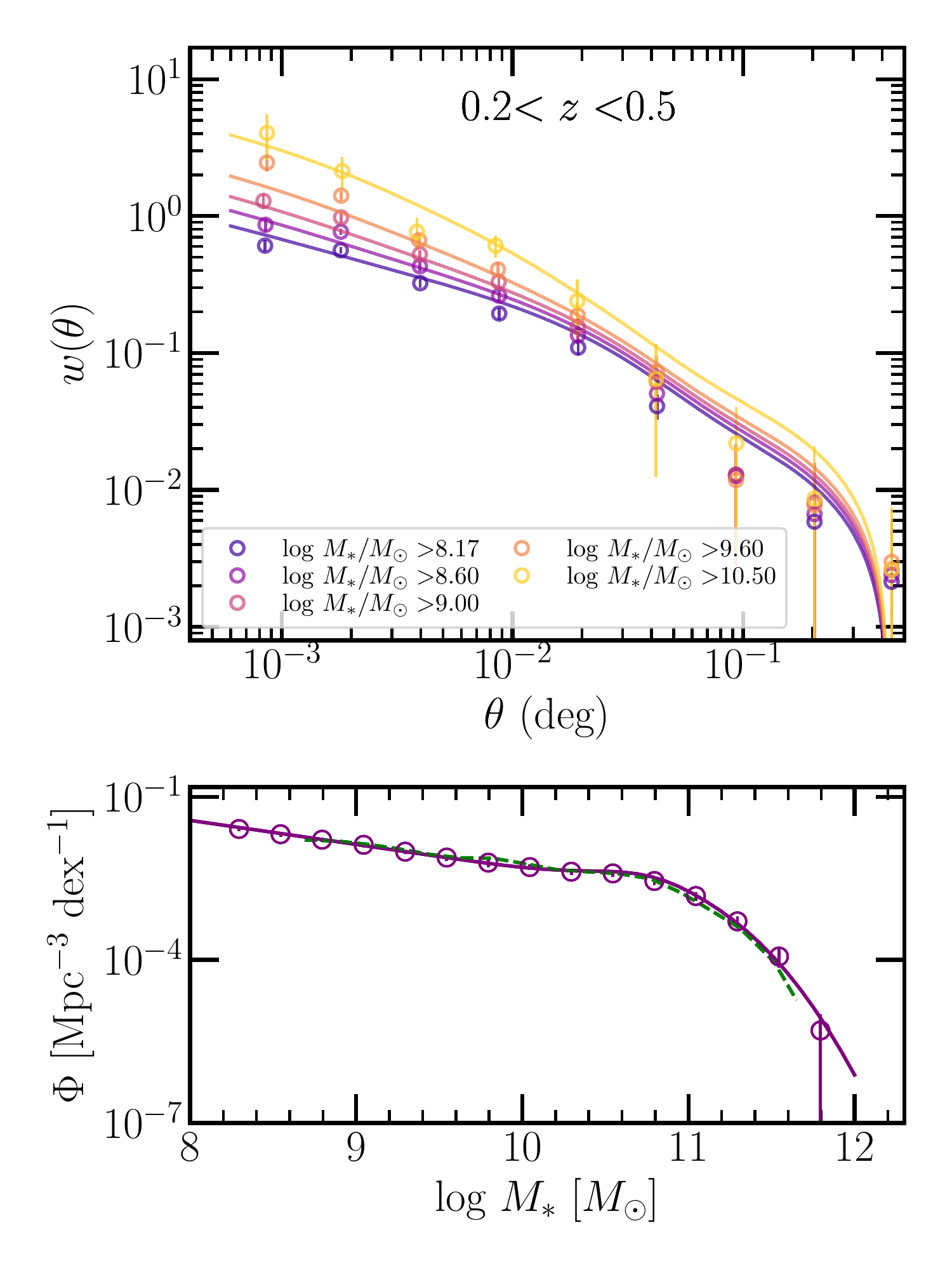}     
   \end{subfigure}
   \hfill
   \begin{subfigure}[b]{0.32\textwidth}
            \includegraphics[width=1\hsize]{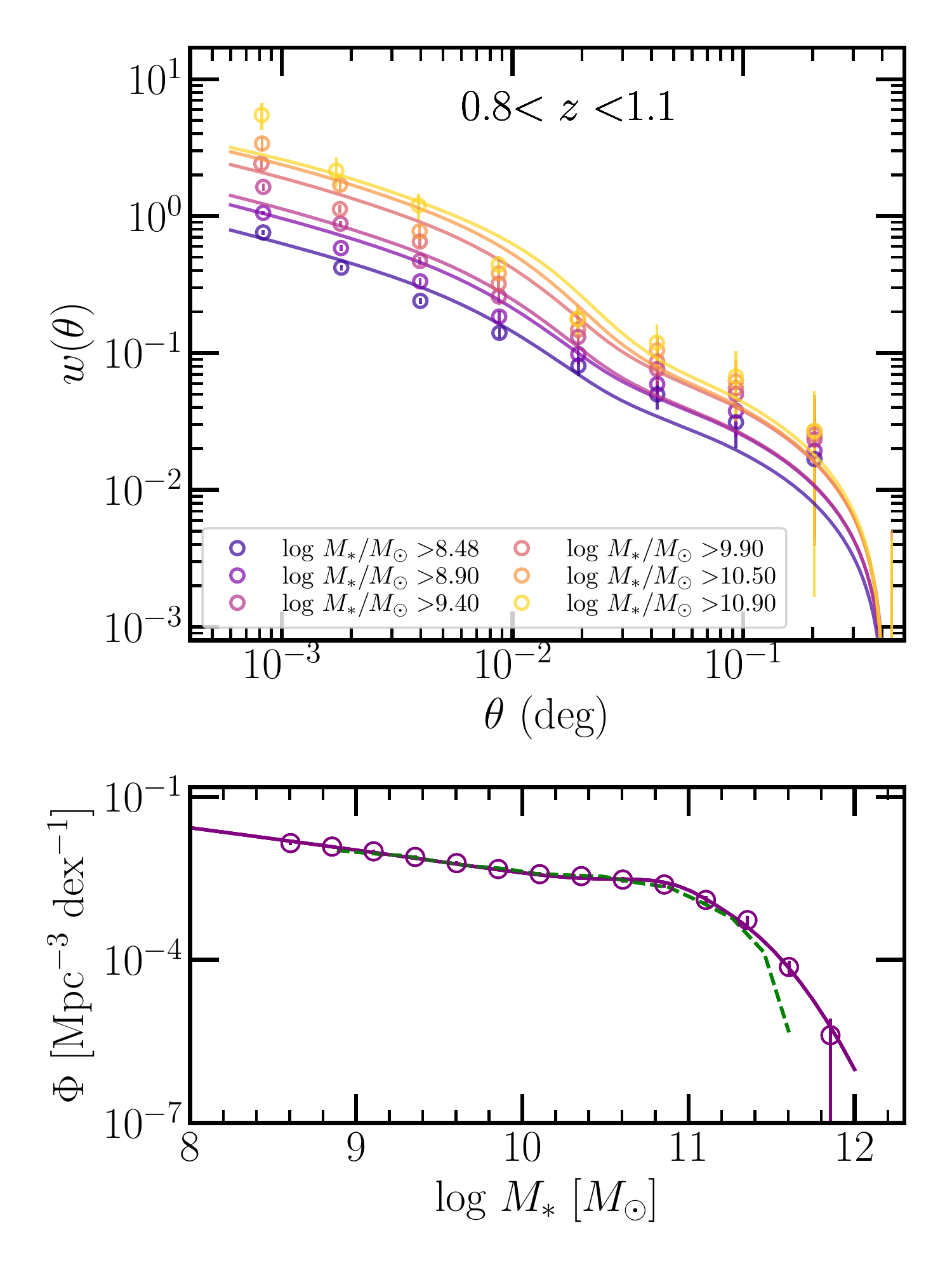}     
   \end{subfigure}
   \hfill
   \begin{subfigure}[b]{0.32\textwidth}
            \includegraphics[width=1\hsize]{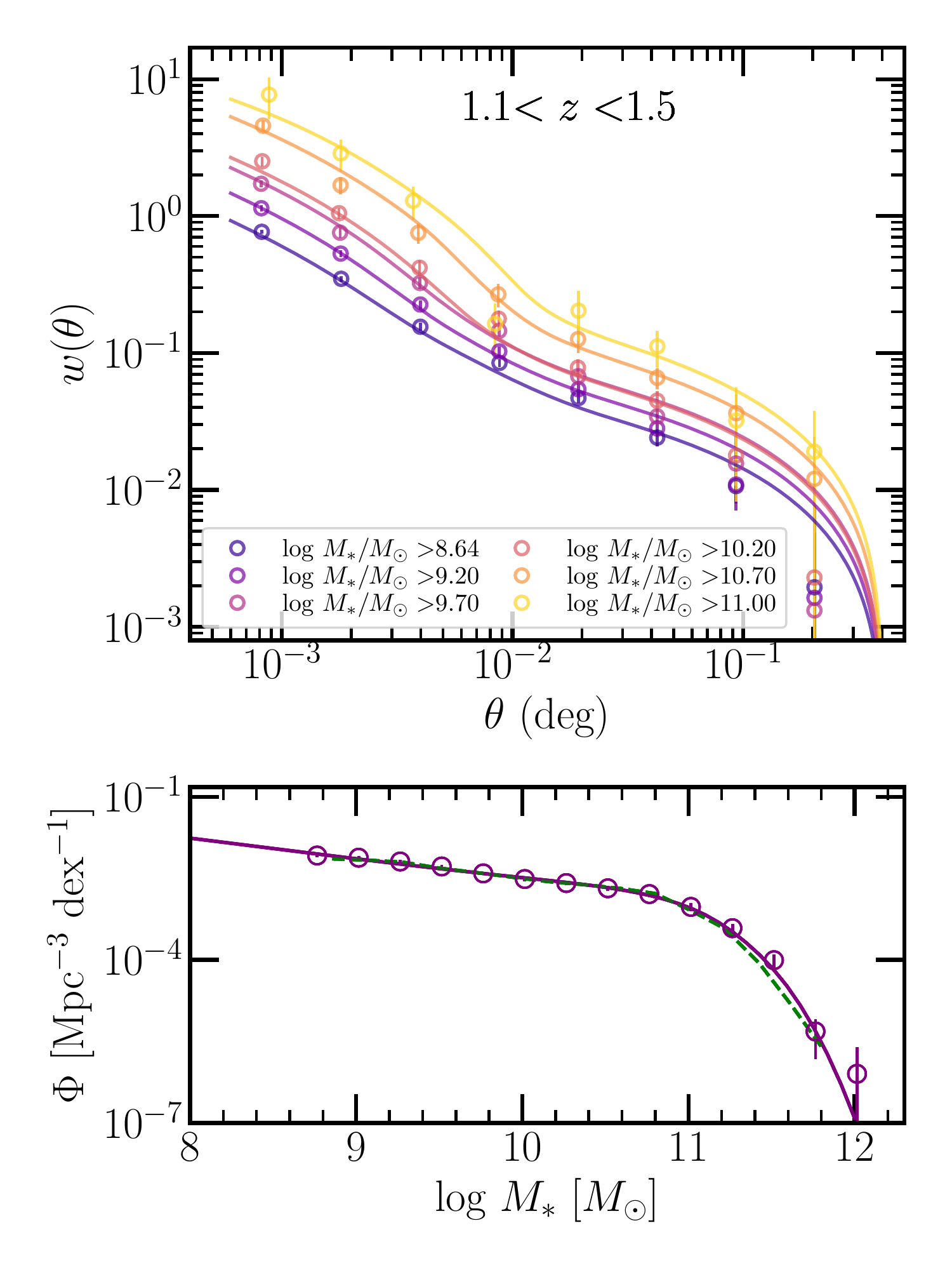}     
   \end{subfigure}
   \hfill
   \begin{subfigure}[b]{0.32\textwidth}
            \includegraphics[width=1\hsize]{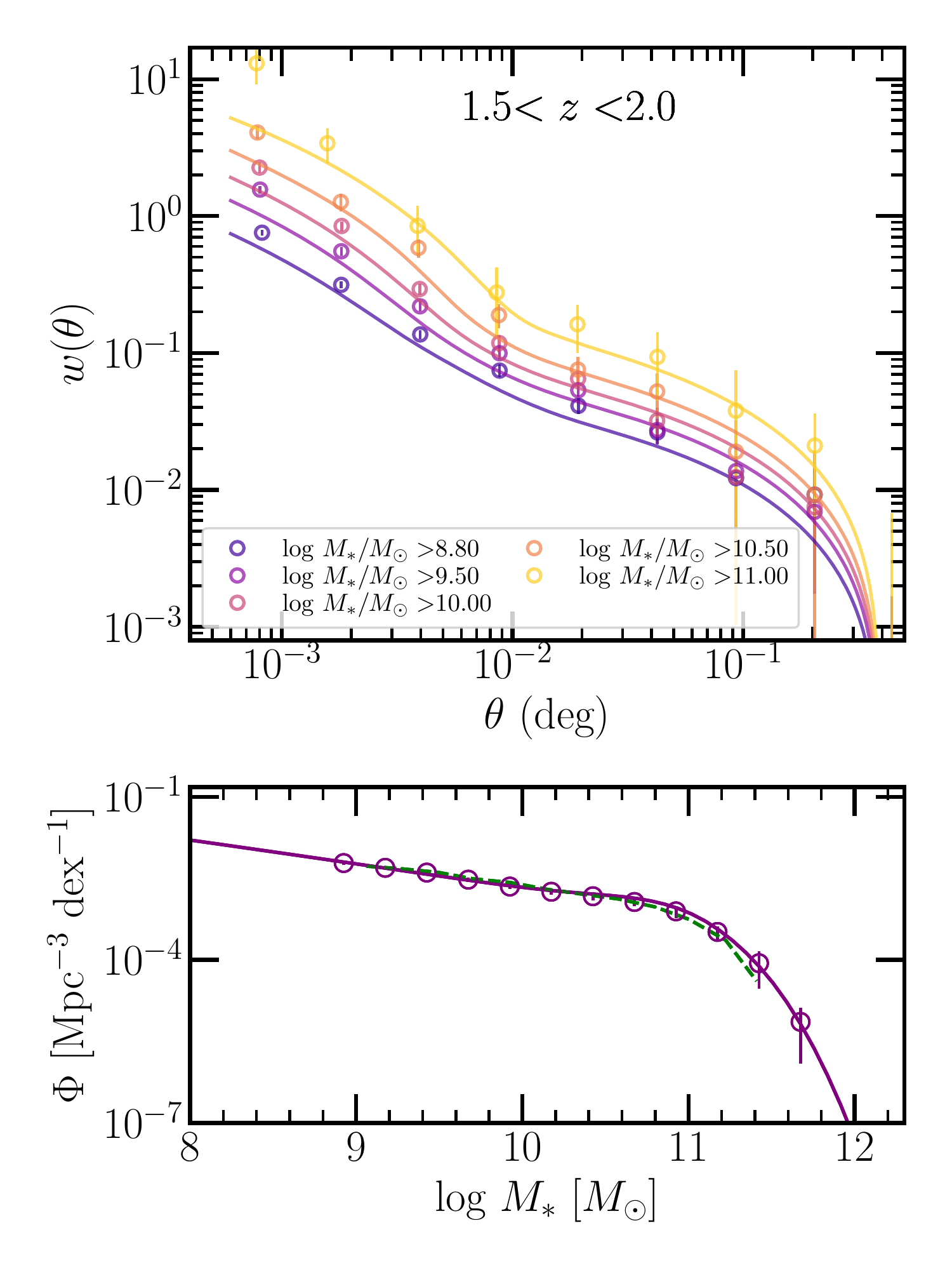}     
    \end{subfigure}
    \hfill
   \begin{subfigure}[b]{0.32\textwidth}
            \includegraphics[width=1\hsize]{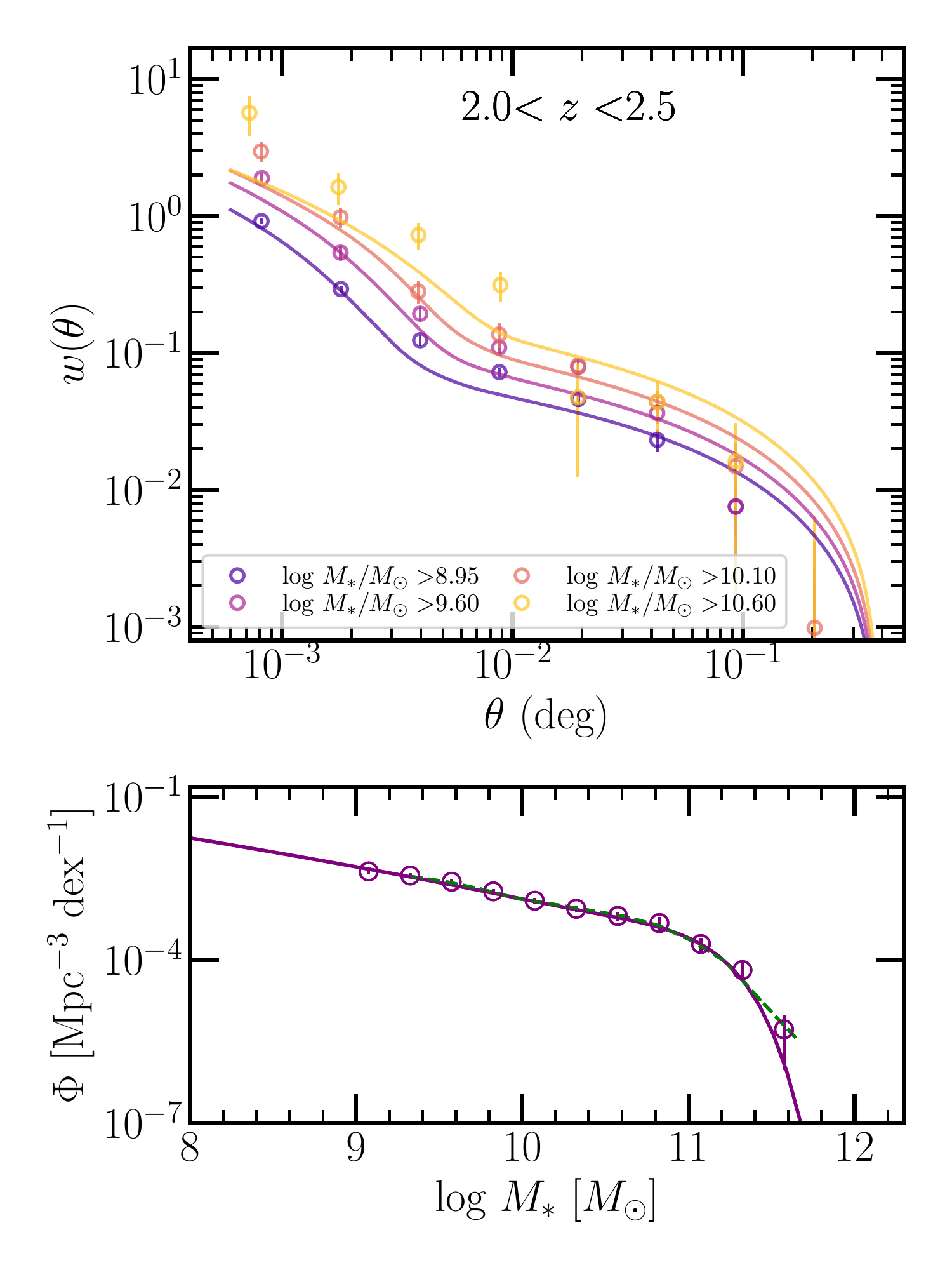}     
   \end{subfigure}
   \hfill
   \begin{subfigure}[b]{0.32\textwidth}
            \includegraphics[width=1\hsize]{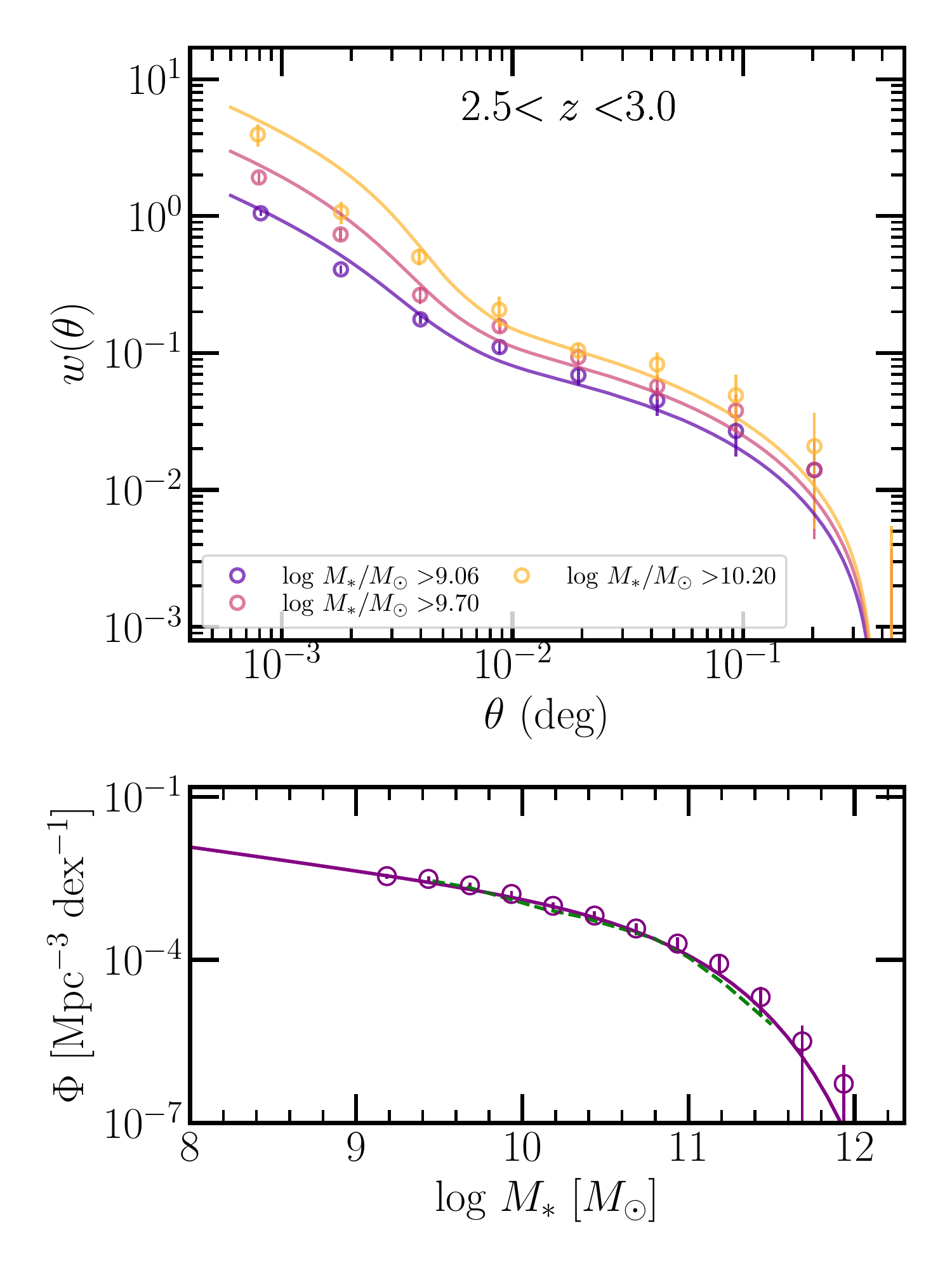}     
   \end{subfigure}
   \hfill
   \begin{subfigure}[b]{0.32\textwidth}
            \includegraphics[width=1\hsize]{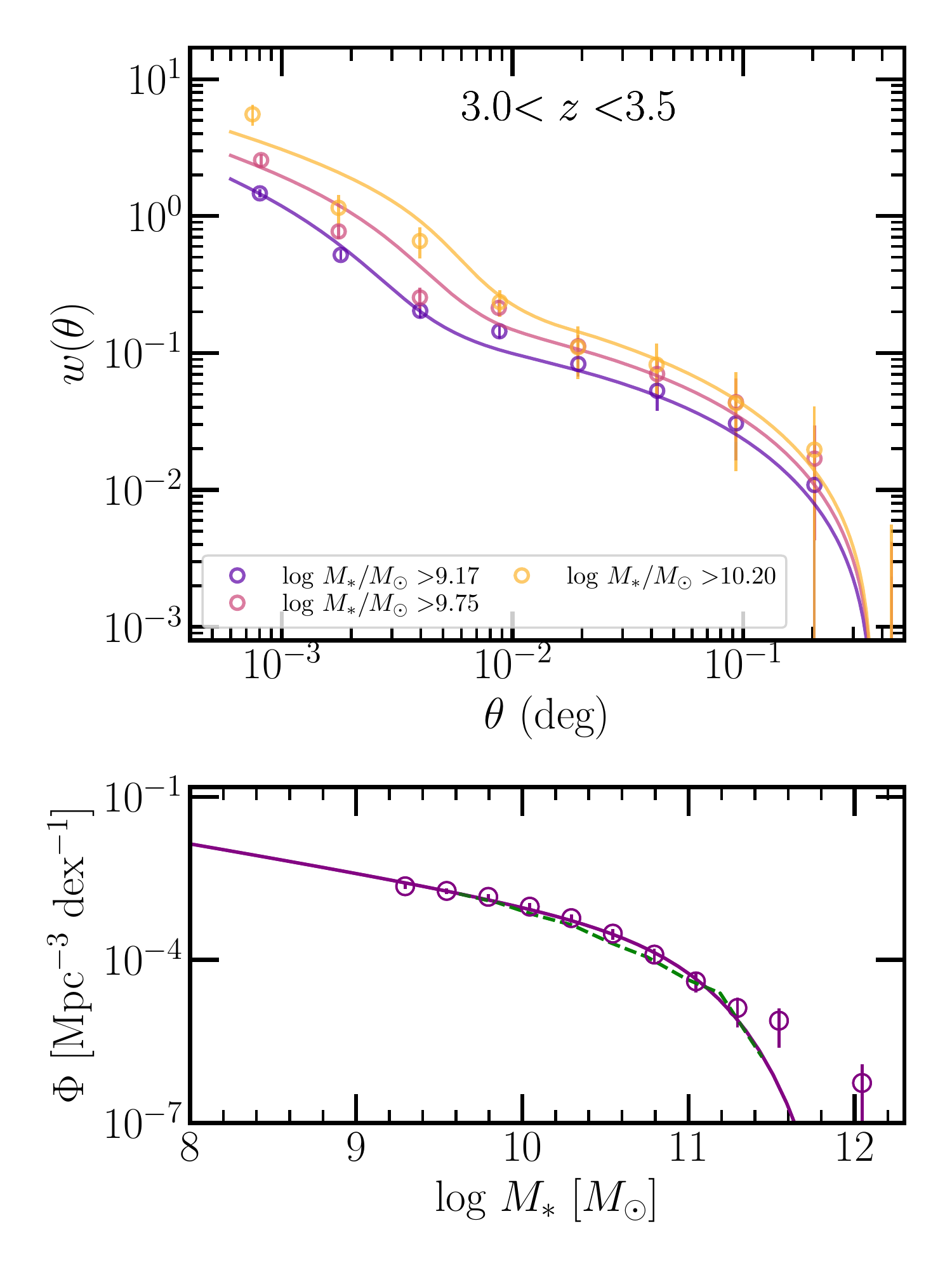}     
   \end{subfigure}
   \hfill
   \begin{subfigure}[b]{0.32\textwidth}
            \includegraphics[width=1\hsize]{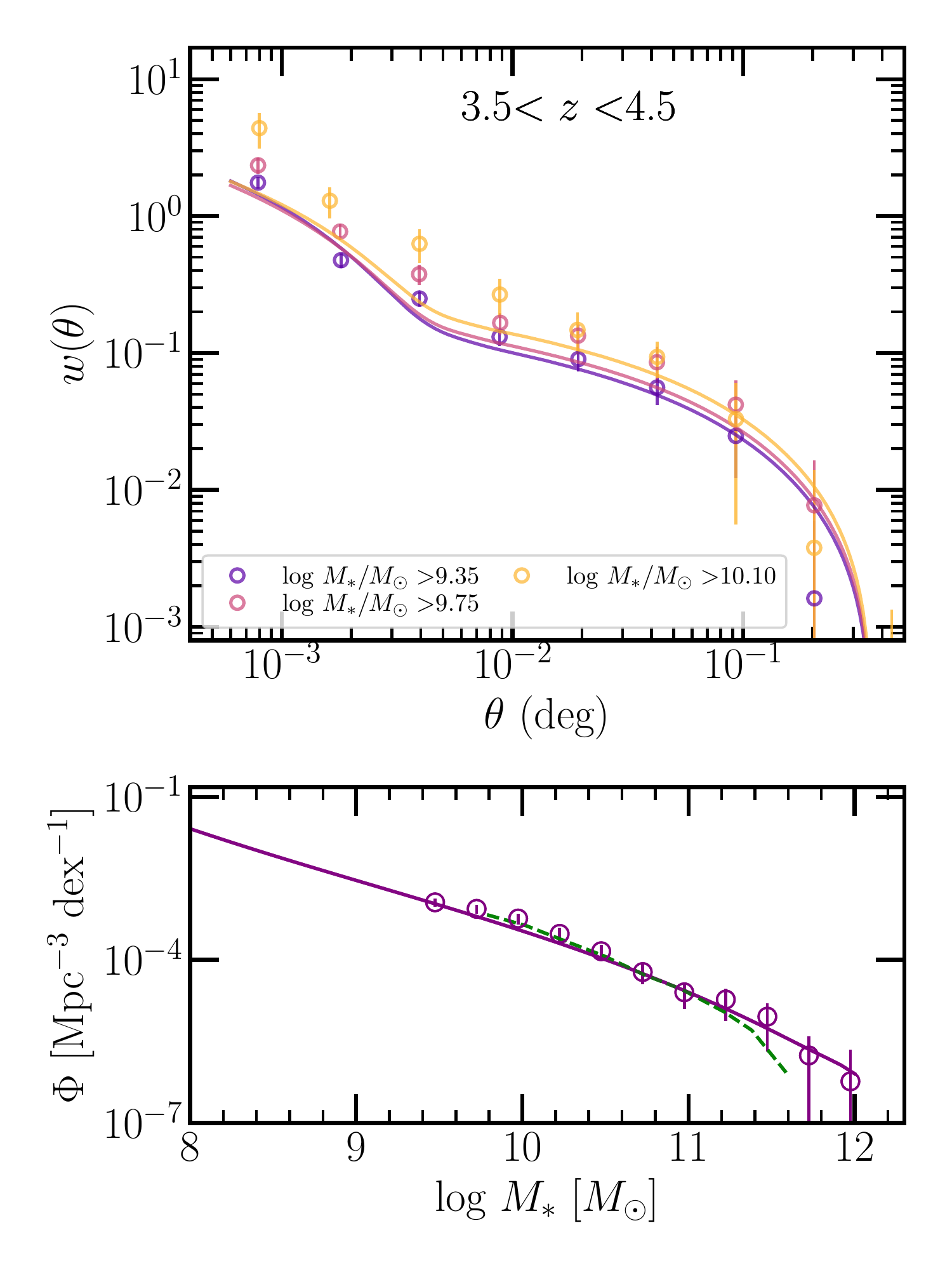}     
   \end{subfigure}
   \hfill
   \begin{subfigure}[b]{0.32\textwidth}
\includegraphics[width=1\hsize]{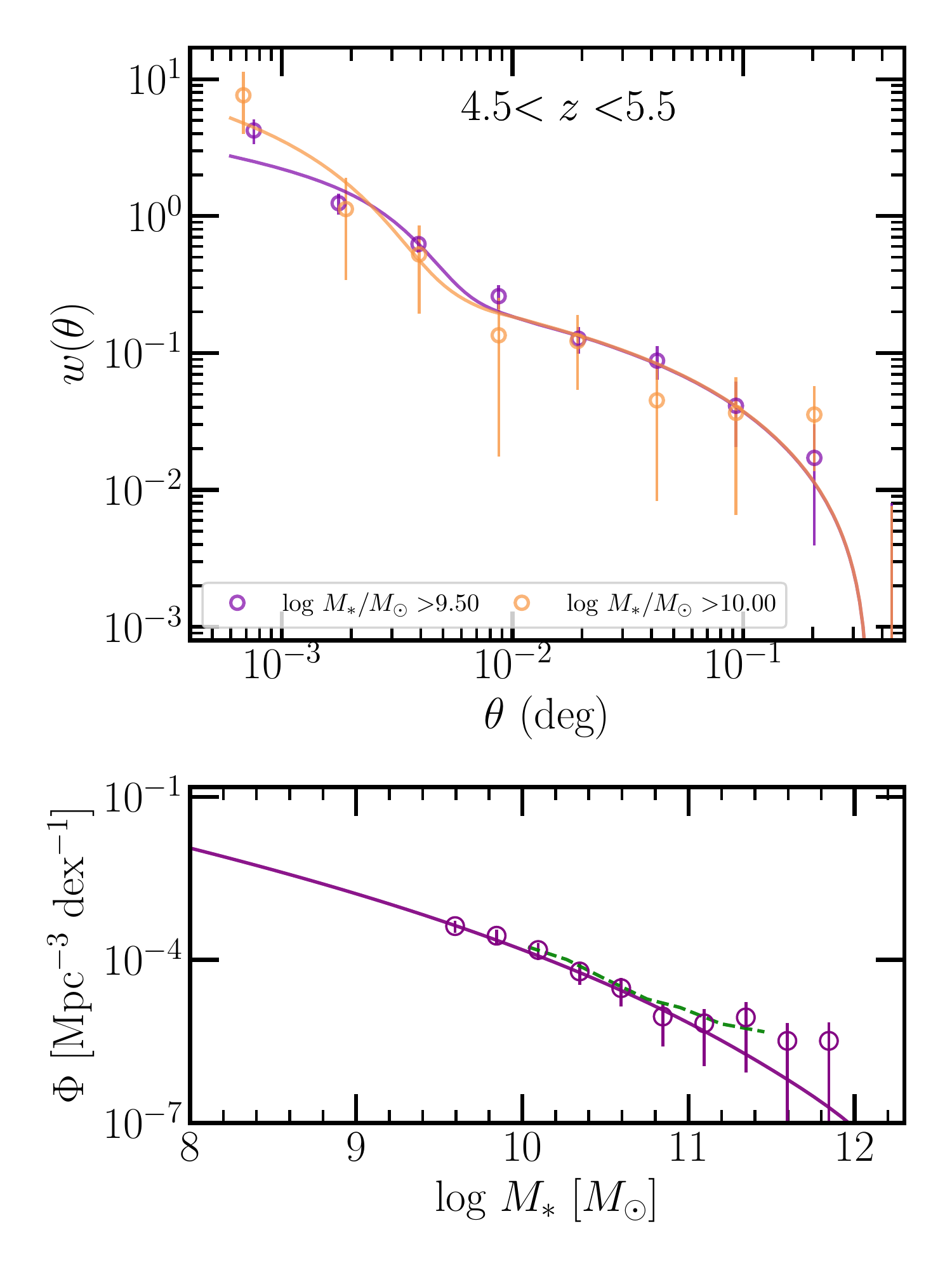}     
   \end{subfigure}
  \caption{Best-fit models of clustering and abundance plotted over the measurements for all $z$-bins apart from $0.5 < z <0.8,$ which is shown in Fig. \ref{fig:data-fits-ex}.}
   \label{fig:measures-fit}
        \end{figure*}

The measurements of the $w(\theta)$ and GSMFs are shown in Fig.~\ref{fig:data-fits-ex}, where we isolate the measurement in $0.5 < z < 0.8$ and compare it with clustering measurements from the literature, as well  as in Fig.~\ref{fig:measures-fit}, where we show the measurements in all the other $z$-bins. In the upper panel, we show the clustering measurements (open circles with errorbars) for different mass-threshold samples and in the bottom panel the SMF measurement. The solid lines show the best-fit models with the color code corresponding to the mass-threshold measurement (in the top panel). Table \ref{apdx:bestfit-table} also shows the reduced chi-squared value $\chi^2_{\rm reduced}$ for the-best fit parameters. Their values range from $2.5$-$6$ for most bins except for $0.8 < z < 1.1,$ where $\chi^2_{\rm reduced} = 10.2$. Given the number of data points, ranging from $75$ at $ 0.8 < z <  1.1$ to $30$ at  $4.5 < z <  5.5$, these values of $\chi^2_{\rm reduced}$ indicate a reasonably good fit.  One possible explanation can be the greater complexity of the data, which might not be completely captured by the fits. For example, we are simultaneously fitting for several mass-selected clustering measurements. Due to uncertainties in the stellar masses, there some super-covariance may exist between all the measurements, however, this type of modeling is beyond the scope of this work.

\paragraph*{\textbf{Description of the clustering measurements.}}
The clustering measurements exhibit the usual behavior, with $w(\theta)$ following a power-law at small scales that breaks at intermediate scales ($\sim 1 \arcmin$). The origin of this break comes from the fact that the power at small scales is dominated by galaxies residing in the same halo (1-halo term), which drops off quickly at intermediate scales; at larger scales, the power mainly comes from large-scale clustering of halos (two-halo term) and the transition between these two regime creates the characteristic shape \citep{Zehavi2004}. The relative contribution from these two terms becomes more apparent at high masses and high redshifts, with the one-halo term dominating the power at small scales with a steep slope and the two-halo term dominating the large scales with a shallower slope at the transition. The clustering amplitude increases with increasing mass threshold -- a familiar behavior based on the fact that  massive galaxies trace high-density and more clustered regions. 

At scales larger than $0.1 \, \rm deg$, there is a sharp drop in power due to the effects of the integral constraint. The best-fit models shown in solid lines generally agree well with the measurements. It should be noted that the measurements show an excess of power at scales of $\gtrsim 0.02\, \rm deg$ ($\sim 0.5$ Mpc at $z\sim7$) and the fits are systematically below the data points.
This may be due to the effects of the non-linear halo bias effect (scale-dependent halo bias). While in this work we use the scale-independent halo bias of \cite{tinker_large_2010}, some studies suggest that the halo bias is scale dependent in the quasi-linear regime at scales of about \st{$50\, \rm Mpc$} {$1\, \rm Mpc$} \citep{jose_understanding_2017}. This will add power in the correlation at scales of $\sim 0.04\, \rm deg$. Furthermore, at  $0.5 < z < 1.5$, a contribution may also come from the known overabundance of rich structures in the COSMOS field at these redshifts, as discussed by \cite{mccracken_angular_2007}, \citep{meneux_zcosmos_2009}, and \cite{mccracken_probing_2015}. This excess of power will decrease the SHMR, indicating an even lower efficiency of converting baryons to stars. Clustering is particularly sensitive to the satellite content within the halo, therefore the parameters regulating the satellite HOD will be constrained by clustering.

\paragraph*{\textbf{Literature comparison.}}
In Fig.~\ref{fig:data-fits-ex}, we show COSMOS clustering measurements at  $0.5 < z <0.8$ from \cite{mccracken_probing_2015} and \cite{leauthaud_new_2012}, and in the HSC-SSP Wide survey from \cite{ishikawa_subaru_2020}. \cite{mccracken_probing_2015} used the $1.5 \ \rm deg ^2$ COSMOS footprint of UltraVISTA DR1 \citep{mccracken_ultravista_2012} to measure clustering for mass-threshold selected samples. We also show the correlation function for galaxies with log$M_*/M_{\odot} > 9.4$ and log$M_*/M_{\odot} > 11.0$ in black wedges and triangles. Qualitatively, measurements from the literature are in agreement with our work, although the 2PCF for the log$M_*/M_{\odot} > 11.0$ sample has a slightly higher amplitude, especially at small scales. \cite{leauthaud_new_2012} used the $1.64 \ \rm deg ^2$ of COSMOS, as imaged by HST/ACS in F814W \citep{Koekemoer2007}, to measure the 2PCF for mass-threshold samples at $0.48 < z < 0.74$. The measurement for log$M_*/M_{\odot} > 11.1$ is shown as  gray hexagons in Fig. \ref{fig:data-fits-ex}, which are consistent with our measurements and those of \cite{mccracken_probing_2015}. Finally, in green diamonds, we show the measurements from \cite{ishikawa_subaru_2020} in $145 \ \rm deg ^2$ in the HSC-SSP Wide for a sample of log$M_*/M_{\odot} > 10.1$ galaxies at $0.55 < z < 0.80$. The amplitude of \cite{ishikawa_subaru_2020} 2PCF corresponds to what we measure in this work between log$M_*/M_{\odot} > 10.3$ and log$M_*/M_{\odot} > 11.0$. This could come from incompleteness in their samples, or/and uncertain stellar masses that were estimated with optical ($grizy$) bands only. 

\begin{figure*}[h]
    \centering
    \includegraphics[width=0.9\textwidth]{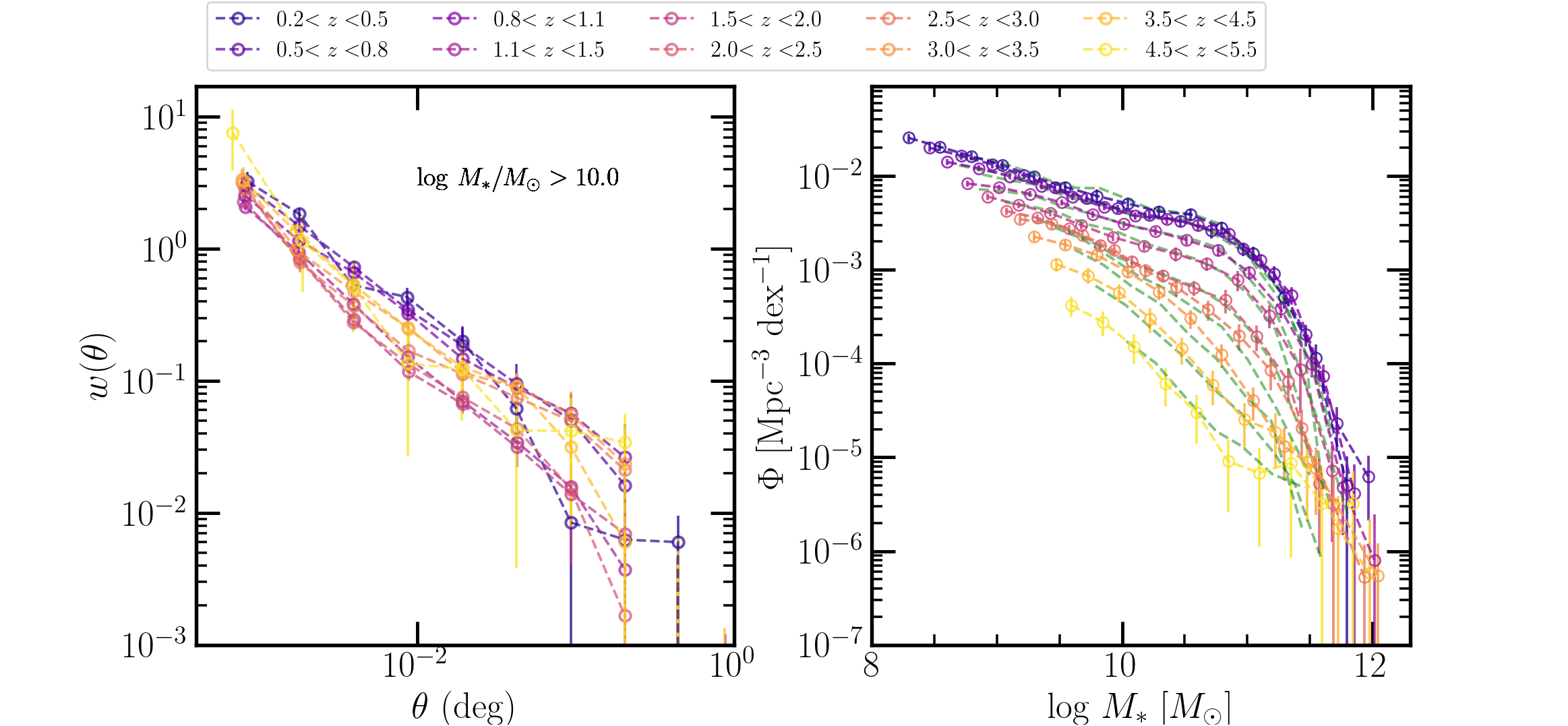}
    \caption{
    Correlation of galaxies with $ M_* > 10^{10} \, M_{\odot}$ (left panel) and GSMF (right panel) for all ten redshift bins. The green dashed lines in the right panel correspond to the GSMF of \cite{davidzon_cosmos2015_2017}.
    }
    \label{fig:w-th_SMF-z}
\end{figure*}

\paragraph*{\textbf{Redshift evolution of the clustering.}}
With respect to redshift, to show a possible evolution of the clustering amplitude, in Fig.~\ref{fig:w-th_SMF-z} we recompute the correlation function for galaxies selected above the same mass threshold in all $z$-bins: $M_* > 10^{10}\, M_{\odot}$. Although the clustering amplitude of dark matter decreases with increasing redshift, the evolution of the clustering amplitude for galaxy samples selected at the same mass-threshold depends on the galaxy formation model. The clustering of galaxies depends on how galaxies occupy DM halos, which can change with redshift. N-body simulations combined with semi-analytical models of galaxy formation indicate that the clustering amplitude of similarly selected galaxies first decreases from $z = 0$ to $z = 1.5$, remains constant up to $z=2.5$, and then increases again at higher redshifts \citep{kauffmann_clustering_1999}. Qualitatively, this behavior can be observed in our measurements in Fig.~\ref{fig:w-th_SMF-z}: the correlation amplitude is the highest in the lowest redshift bins, reaches the lowest amplitude for intermediate $z$-bins of about $z\sim1.5$ and then increases again at $z>2.0$.


\paragraph*{\textbf{Redshift evolution of the SMF.}}
The SMF measurements in Fig.~\ref{fig:w-th_SMF-z} also show the usual evolution with redshift \citep[see e.g.,][]{ilbert_mass_2013, davidzon_cosmos2015_2017}: the normalization decreases and the knee at $M_* \sim 10^{11} \, M_{\odot}$ becomes less and less prominent with increasing redshift; the slope of the low-mass end remains constant up to $z=2$ but steepens at higher redshifts where the SMFs resemble more a power-law (e.g., in the $z>4.5$ bin); the redshift evolution is strongly dependent on mass: the low-mass end evolves more rapidly than the high-mass end. The SMFs, having the most constraining power over the model parameters (due to the small measurement errors and sensitivity of the model), show an excellent fit of the models to the measurements. The dashed lines in Fig.~\ref{fig:measures-fit} show the SMFs measured by \cite{davidzon_cosmos2015_2017} using the previous version of the catalog, COSMOS2015. Overall, they are in agreement with our measurements over the whole redshift range.

\begin{figure}[h]
    \centering
    \includegraphics[width=0.99\columnwidth]{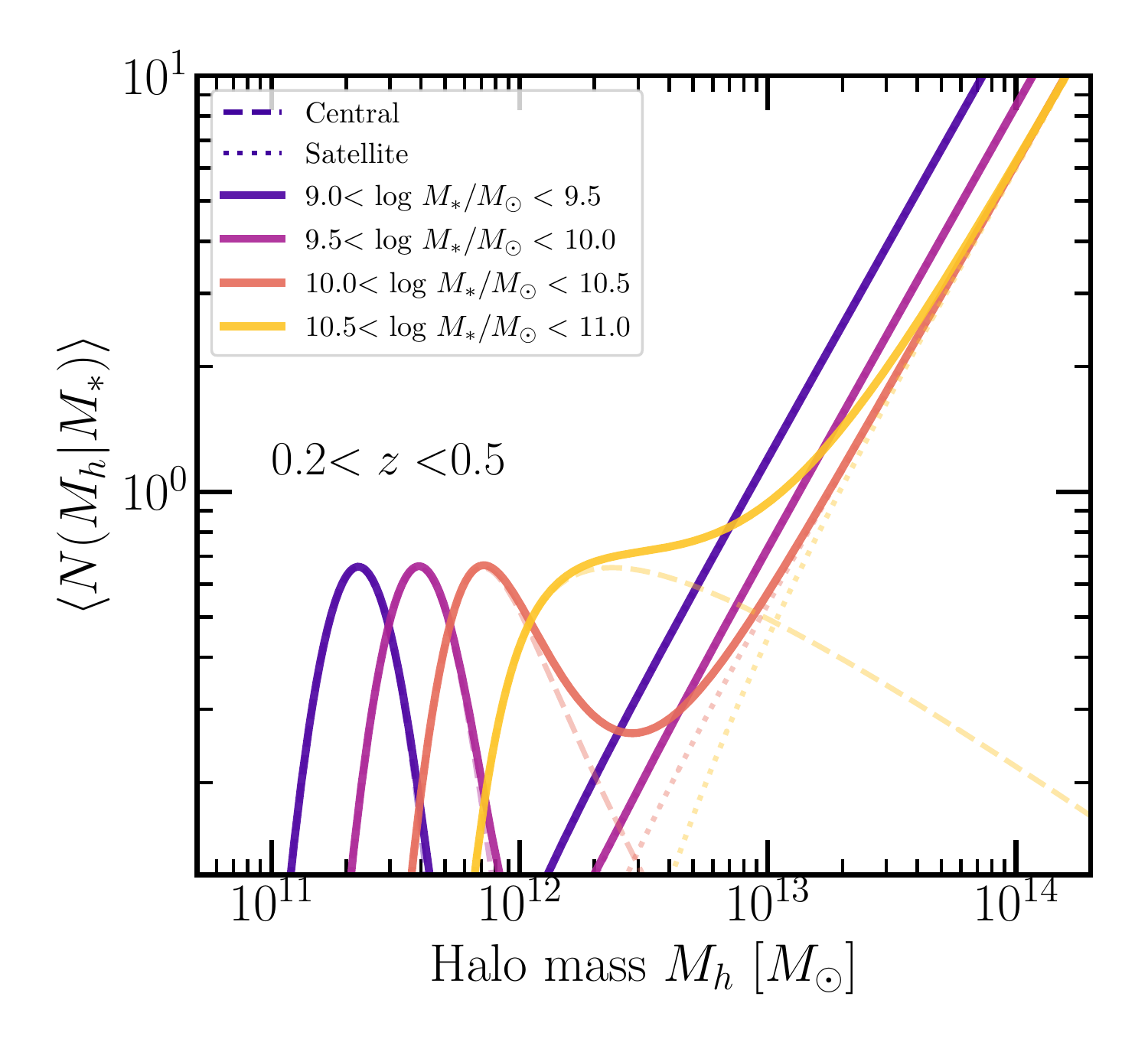}
    \caption{Mean number of galaxies with stellar masses in a given mass bin as a function of the mass of the halo that they occupy. We show the mean halo occupation function for galaxies in 4 stellar mass binscolorur coded accordingly) at $0.2 < z < 0.5$. The thick solid lines show the total $\langle N_{\rm tot} \rangle$ and the dashed and dotted lines show the centrals and satellites.
    }
    \label{fig:NofMh-z3}
\end{figure}

\begin{figure*}[h]
    \centering
    \includegraphics[width=0.99\textwidth]{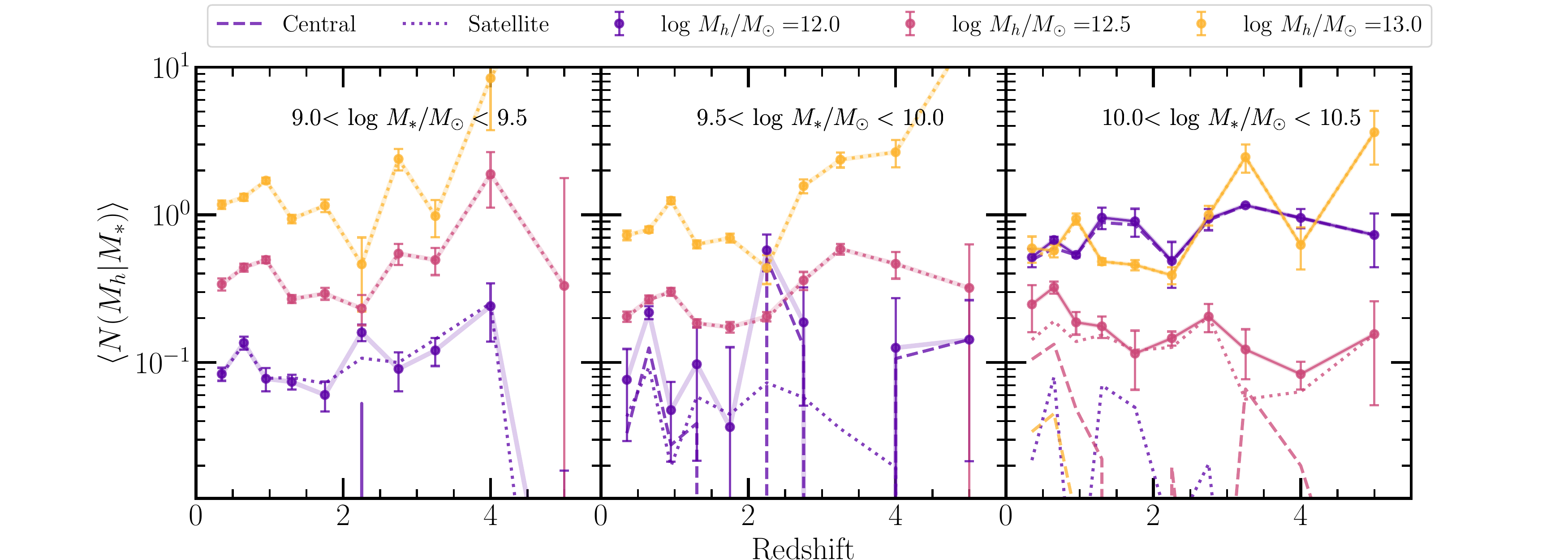}
    \caption{Mean halo occupation in halos of a given mass as a function of redshift. In each panel, we show the $\langle N_{\rm tot}(M_h) \rangle$ at ${\rm log} \,M_h/M_{\odot} = [12.0, \, 12.5. \, 13.0]$. The three different panels show the mean occupations by galaxies in three different stellar mass bins ${\rm log} \,M_*/M_{\odot} = \{[9.0, \, 9.5], \, [9.5, \, 10.0], \, [10.0, \, 10.5] \}$. The dashed and dotted lines show the central and satellite occupations, while the points connected with transparent solid line show the total.}
    \label{fig:NofMh-vs-z}
\end{figure*}


\subsection{Evolution of the mean halo occupation with redshift} \label{sec:HODparams-vs-z}

The mean halo occupations, as defined by Eq.~\ref{eq:Ncent} and Eq.~\ref{eq:Nsat},  are shown in Fig.~\ref{fig:NofMh-z3} for $0.2 < z < 0.5$.
We show the mean number of galaxies in four mass bins: log$\,M_*/M_{\odot} = \{[9.0, \, 9.5], \ [9.5, \, 10.0], \ [10.0, \, 10.5], \ [10.5, \,11.0]\}$, as a function of halo mass for all galaxies in the thick solid lines, and for satellites and centrals in dotted and dashed lines, respectively. It is immediately evident that the mean halo occupation shifts toward high halo masses for more massive galaxies, as it requires more massive halos to host more massive galaxies. 
Furthermore, the central occupation peaks at some characteristic mass. Halos that have this characteristic mass can be considered as most likely to host a central galaxy in a given stellar mass bin. 

As the halo mass increases, the number of satellites starts to increase sharply. The mean occupation for low-mass galaxies shows that there can be halos of intermediate mass that do not host any low-mass galaxies. For example, halos of $M_h \sim 10^{12} \, M_{\odot}$ have a very low probability of hosting of $10^{9.5} < M_*/M_{\odot} < 10^{10}$ galaxies. The central and satellite decompositions (dashed and dotted lines) show that this is because galaxies in this mass bin cannot be centrals in $M_h \sim 10^{12} \, M_{\odot}$ halos and can only be satellites in even more massive halos. 
We also note that as their stellar mass increases, central galaxies are more likely to occupy halos with a larger variety of masses (looking at the dashed line, for higher mass bins there is shallowing of the slope at which the central occupation decreases with halo mass). This behavior can come from a quenching of massive galaxies -- as their stellar mass growth stops, the halo they inhabit continues to grow in mass.
Finally, in clusters ($M_h > 10^{13}\, M_\odot$), low-mass satellites dominate the number of galaxies in the halo. This can also be seen as a consequence of quenching: satellites stop their growth because of quenching in the halo and remain less massive, while the halo can grow by merging with other halos containing more satellites of low masses.

To investigate the redshift evolution, in Fig. \ref{fig:NofMh-vs-z} we show the mean occupation distribution for galaxies with $9.0 < {\rm log } M_*/M_{\odot} < 9.5$ (left panel), $9.5 < {\rm log } M_*/M_{\odot} < 10$ (middle panel) and $10.0 < {\rm log } M_*/M_{\odot} < 10.5$ (right panel) as a function of redshift at three different halo masses log $M_h/M_{\odot} = [12.0, \, 12.5, \, 13.0]$. Dashed and dotted lines show the central and satellite mean halo occupations, while the points connected with transparent solid line show the total. The panels show that the total $\langle N(M_h/M_*)\rangle$ of $M_h \leq 10^{12} \, M_{\odot}$ halos is dominated by centrals at all redshifts, whereas satellite dominate at higher halo masses at all redshifts. An exception are galaxies with $9.0 < {\rm log }M_*/M_{\odot} < 9.5$ (left panel) which are found as satellites in halos of $M_h \geq 10^{12} , M_{\odot}$ and at all redshifts.
In each panel and for every halo mass, we detect little-to-no evolution of the mean occupation number, 
in accordance with previous findings based on N-body simulations \citep[e.g.,][]{kravtsov_dark_2004}. At $z> 2.5$ there are variations toward higher mean halo occupation number but with overly large uncertainties to be significant. These results indicate that (statistically) in terms of mean occupation numbers, galaxies populate DM halos in the same way throughout cosmic time. 

\subsection{Satellite fraction and its evolution with redshift} \label{sec:sat-frac}

\begin{figure*}[t]
    \centering
    \includegraphics[width=1.0\textwidth]{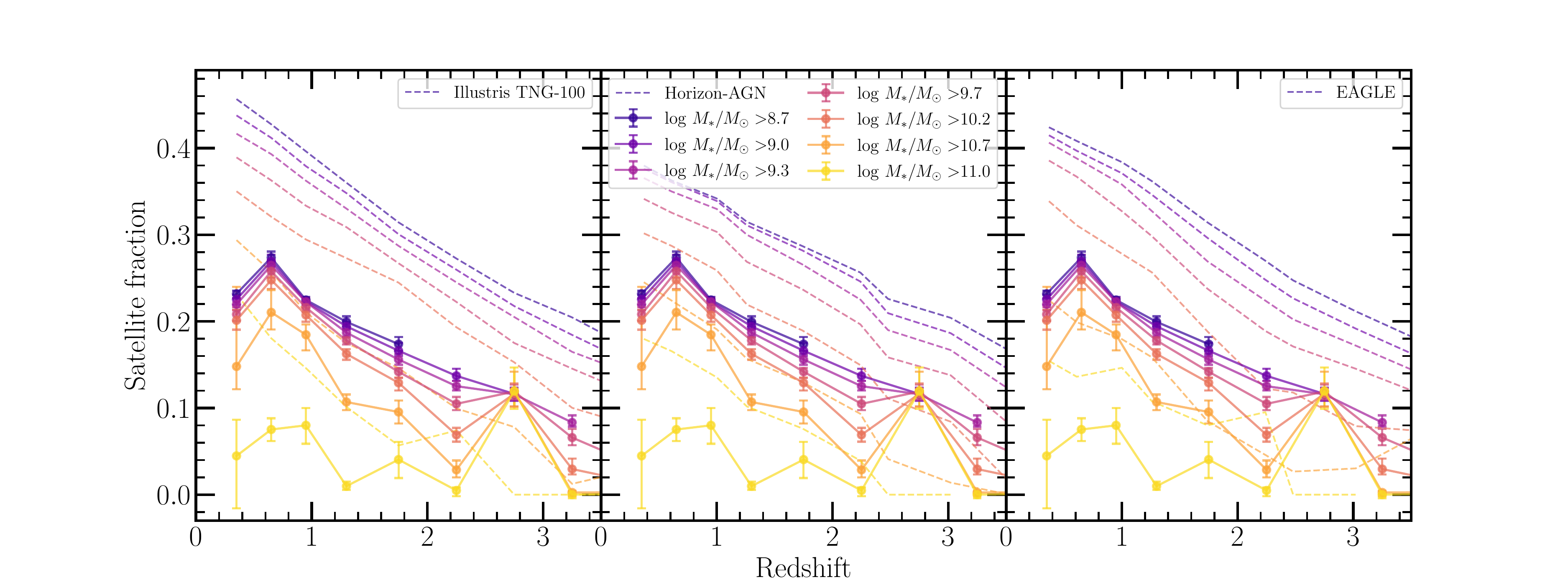}
    \caption{Fraction of satellite galaxies with masses above a given threshold as a function of redshift (solid lines). In each of the three panels, we compare with the satellite fractions measured in the hydrodynamical simulations ({dashed lines}) \textsc{TNG100} (left), \textsc{Horizon-AGN} (center), and \textsc{EAGLE} (right). The redshift evolution of the satellite fraction is shown for galaxies with masses above log$\,M_*/M_{\odot} > [8.7, \ 9.0, \ 9.3, \ 9.7, \ 10.2, \ 10.7, \ 11.0 ] $. }
    \label{fig:fsat-vs-z}
\end{figure*}

Dark matter halos are usually inhabited by a massive central galaxy and a number of smaller satellite galaxies orbiting the potential well of the halo. At a fixed stellar mass, a galaxy can be either a central in a relatively low-mass halo or a satellite in a massive one. The number of satellite galaxies in a halo and its evolution with redshift reflects the halo's evolutionary history in terms of its hierarchical merger tree, but it also reflects the physical processes and environmental effects that can affect the assembly of satellites.
Using our constraints on the HOD in the broad redshift span up to $z \sim 5,$ we can study the evolution of the satellite fraction and get insights into the halos' evolutionary history. 

Within the HOD framework, we can compute the fraction of satellite galaxies, summed over all halos and with masses above a given stellar mass threshold; then, using our best-fit parameters in the ten $z$-bins reconstruct its evolution with redshift. To compute the satellite fraction, we perform the following integration:
\begin{equation} \label{eq:sat_frac}
    f_{\rm sat}(z | > M_*^{\rm th}) = 1 - \dfrac{1}{\bar{n}_g}\displaystyle\int \diff M_h \dfrac{\diff n}{\diff M_h} {\left \langle N_{\rm cent} (M_h | > M_*^{\rm th}) \right \rangle},
\end{equation}
where, as before, ${\diff n}/{\diff M_h}$ is the halo mass function, $\bar{n}_g$ is the mean number density of galaxies with $M_{*} > M_*^{\rm th}$ and $\langle N_{\rm cent} (M_h | > M_*^{\rm th}\rangle$ is the mean occupation function for centrals with the best-fit parameters

Our results on satellite fraction of galaxies with masses above log$\,M_*/M_{\odot} > [8.7, \ 9.0, \ 9.3, \ 9.7, \ 10.2, \ 10.7, \ 11.0 ] $ as a function of redshift are shown in Fig. \ref{fig:fsat-vs-z}. The general trend at all mass thresholds is an increase of the satellite fraction as cosmic time flows. For example, galaxies with masses log$\,M_*/M_{\odot} > 9.7$ see an increase from about $10 \%$ at $z \sim 3$ to $\sim 18 \%$ at $z \sim 1.5$ all the way up to $\sim 25 \%$ at $z \sim 0.9$. In the lowest bin $0.2 < z < 0.5$, $f_{\rm sat}$ appears to systematically drop by about 3-4\% for all stellar mass thresholds. This is likely results from a feature in the data, since the survey is not optimized for low redshifts. The fraction of satellites depends on the stellar mass threshold -- at all redshifts there are more low-mass satellites than high-mass ones. Furthermore, the increase with redshift is different with respect to the stellar mass threshold -- the fraction of high mass satellites increases more slowly, only to reach $\sim 8 \%$ at $z \sim 0.6$. 
We note that the $f_{\rm sat}$ in $2.5 < z < 3.0$ have all very similar values, which is an artifact arising from systematic errors in the HOD parameters. We investigated that this is mainly driven by $\beta_{\rm cut}$ parameter which is poorly constrained.


The satellite fraction $f_{\rm sat}$ as a function of stellar mass threshold rises sharply from very massive to intermediate-mass satellites but then reaches a plateau for intermediate to low-mass thresholds, especially at low $z$. This can be explained by the fact that low-mass galaxies are preferentially central galaxies in smaller halos rather than being satellites in more massive halos. This can be understood considering the halo mass function and the halo occupation function: even though the number of satellites increases as a power law with halo mass, there are simply more low-mass halos that can host a lower mass central; furthermore, the exponential high-mass cut-off of the halo mass function means that high mass halos that can host many low-mass satellites are very rare. Therefore, at a fixed low redshift the satellite fraction increases with decreasing stellar mass threshold and reaches a plateau at about $30 \%$.




\subsection{Inferred SHMR for centrals} \label{sec:inferred-SHMR}

\begin{figure}[t]
    \centering
    \includegraphics[width=0.99\columnwidth]{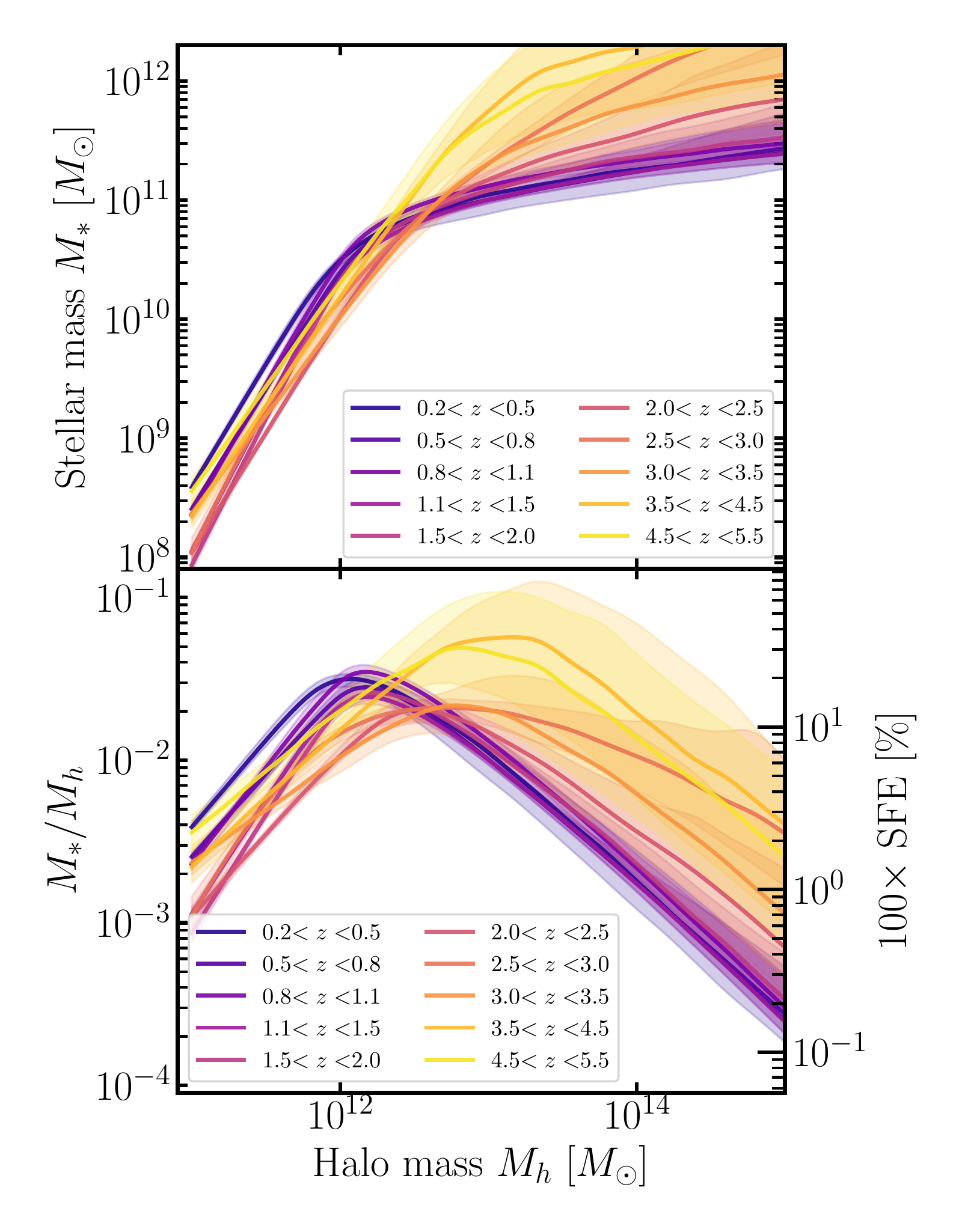}
    \caption{Stellar-to-halo mass relation (top) and $M_*/M_h$ ratio (bottom) in the ten redshift bins. The solid lines and shaded regions show our inferred SHMR and $1 \sigma$ confidence interval color-coded according to the redshift bin.}
    \label{fig:SHMR-all-vs-lit}
\end{figure}

The SHMR and $M_*/M_h$ ratio for centrals are shown in Fig. \ref{fig:SHMR-all-vs-lit} at all  $z$ in the top and bottom panels, respectively. The shaded region envelops the $16^{\rm th}$ and $84^{\rm th}$ percentiles of the distribution of $M_*$ at a given $M_h$ that is obtained by plugging in the parameters of the MCMC chain in Eq. \ref{eq:shmr-cent}. The solid line corresponds to the $50^{\rm th}$ percentile of this distribution. In the remainder of the paper, the $1\sigma$ confidence intervals are always computed in this way, unless stated otherwise. On the right-hand side of the $y-$axis, we show the corresponding halo star-formation efficiency (SFE) in percentages.

The SHMR increases monotonically with halo mass, changing slope at $M_h \approx 10^{12} \, M_{\odot}$ and $M_* \approx 5 \times 10^{10} \, M_{\odot}$. Below this pivot mass, the SHMR increases steeply with a slop that remains constant with redshift. Above the pivot mass, the slope suddenly decreases and the stellar mass increases more slowly with halo mass. The SHMR is higher at low-$z$ for masses below the pivot, and lower at low-$z$ for masses above the pivot. 

The $M_*/M_h$ ratio, which can be considered as the star-formation efficiency integrated over the halo's lifetime, strongly depends on halo mass. The SFE can be defined as $\epsilon = f_b^{-1} \, M_*/M_h$ to quantify how efficiently baryons are converted into stars in galaxies residing in halos of a given mass -- it is essentially a ratio between the star-formation rate and halo growth rate multiplied by the universal baryonic fraction.
Our results, in line with previous findings, show that at all halo masses and at redshifts at least up to $z\sim 3$ the SFE is lower than $20 \%$, indicating a globally inefficient galaxy formation process. In the last three $z$-bins above $z>3$, our results become very uncertain -- the large error bars on the fitted parameters propagate into large uncertainties on the SHMR that make the interpretation difficult.  This can be due to increasingly smaller sample, especially of high-mass galaxies, as well as uncertainties in the physical parameters and possible cosmic variance effects.

The SFE peaks at  $17 \%$ occurs at halo masses of about $M_h = 2 \times 10^{12} M_{\odot}$. It then decreases rapidly at lower and higher halo masses -- about a $15 \%$ decrease in SFE for a decrease of 1 dex in halo mass, and a $\sim 10\%$ decrease for an increase of 1 dex in halo mass. This behaviour indicates that the majority (around two-thirds) of star-formation occurs in a relatively narrow range of halo masses around this peak \citep[see e.g.,][]{behroozi_average_2013, behroozi_universemachine_2019}. This peak corresponds to stellar mass of about $M_* = 5 \times 10^{10} \, M_{\odot},$ which is the typical $M_*$ mass scale of Milky Way-like galaxies \citep[$\sim 6 \times 10^{10} \, M_{\odot}$,][]{licquia_improved_2015}. With respect to redshift, the peak SFE shows only a mild evolution, generally toward lower values with increasing redshift. This is further discussed in the next subections.

\subsection{Redshift evolution of the peak mass quantities} \label{sec:pivot-mass}

\begin{figure*}[t]
    \centering
    \includegraphics[width=0.99\textwidth]{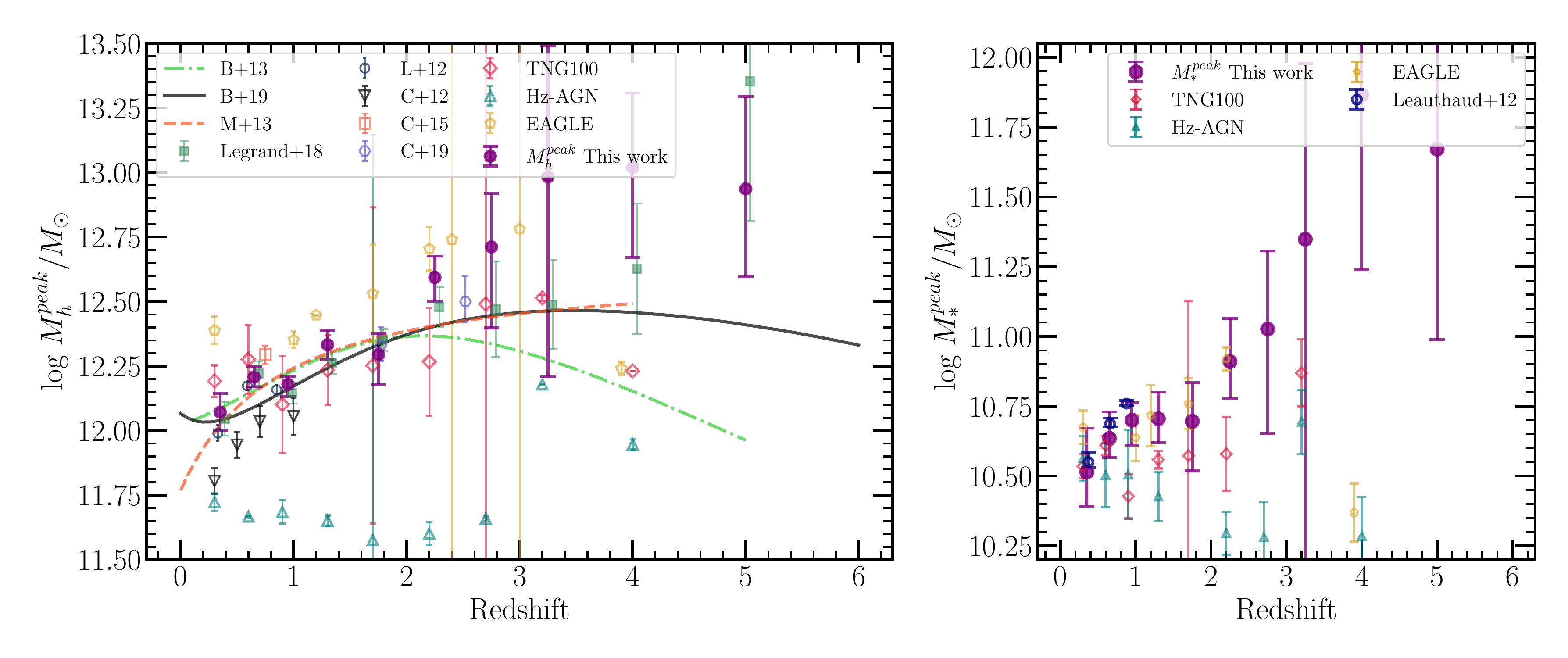}
    \caption{Evolution of the peak halo (left panel) and peak stellar mass (right panel) with redshift. The results from our analysis are shown in purple points. For comparison, we show measurements from the literature rescaled to match our chosen value for $H_0 = 70\, \rm km\, \rm s^{-1} \, \rm Mpc^{-1}$. The literature measurements include \cite{legrand_cosmos-ultravista_2019}, \citet[][L+12]{leauthaud_new_2012}, \citet[][C+12]{coupon_galaxy_2012}, \citet[][C+15]{coupon_galaxy-halo_2015}, \citet[][C+18]{cowley_galaxyhalo_2018}, \citet[][M+13]{moster_galactic_2013} , \citet[][B+13]{behroozi_average_2013}, \citet[][B+19]{behroozi_universemachine_2019}, and from the hydrodynamic simulations {\sc Horizon-AGN}, {\sc TNG100,} and {\sc EAGLE} (references in the main text).
    }
    \label{fig:Mpeak_vs_z}
\end{figure*}

The peak halo mass ($M_h^{\rm peak}$) in the $M_*/M_h$ ratio represents the mass at which the galaxy formation process, integrated over the entire history of the halo, has been most efficient.
Since the feedback mechanisms also depend on halo mass, the redshift evolution of $M_h^{\rm peak}$ informs us about the halo mass scales at which different feedback mechanisms become more important throughout cosmic time. We compute the peak SFE from the central $M_*/M_h$.



\paragraph*{\textbf{Redshift evolution of the peak halo mass.}}
Figure \ref{fig:Mpeak_vs_z} shows the redshift evolution of $M_h^{\rm peak}$ inferred from our analysis, compared to a compilation of measurements from the literature. To obtain  $M_h^{\rm peak}$ and its error bars we compute the peak mass for each parameter set of the MCMC samples; then from this distribution we compute the median, $16^{\rm th}$ and $84^{\rm th}$ percentile. Our results show that the peak halo mass increases with redshift from $M_h^{\rm peak} = 1.43 \times 10^{12} \, M_{\odot}$  at $z=0.35$ to $M_h^{\rm peak} = 4.89 \times 10^{12} \, M_{\odot}$ at $z=2.75$. The peak halo mass continues to increase up to $M_h^{\rm peak} = 7.25 \times 10^{12} \, M_{\odot}$ in our highest bin at $z=5$. At $z>3$, the uncertainty of the peak position increases due to the large uncertainties in the SHMR. While at low redshifts the peak value and evolution is in agreement with the literature, at $z>3$ there is a large scatter in the literature with $M_h^{\rm peak}$ values ranging from $M_h \sim 10^{12} \, M_{\odot}$, as found by \cite{behroozi_average_2013}, to  $M_h \sim 2.5 \times 10^{13} \, M_{\odot}$, as found by \cite{legrand_cosmos-ultravista_2019}.

\paragraph*{\textbf{Redshift evolution of the peak \emph{stellar} mass.}}
The right panel of Fig. \ref{fig:Mpeak_vs_z} shows the evolution of the peak stellar mass $M_*^{\rm peak}$. At the peak stellar mass, galaxies can be considered to have been the most efficient in converting baryons to stars. We find an increase of $M_*^{\rm peak}$ from $3.1 \times 10^{10} \, M_{\odot}$ at $z=0.35$ to $8.7 \times 10^{10} \, M_{\odot}$ at $z=2.75$. This increase means that at earlier times more massive galaxies have been more efficient in the star-formation process; as time elapses, this efficiency moves toward lower mass galaxies. However, the co-evolution of both the peak halo and stellar mass leaves the $M_*/M_h$ ratio nearly constant with time (further discussed in Section \ref{sec:SHMR-at-Mh}).

The trends of increase with increasing redshift of both the peak halo and peak stellar mass is a signature of the downsizing scenario. Downsizing, in its most general sense, refers to the decrease with time of some mass scale parameter that is related to stellar growth or star-formation \citep{1996AJ....112..839C}. In our case, these mass scale parameters are the peak halo and stellar mass, which are related to the efficiency of the star-formation process. Their increase with redshift means that higher mass halos and galaxies were more efficient in converting the baryon reservoir to stars at higher redshifts. Consequently, the feedback mechanisms, especially the ones active at the massive end, were less efficient in the past.

\paragraph*{\textbf{Literature comparison.}}
From the  literature compilation, we remark on comparisons with the following works. \cite{legrand_cosmos-ultravista_2019} used the previous iteration of the photometric catalog in the COSMOS field -- COSMOS2015 -- to infer the SHMR by fitting the same functional form using parametric sub-halo abundance matching. Their results (shown in green squares) are in close agreement with our results. Next, in three $z$-bins up to $z<1,$ we include the results from \cite{leauthaud_new_2012}, which serve as our main reference for the theoretical modeling. Their analysis is based on a joint abundance, clustering, and galaxy-galaxy lensing fit on measurements done in the COSMOS field; their results are shown in dark blue circles. Unsurprisingly, our results are in agreement with the trend. The higher value in the $z\sim0.65$ bin found by our work, \cite{leauthaud_new_2012} and \cite{legrand_cosmos-ultravista_2019}, is likely a feature of the COSMOS field. Indeed, using $10\,000$ spectroscopic redshifts, \cite{Kovac2010} reported a very large overdense structure in COSMOS at these redshifts.
\cite{behroozi_average_2013, behroozi_universemachine_2019} used empirical modeling where galaxies are populated in dark matter halos and are traced within their halos over time; the models are constrained to match a set of observables such as the SMF, luminosity function, and cosmic star formation rate, among others. Their results are in general agreement at low redshifts, but show that the peak halo mass turns over and starts to decrease at $z \sim 2$ or $z\sim3$. Our results, as well as \cite{legrand_cosmos-ultravista_2019}, on the other hand, show a peak halo mass increasing up to $z\sim4$. It is possible that this effect is driven by effectively the same COSMOS dataset used in ours and \cite{legrand_cosmos-ultravista_2019} analyses. 
\cite{behroozi_universemachine_2019}, for example, constrained the SHMR at high redshifts using GSMF derived from \cite{Song2016} data. It is important to note that they used UV-to-$M*$ conversions to estimate stellar masses, which comes with some caveats. The SHMRs at $z\sim3-4$ of \citeauthor{behroozi_universemachine_2019} method are sensitive to the GSMF at $z>4$, and clearly sensitive to the choice of observational constraints \citep[see e.g.,][]{Behroozi2015}.

Finally, in the left panel of Fig.~\ref{fig:Mpeak_vs_z}, we compare the peak stellar mass with the literature. Our results are in good agreement with \cite{leauthaud_new_2012}. Interestingly, no hydrodynamical simulations show a clear downsizing trend in the peak stellar mass.

In summary, our results show that the peak halo mass increases monotonically with redshift in agreement with the literature up to $z<3$, including findings in hydrodynamic simulations. At higher redshift, the literature suggests a turnover and decrease of the peak halo mass, which is not captured by our analysis. The increase of both peak halo and peak stellar mass with redshifts is in accordance with the downsizing scenario, where the more massive halos and galaxies were more efficient in star-forming earlier in the universe and the peak efficiency shifts toward lower mass halos.


\subsection{Total SHMR} \label{sec:res-total-shmr}

\begin{figure}[h]
    \centering
    \includegraphics[width=0.99\columnwidth]{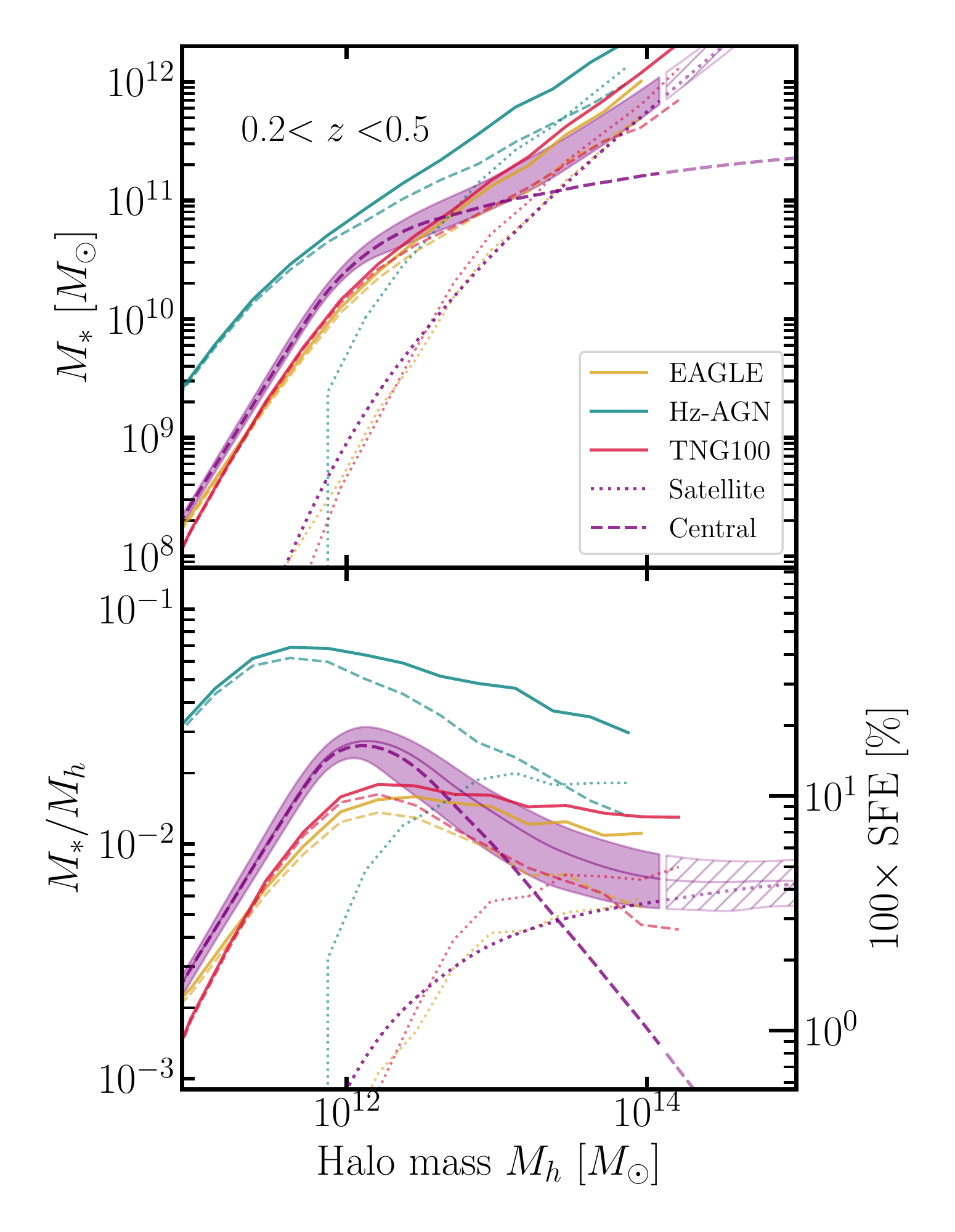}
    \caption{Total stellar-to-halo mass relation (top panel) and total $M_*/M_h$ (bottom panel) at $0.2<z<0.5$ compared to hydrodynamical simulations. The purple dashed and dotted lines show our central and satellite contribution to the total SHMR respectively, with the shaded region showing $1 \sigma$ confidence interval of the sum of the two. The break in solid purple lines and shaded regions indicate the highest stellar mass probed in our analysis, which we take to be the highest mass bin in the SMF. The transparent purple lines and hatched region is an extrapolation at higher masses. We overplot the SHMRs measured in the hydrodynamical simulations {\sc Horizon-AGN} in teal,  {\sc TNG100} in red, and {\sc EAGLE} in dark yellow, where the dashed dotted and solid lines show the central, satellite, and total SHMR.}
    \label{fig:SHMR-tot-z3-vs-lit}
\end{figure}

\begin{figure*}[h]
    \centering
    \includegraphics[width=0.99\textwidth]{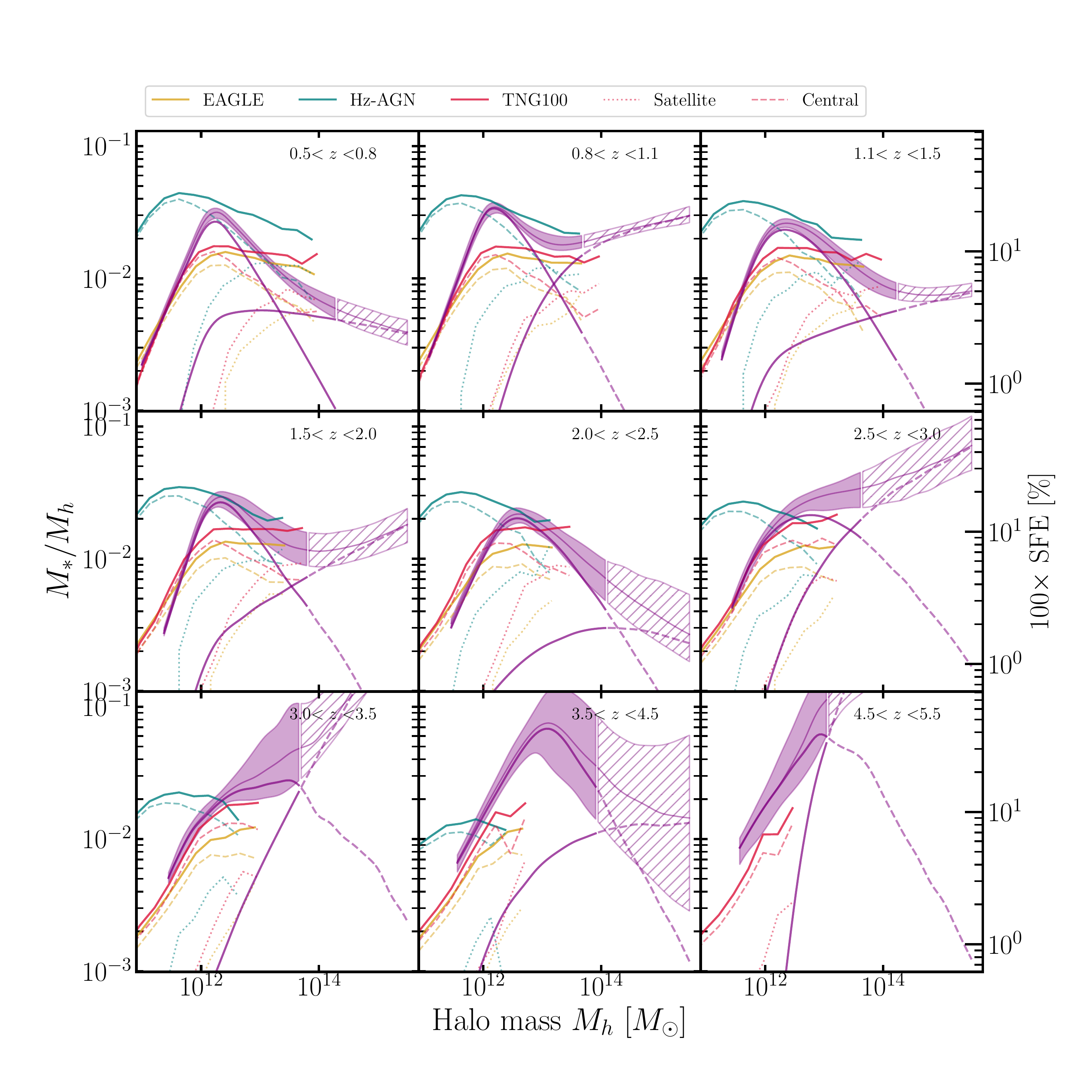}
    \caption{Total $M_*/M_h$ in the different redshift bins compared to hydrodynamical simulations. The purple lines  show our inferred central and satellite contribution to the total SHMR with the shaded region showing a $1 \sigma$ confidence interval of the sum of the two. The break in solid purple lines and shaded regions indicate the highest stellar mass probed in our analysis, which we take to be the highest mass bin in the SMF. The dashed purple lines and hatched region is an extrapolation at higher masses. The comparison includes total $M_*/M_h$ found in the hydrodynamical simulations {\sc Horizon-AGN} in teal,  {\sc TNG100} in red and {\sc EAGLE} in dark yellow. The dashed and dotted lines for the simulations indicate the central and satellite contributions, respectively.}
    \label{fig:SHMR-tot-all-vs-lit}
\end{figure*}

As the mass of the halo increases, the number of satellite galaxies that occupy it also increases and, naturally, their contribution to the total stellar mass budget of the halo becomes important. In massive halos, stellar mass is assembled from in-situ star-formation and from mass accretion via merging of halos, while the growth is regulated by various quenching mechanisms.
The ratio between the total stellar mass and the halo mass (total SHMR) can then inform us about the efficiency of the combination of both effects. The model adopted in this study allows us to compute the total stellar mass contained in a halo of a given mass using Eq.~\ref{eq:total-SHMR}. To obtain the total stellar mass, an integration is carried out over the stellar masses with lower and upper mass limits. Ideally, we would integrate over the whole range of possible stellar masses, but in our case that would mean an extrapolation of the models beyond the stellar masses probed in our analysis. This can introduce inaccuracies, especially for computing the satellite contribution. However, we checked that most of the contribution to the total stellar mass content in $M_h > 10^{12}\, M_{\odot}$ halos comes from satellites in the mass range of $10^{10}\, M_{\odot} < M_* < 10^{11}\, M_{\odot}$, well within the mass scales probed by our analysis; this is also stated in \cite{leauthaud_new_2012}. For our purposes, to compute $M_*^{\rm tot,sat}$, we set the lower integration limit to $M_* = 10^{8.5}\, M_{\odot}$. This lower stellar mass limit is below our completeness limit at $z>2$, but we expect that the extrapolation at lower masses is not inaccurate enough to to bias the results.

Figure \ref{fig:SHMR-tot-z3-vs-lit} shows our results on the total stellar content as a function of the host halo mass in the top panel and $M_*^{\rm tot}/M_h$ ratio in the bottom panel in $0.2 < z <0.5$ and for the other nine $z$-bins in Fig. \ref{fig:SHMR-tot-all-vs-lit}. We show the central SHMR in dashed purple, while the dotted purple line shows the satellite contribution. The shaded purple area envelops the $1\sigma$ uncertainty in the total SHMR. In Fig. \ref{fig:SHMR-tot-all-vs-lit}, both the central and satellite are displayed in solid purple with the $1\sigma$ envelope of the total $M_*^{\rm tot}/M_h$ ratio. The break in the lines and in the shaded region indicates the upper stellar mass limit probed by our analysis, as well as the extrapolation to higher masses is shown in transparent purple lines and hatched region. On the right-hand side $y$-axis, we show the integrated SFE. We recall here that by definition, the integrated SFE describes how efficiently stellar mass has been assembled in halos integrated over the halo's lifetime. This inevitably mixes various assembly paths, such as stellar mass formed in low-mass halos (where different mechanisms regulate growth) that are later accreted as satellites in massive halos. 

At masses below $M_h \lesssim 10^{13}\, M_{\odot}$, the total stellar mass content is completely dominated by central galaxies. It rises sharply up to the peak halo mass at around $M_h \sim 10^{12}\, M_{\odot}$, and then falls more gradually with increasing halo mass. The peak of the total $M_*^{\rm tot}/M_h$ ratio is completely set by the centrals, meaning that the physical processes  shaping the peak have to be related to the quenching of the central galaxy. Since the satellite contribution only becomes important at halo masses almost one order of magnitude higher, the accumulation of stellar mass in satellites (instead of the central growth or mergers), cannot be responsible for setting the peak efficiency.

For masses higher than $M_h \sim 10^{12} \, M_{\odot}$, the satellite contribution to $M_*^{\rm tot}$ rise as the central contribution drops and a transition occurs at about $M_h = 2 \times 10^{13}\, M_{\odot}$ where satellites start to dominate. Going to higher masses, the satellite $M_*^{\rm tot}/M_h$ starts to flatten out, meaning that the stellar mass keeps up with the growth rate of halos in group- and cluster-scale halos. Excluding the $2.5 < z < 3.5$ and $4.5 < z < 5.5$ bins, the total $M_*^{\rm tot}/M_h$ ratio always remains below the peak set by the centrals at an SFE below $5 \%$. This suggests that even if all the satellites were to merge into the central, the SFE would still be lower than the peak, indicating strong environmental quenching mechanisms in massive halos. In other words, the accumulation of mass in satellites is not responsible for the low $M_*/M_h$.


\subsection{Redshift dependence of $M_*/M_h$ at fixed halo mass} \label{sec:SHMR-at-Mh}

\begin{figure}[h]
    \centering
    \includegraphics[width=1.0\columnwidth]{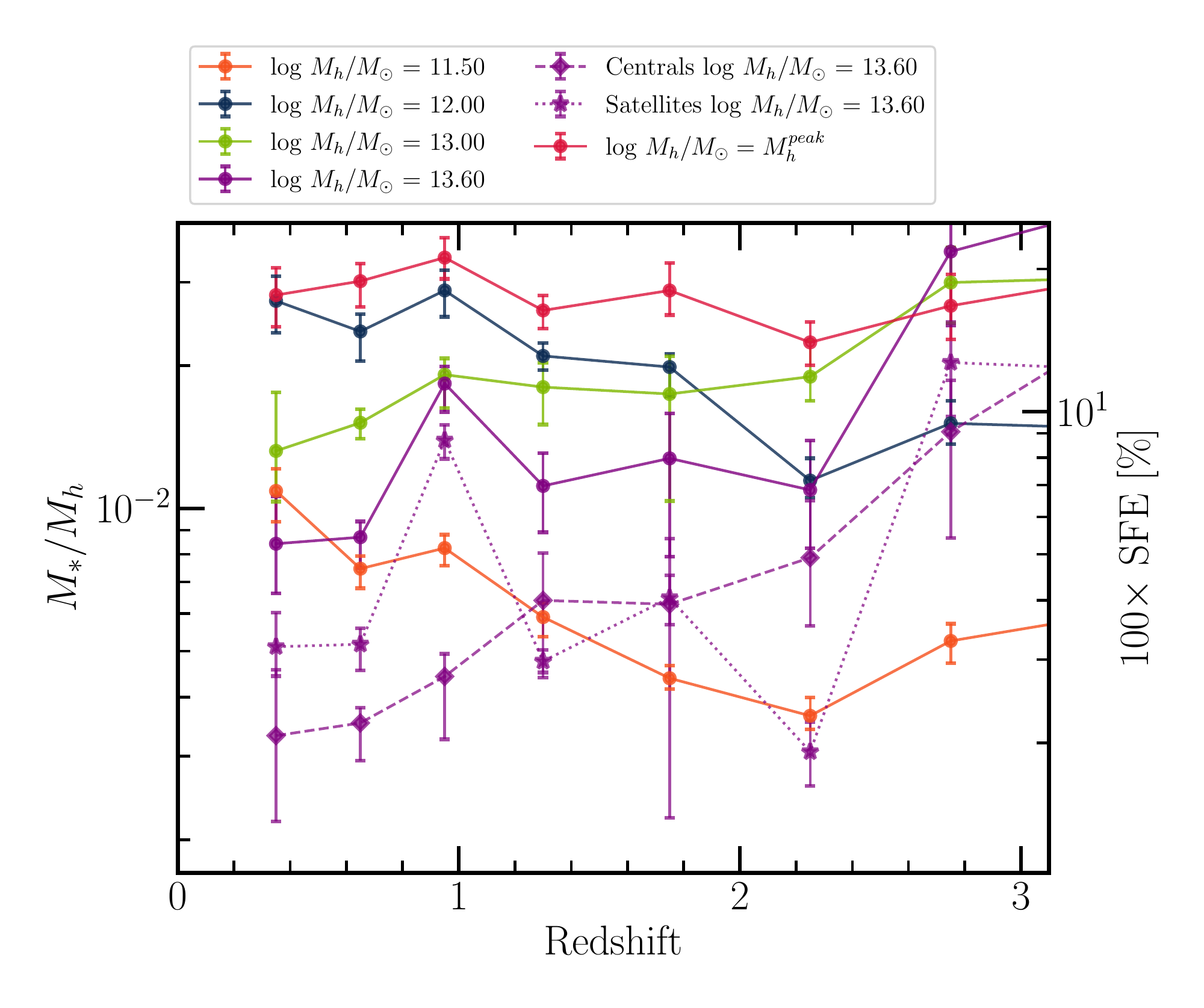}
    \caption{Redshift dependence of $M_*/M_h$ at fixed halo mass. We show the total (centrals and satellites) $M_*/M_h$ fixed at log$\,M_h/M_{\odot} = [11.50, \ 12.00, \ 13.00, \ 13.60, \ M_h^{\rm peak}]$, while the $M_*/M_h$ for log $M_h/M_{\odot} = 13.60$ is also shown for centrals and satellites separately with star and diamond symbols, respectively.}
    \label{fig:SHMR-at-Mh}
\end{figure}

The $M_*/M_h$ ratio (i.e., the integrated star-formation efficiency) might evolve with redshift differently depending on the mass of the halo. This dependence, especially when considering the contributions from both centrals and satellites, could shed light on the importance of the feedback mechanisms that regulate star-formation at different halo mass-scales. 
Figure~\ref{fig:SHMR-at-Mh}  shows the $M_*/M_h$ ratio as a function of redshift at different halo masses log$\,M_h/M_{\odot} = [11.50, \ 12.00, \ 13.00, \ 13.60]$, as well as for $M_h^{\rm peak}$. For the most massive halos at $M_h = 10^{13.6} \, M_{\odot}$, we also decomposed it in the contributions from satellites and centrals. We restricted our analysis to $z<2.5$ since the large uncertainties at higher $z$ prohibit a meaningful quantitative analysis.

Low-mass halos, below the peak halo mass ($M_h < 10^{12} \, M_{\odot}$), show their SFE steadily increasing from $z\sim 2.3$ up to present day. For $M_h = 10^{11.5} \, M_{\odot}$ the SFE goes from $\sim 2.5 \%$ at $z=2.3$ to $\sim 7\%$ at $z=0.3$. For slightly more massive halos of $M_h = 10^{12} \, M_{\odot}$, the SFE increases even faster from $\sim 8 \%$ at $z=2.3$ to $\sim 17\%$ at $z=0.3$. In contrast, for high-mass halos above the peak halo mass ($M_h > 10^{13} \, M_{\odot}$), the SFE remains almost constant, with a slight decrease with decreasing redshift. Furthermore, more massive halos show an even lower SFE by several percents. 

These trends are a signature of the downsizing scenario, which was already mentioned in Section ~\ref{sec:pivot-mass}. Downsizing refers to the observation that, contrary to the hierarchical formation scenario in which small halos are formed first and subsequently grow by merging and accretion, more massive and early-type galaxies have stellar populations that are formed earlier \citep{de_lucia_formation_2006, thomas_environment_2010}. 
This downsizing is observed in our results from the increase of the SFE of low-mass halos and the slight decrease in the efficiency of high-mass halos with decreasing redshift. Low-mass halos having been more efficient in forming stars at later times means that the stellar populations of lower mass galaxies inhabiting them would also be younger. 

The SFE at the peak halo mass shows only a weak evolution with redshift, remaining constant to $z=0.95$ then dropping from $20\%$ to $16\%$ at $z=2.5$. Previous findings also point to a peak efficiency constant with redshift \citep{behroozi_universemachine_2019, moster_emerge_2018}.  This behavior suggests that the $M_h^{\rm peak}$ can be considered as the halo mass scale at which the integrated SFE history remains constant with redshift.

Satellites dominate the mass budget in high-mass halos, as can be seen in Fig.~\ref{fig:SHMR-at-Mh} for $M_h = 10^{13.6} \, M_{\odot}$ where we also show $M_*/M_h$ for centrals and satellites separately. The satellite dominance is stronger at lower redshift, whereas from $z > 1.2 $, the central contribution catches up and both remain comparable. This is unsurprising given the fact that the satellite fraction increases with decreasing redshift.

\section{Discussion}\label{sec:discussion}

\subsection{Physical mechanisms that regulate the stellar mass assembly of centrals}

The shape of the SHMR as a function of halo mass can provide us with qualitative information on the stellar mass growth mechanisms. For example, at $M_h < 10^{12} \, M_{\odot}$, the steep increase in the SHMR with halo mass tells us that the stellar mass growth in galaxies is mainly driven by in situ star formation, as opposed to mergers \citep{conroy_connecting_2009, leauthaud_new_2012}. The explanation comes from the fact that halos grow by merging with lower mass halos, in which the stellar mass drops as a power law (very low mass halos can even be devoid of stars). Since merging will increase the halo mass more significantly compared to the stellar mass, the stellar mass growth in low-mass halos has to be driven by star formation in order to obtain the steep rise of the SHMR. 

The shape of the SFE with its characteristic peak is mostly driven by  different feedback processes, whose efficiencies are dependent on halo mass. At low masses, stellar feedback from supernovae (SNe), stellar winds, radiation pressure, and photoheating can heat up and prevent the baryons from collapsing into stars \citep[e.g.,][]{dekel_origin_1986,dubois_teyssier_08, hopkins_galaxies_2014,kimm_towards_2015}. As the halo mass increases, its potential well deepens and cold flows feed the halo with gas that fuels star-formation. The SNe are not powerful enough to prevent these cold flows, so the SFE increases with halo mass. As the halo mass increases even further, virial shock heating and AGN activity heat up the gas and quench star formation, thus driving the SFE back down \citep[see e.g., ][for a review]{silk_galaxy_2014}. \cite{vogelsberger_model_2013} show in their hydrodynamical simulations that the mechanical, radio-mode AGN feedback is the most responsible for the suppression of star formation in these massive halos \citep[see also e.g.,][]{dubois_jet-regulated_2010}. Similar conclusions have been reached from AGN observations in the COSMOS field, where \cite{vardoulaki_m-mhalo_2021} show that the radio-mode AGN feedback plays a significant role in shaping the SHMR at lower redshifts. Additionally, \cite{gabor_hot_2015} in their hydrodynamical simulations, these authors show that galaxies above the peak mass indeed tend to be quenched. These quenched centrals can still continue to grow in stellar mass via dry merging, but halo mass growth being faster -- since halos keep accreting from their large-scale environment, while the gas accreted by galaxies is not efficiently turned into stars any longer due to feedback -- leads the $M_*/M_h$ ratio to decrease.

In Fig.~\ref{fig:Mpeak_vs_z}, we showed that more massive halos are more efficient in forming stars at higher redshifts -- the peak halo mass increases with $z$. This peak in the central SHMR is shaped by the interplay between cold flows being more efficient as the halo mass increases and shock heating in the hot halo. The increase of the peak halo mass is a signature that cold flows become more important in driving star formation at $z>1.5$ in $M_h>10^{12} \, M_{\odot}$ \citep{dekel_cold_2009,oser_2010,dubois_2013}.

With respect to redshift, we observe in Fig.~\ref{fig:SHMR-tot-all-vs-lit} that the peak generally flattens out at higher redshift. This is directly connected to the fact that the SMF at higher redshifts increases its low mass slope and the knee is smoothed (i.e., the SMF and HMF become more similar in shape). This indicates that the high mass halos decrease in their star-forming efficiency with time and that the AGN feedback has a larger impact at later times.


\subsection{Physical mechanisms that regulate the stellar mass assembly in satellites}

In Section \ref{sec:SHMR-at-Mh} and Fig. \ref{fig:SHMR-at-Mh}, we presented how the total SHMR ratio evolves with redshift at fixed halo mass. Comparing the contributions of centrals and satellites to the total stellar mass content at different $M_h$ can shed some light on the relative importance of the quenching mechanisms acting on the central and satellite galaxies in a halo.
Figure \ref{fig:SHMR-at-Mh} shows that in high-mass halos of about $M_h = 10^{13.6} \, M_{\odot}$, the $M_*/M_h$  is about $0.2 \, \rm dex$ lower than $M_*/M_h$ at the peak halo mass.
At least up to $z<2.5$, the total $M_*/M_h$ ratio for high-mass halos is always lower than $M_*/M_h$ at the peak halo mass. This suggests that the low $M_*/M_h$ ratio for centrals is not due to the stellar mass being accumulated in satellites instead of the central because, even if all satellites were to merge with the central, the total stellar mass content would still be lower than at the peak. This is a clear indication of powerful feedback present in massive halos that prevents gas from cooling and forming stars, and these feedback processes are acting on both the central and the satellites. 

In halos more massive than $\sim 10^{12} \, M_{\odot}$, which includes group- and cluster-scale halos, the virial shock heats the halo gas \citep{birnboim_virial_2003}. Naturally, then, satellite galaxies that reside in these massive halos have their SF impeded by the hot gas. This hot-halo mode is likely to be responsible for the satellite quenching. This mechanism is also called environment quenching \citep{peng_mass_2010, gabor_hot_2015} and is found to be independent of stellar and halo mass -- galaxies of all stellar masses reside in the densest environments in halos of $M_h > 10^{12}$. Additional environmental effects such as strangulation \citep{1980ApJ...237..692L, 2000ApJ...540..113B}, ram-pressure stripping \citep{1972ApJ...176....1G}, and harassment \citep{moore_galaxy_1996} further prevent the stellar mass assembly in satellites. Altogether, these processes could explain the low SFE at high halo masses in our results. Our results at high masses remain too uncertain to unveil a consistent picture of the redshift evolution of the efficiency of the satellite quenching mechanisms. Incorporating lensing measurements to better constrain the satellite contribution and separating active versus passive populations could shed more light on this aspect. This is a natural extension to this work that we plan to carry out in the future.


\subsection{Comparison with hydrodynamical simulations} \label{sec:comparison-hydro}
We compare our results with several state-of-art cosmological hydrodynamical simulations of galaxy formation: {\sc Horizon-AGN}\footnote{\url{www.horizon-simulation.org}} \citep{dubois_dancing_2014,kaviraj_horizon-agn_2017}, {\sc EAGLE}\footnote{\url{http://eagle.strw.leidenuniv.nl/}} \citep{crain_eagle_2015,schaye_eagle_2015}, and {\sc TNG100} of the {\sc IllustrisTNG} project \footnote{\url{https://www.tng-project.org/}}\citep{Nelson2019-release,springel_first_2018,nelson_first_2018,Pillepich2018_presentation,marinacci_first_2018,naiman_first_2018}. A brief recap of the main features of these simulations is given in Appendix~\ref{apdx:sims-described}. Once central and satellite galaxies are matched with DM halos as described in Appendix~\ref{apdx:sims-described}, the SHMR is directly measured in halo mass bins at different redshifts, as the mean of the ratio between the stellar and the halo mass. Baryonic physics has a small but real impact on the halo mass function.  \cite{beltz-mohrmann_impact_2021} showed that the halo mass function in DM-only simulations is overestimated (with respect to hydrodynamical simulations) by a factor of about $\sim$1.1-1.2 depending on the mass range at $z=0$, though this effect disappears at high redhsift and high mass \citep[see also e.g.,][]{bocquet_halo_2016,desmond_galaxyhalo_2017,castro_impact_2020}. This aspect should be kept in mind, knowing that the modeling presented in Sec.~\ref{sec:modeling} relies on DM-only halo mass function, while for the hydrodynamical simulations, the mass of the DM host halos is naturally impacted by baryonic physics. This could lead to a small systematic underestimation of the SHMR in the observational data with respect to the simulated ones.

\paragraph*{\textbf{Redshift evolution of the peak halo mass.}}
In Fig.~\ref{fig:Mpeak_vs_z}, we show the evolution of the peak halo mass in the three hydrodynamic simulations {\sc Horizon-AGN}, {\sc TNG100,} and {\sc EAGLE}. We directly compute the peak halo mass using the measured SHMR in the simulations. The errorbars are obtained by computing the peak of the lower and upper $1\sigma$ values of the $M_*/M_h$ relation as measured in the simulations.
Here,{\sc \ TNG100} shows the best agreement with both our results and the literature, whereas {\sc EAGLE} shows slightly higher peak halo masses up to $z\sim3$. We also note that in both simulations, the peak halo mass decreases from $z\sim3$ to $z\sim4$, which is a priori in disagreement with our results. That being said, the ratio $M_*/M_h$ at $z>3$ is also highly uncertain  (as seen in Fig.~\ref{fig:SHMR-tot-all-vs-lit}) because the simulations start to miss massive halos with masses comparable to the peak due to their fixed small volume and the rarity of these halos. {\sc Horizon-AGN} shows the biggest discrepancy with our results and the literature. In this simulation, the peak efficiency is reached in much lower mass halos: $M_h^{\rm peak} = 10^{11.75} \, M_{\odot}$ at $z \sim 0.35$ that decreases up $10^{11.55} \, M_{\odot}$ to $z\sim 2$ and then increase at higher redshifts. This discrepancy can be explained by the inefficiency of the SN feedback in low mass galaxies as implemented in the simulation (see e.g., \cite{hatfield_comparing_2019} or \cite{kaviraj_horizon-agn_2017} for further discussions).


\paragraph*{\textbf{The simulated SHMR in the low-mass regime.}} 
We present the $M_*^{\rm tot}/M_h$ for {\sc Horizon-AGN} (in teal), {\sc TNG100} (red), and {\sc EAGLE} (dark yellow color) in Figs. \ref{fig:SHMR-tot-z3-vs-lit} and \ref{fig:SHMR-tot-all-vs-lit}. 
At the low-mass end, where $M_*^{\rm tot}/M_h$ increases with halo mass, {\sc TNG100} and {\sc EAGLE} show a reasonable agreement with our results with a similar normalization, slope and position of the peak, albeit the peak SFE is lower by a few percent. 
{\sc Horizon-AGN}, on the other hand, has a higher normalization of the $M_*^{\rm tot}/M_h$ and a peak efficiency at lower halo masses (as already noted above) probably driven by insufficient SN feedback especially at high redshift. This issue is also responsible for the overall overestimation of the stellar mass in low-mass galaxies. 
The good agreement of {\sc TNG100} and {\sc EAGLE} with our results in this regime suggests that the implementation of feedback from SNe and stellar winds in these simulations is well calibrated to reproduce observational data.  We note that we cannot conclude from this measurement that the star-formation model at the subgrid scale is realistic nor that stellar feedback in these simulations is physically meaningful (some feedback mechanisms could be missing, such as radiation from young stars that suppresses star formation in low-mass galaxies, and others could be overestimated). What we can conclude is simply that the cumulative strength of the implemented feedback  processes leads to realistic SHMR relations (see the discussion on subgrid models in Appendix~\ref{apdx:subgrid}). 


\paragraph*{\textbf{Simulated SHMR in the high-mass regime.}} 
The most significant discrepancies between the observational data and all simulations appear for masses above the peak. At least up to $z=3.5$ the central $M_*^{\rm cent}/M_h$ for all the simulations shows a peak, meaning that AGN feedback are powerful enough to quench the central galaxy growth compared to the halo growth. However, at all redshifts in {\sc TNG100} and {\sc EAGLE}, the central $M_*/M_h$ decreases with increasing halo mass more gradually (shallower slope) than in observations, while {\sc Horizon-AGN} shows this tendency only at $z<1$. Furthermore, the contribution of satellites relatively to central at all redshifts is higher in the simulations. These two facts contribute to a flattening of the peak and a higher SFE with increasing mass in the simulations, especially in {\sc TNG100} and {\sc EAGLE}.

\paragraph*{\textbf{Redshift evolution of the satellite fraction.}}
In Fig.~\ref{fig:fsat-vs-z} we also compare the satellite fractions against the three hydrodynamical simulations. The dashed lines in the left, middle, and right panels show the $f_{\rm sat}$ in {\sc TNG100}, {\sc Horizon-AGN}, and {\sc EAGLE,} where the different color-coded lines correspond to the mass thresholds indicated in the legend. The satellite fraction in all simulations was computed with the same criterion as in Eq.~\ref{eq:sat_frac} -- the ratio between the number of $M_* > M^{\rm th}$ galaxies that are satellites and the total number of $M_* > M^{\rm th}$ galaxies for all the halos in the simulation. 
The satellite fraction in  all the simulations increases linearly with decreasing redshift, a trend which is in agreement with our results. However, at fixed stellar mass thresholds, the simulations exhibit higher numbers of satellites at all redshifts compared to our analysis. In {\sc TNG100} and {\sc EAGLE}, $f_{\rm sat}$ goes from $\sim 20 \%$ at $z\sim3$ to $\sim 42 \%$ at $z\sim0.3$ for satellites with $\log M_*/M_{\odot} > 9.0$. Compared to our results, the satellite fraction is higher by about a factor of 2 for the lowest mass thresholds and a factor of 1.2 for the highest mass thresholds. In {\sc Horizon-AGN,} the satellite fraction is lower by about $0.5$ compared to the other two simulations, but still a factor of $1.4$ higher than our results for the lowest mass thresholds.  

The difference in the satellite fractions can be due to a degeneracy between the satellite fraction and the spatial distribution of satellites within halos \citep{Xu2016} -- steeper satellite profiles at small scales increase the correlation, thus decreasing the need for high satellite fraction. This raises the possibility that the higher satellite fraction in the simulations is due to flatter satellite profiles compared to our models which assume an NFW. \cite{Bose2019} show that the mean radial number density of luminous satellites in {\sc TNG100} matches the NFW profile. Furthermore, we compared the radial satellite density distribution in all three simulation and find that they follow the NFW profiles reasonably well, with {\sc TNG100} showing the best match, while the {\sc EAGLE} and {\sc Horizon-AGN} show flatter profiles that are consistent with NFW but for lower concentration parameters. The agreement in {\sc TNG100} suggests that satellite profiles cannot serve as the explanation behind the different satellite fractions. In {\sc EAGLE} and {\sc Horizon-AGN}, on the other hand, the flattening of the profiles can be interpreted as a consequence of AGN feedback \citep[see e.g.][for comparison of DM halo profiles with and without AGN]{Peirani2017}.

Excess of satellites in halos "above fixed mass thresholds" would translate into a higher level of small-scale clustering, which is measurable by the two-point correlation function of low-mass and red galaxies. \cite{artale_small-scale_2017} compares the small-scale correlation function of galaxies in {\sc EAGLE} and the GAMA survey at $z=0.1$. They confirm that low-mass red galaxies have a considerably higher correlation at small scales, which is consistent with high fractions of satellite galaxies. Although they did not investigate this feature specifically, \cite{springel_first_2018} mentioned that the present-day clustering of red galaxies less massive than $10^{10}~M_{\odot}$ was overestimated in {\sc TNG100}  with respect to the SDSS. 
This difference in the satellite fractions corroborates our measurement of the $M_*/M_h$ in Fig.~\ref{fig:SHMR-tot-all-vs-lit}, where the satellite contribution in the simulations is usually higher than our measurements. 
A higher fraction of satellites above a mass threshold could be the consequence of satellite galaxies having grown more massive due to lack of quenching, at least during a certain period of cosmic time. Due to the power-law increase of the SMF at the low-mass end, inefficient satellite quenching would shift the satellite SMF toward higher masses. Then, above a stellar mass threshold, there would be more satellites. This indicates that satellite quenching in the hydrodynamical simulations is inefficient, or that it happens at later times in the halo lifetime, or both. 

It is worth noting that \cite{donnari_quenched_2021} made a thorough comparison of the quenching fraction between {\sc TNG100} simulation and observations up to $z=0.65$. They concluded that TNG quenched fractions of centrals and satellites are qualitatively in agreement with observations. The scenario behind our findings might therefore lie at higher redshift: if strong satellite quenching occurs too late in the simulations, these galaxies would have had time to grow more massive than in the observations, leading to the overestimation of the satellite SHMR at high halo mass. We note that, at least for {\sc TNG100} and {\sc EAGLE}, this quenching inefficiency must be specific to galaxies that have already been accreted as satellites: indeed  the SHMR of low-mass central galaxies at high redshift is in good agreement with our observational measurements\footnote{The issue is different for {\sc Horizon-AGN}, as previously noted. In this simulation, all low-mass galaxies (central and satellites) are concerned by inefficient quenching, which suggests a different type of failure or mis-modeling in the simulation.}. 
It is possible to argue that this issue is related to the resolution limit of the simulations. The coarse resolution reached in the circum-galactic medium might be insufficient to correctly model high gas temperature {inherited from virial shock heating, preventing an overly strong quenching of satellite galaxies in massive halos  \citep{gabor_hot_2015}.} However, this interpretation does not hold to explain the too large contribution of satellites at high redshift in halos less massive than $M_h > 10^{12} \, M_{\odot}$ (which are less prone to shock heating). For those very small satellites, we could suspect that other environmental mechanisms (e.g., ram-pressure and tidal stripping, harassment) are not correctly modeled, also due to the lack of resolution in the simulations. Interestingly, \cite{Costa2019} also shows that the efficiency of satellite tidal stripping depends on the degree of pre-processing of low-mass galaxies by radiative stellar feedback (which are not implemented in the simulations studied here. See also \citealp{Katz2020} for a discussion of stellar radiation during reionization). Therefore, the lack of satellite quenching at high redshift in the cosmological simulations  studied here could also be a consequence of the absence of pre-processing by radiative stellar feedback. 

\subsection{Sources of uncertainties and the effects of model assumptions on the inferred SHMR} \label{sec:hmf-on-inferredSHMR}

The measurements presented here can suffer from a number of systematic errors. Firstly, due to the relatively small volume probed by COSMOS, the effects of sample variance can lead to biased measurements. This might be the case at $z\sim 1$, where several works indicate an overabundance of rich structures \citep[][and references therein]{mccracken_probing_2015}. Such an overdensity increases the normalization of the SMF and adds extra power on intermediate and large scales in $w(\theta)$. The effect of these overabundances on the inferred SHMR would be to decrease this ratio -- indicating an even lower efficiency of converting baryons to stars -- and a shift of $M_h^{\rm peak}$ toward higher masses. 

A second source of systematic error are uncertainties in the estimation of physical parameters, for instance, stellar masses and redshifts \footnote{for a detailed analysis of the uncertainties affecting the SHMR see \cite{behroozi_comprehensive_2010}}. Stellar mass uncertainties propagate into the Eddington bias at the high-mass end of the SMF. The effect of the Eddington bias on the inferred parameters is an increase in the value of $\sigmaLogMs$ which sets the scatter in stellar mass at fixed halo mass. 

\begin{figure}[h]
    \centering
    \includegraphics[width=1.0\columnwidth]{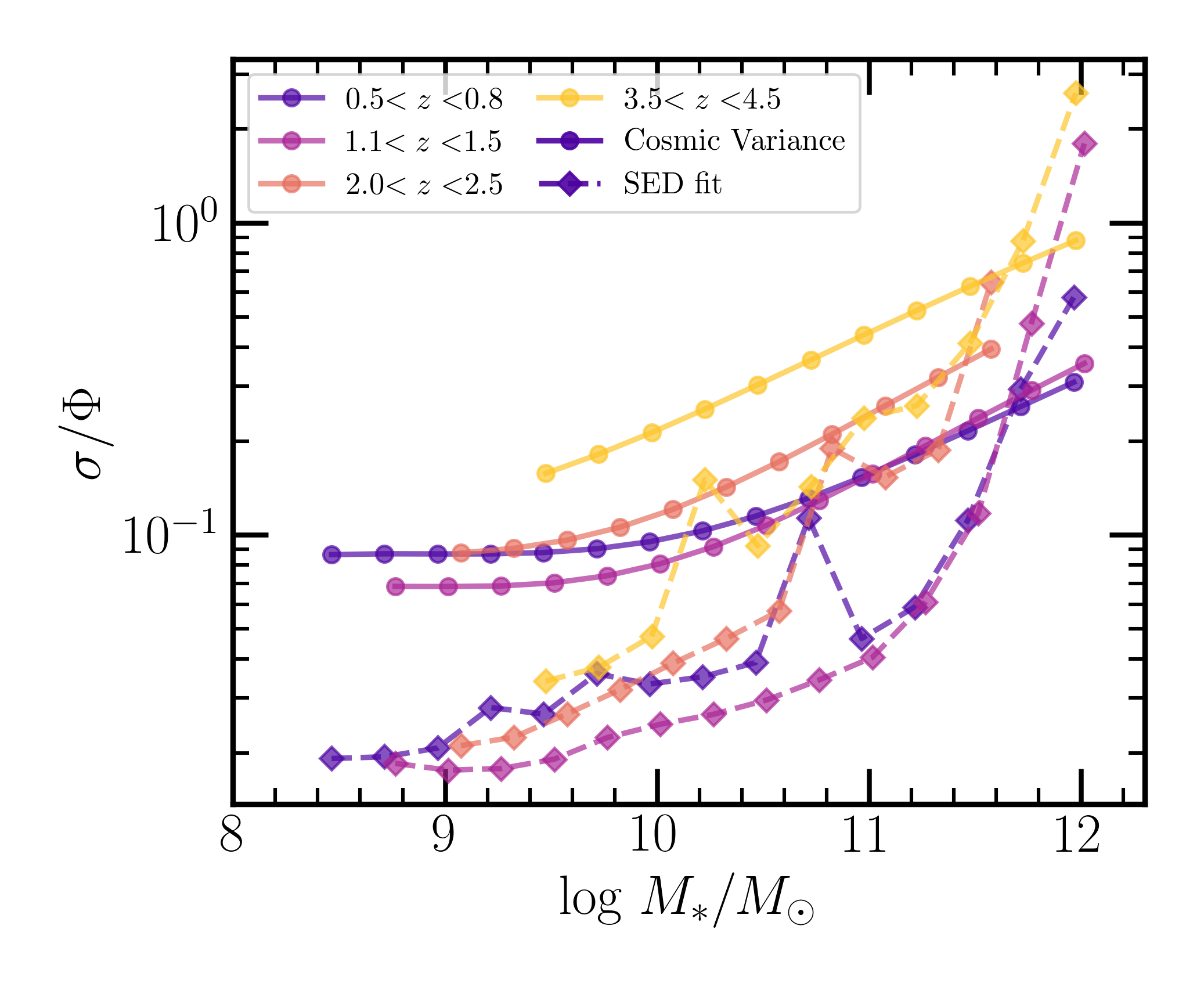}
    \caption{Uncertainties in the SMF expressed as a fractional error $\sigma/\Phi$. We show the uncertainties from cosmic variance with solid line-connected points and the uncertainties from SED fitting in dashed line-connected diamonds. We show the uncertainties for four redshift bins.}
    \label{fig:SMF_errorbudget}
\end{figure}

In Fig.~\ref{fig:SMF_errorbudget}, we show the uncertainties in the SMF as a function of stellar mass and redshift. We show the fractional error ($\sigma/\Phi$) from the cosmic variance (CV) and SED fitting. We do not show the Poisson errors because they are subdominant. Cosmic variance dominates the error budget by about 0.8 dex at $M_* < 10^{11} \, M_{\odot}$ compared to the SED fitting. Both increase with mass, but the SED fitting uncertainties increase exponentially at $M_* > 10^{11} \, M_{\odot}$ and become dominant over CV at the most massive end ($M_* > 6 \times 10^{11} \, M_{\odot}$). The most massive galaxies are rare and reside in the densest regions, therefore, a small survey is more likely to get a biased view of these objects. However, due to photometric errors and degeneracies between different SEDs, their stellar masses come with uncertainties. Due to the exponential shape of the SMF at the high mass end, these uncertainties in the $M_*$ translate into large uncertainties in the SMF (Eddington bias) and become dominant. The SED fitting uncertainties in the $M_*$ uncertainties being amplified by the Eddington bias into large $\sigma_{\rm SED}$ means that it will be very difficult, even for future surveys, to improve on these uncertainties.

Additional systematic errors arise from the number of assumptions in the model. The ingredients that go into the HOD-based model of the observables are the halo abundance (HMF), their clustering properties (halo bias, $b_h$), the radial distribution of dark matter and galaxies within halos, and halo mass definitions. The literature abounds with prescriptions for all of these ingredients, which can all lead to different results for the SHMR. This often makes the comparison with the literature difficult. \cite{coupon_galaxy-halo_2015} has investigated the effect on various model assumptions such as the $\sigma_8$ value, $b_h(M_h)$ relation, assembly bias, mass concentration relation, and halo profiles, on the error budget of several HOD parameters. Their conclusion is that the model systematics can lead to errors comparable to the statistical errors. 

\section{Summary and conclusions} \label{sec:summary-conclusion}

This paper presents the redshift evolution of the stellar-to-halo mass ratio to $z\sim5$, derived from measurements of the galaxy stellar mass function and angular correlation function in the COSMOS2020 catalog fitted to a phenomenological model. The advantages of our work is the use of a single, homogenous dataset to perform all the measurements. Additionally, the HOD-based modeling allows us to consistently probe the contribution to the total stellar mass budget in halos of both central and satellite galaxies over this large redshift range, for the first time in the literature.

Our principal results are as follows: 

\begin{itemize}
    

    \item The mean halo occupation shows little-to-no evolution with redshift at fixed stellar mass, suggesting that galaxies occupy dark matter halos similarly throughout cosmic history.

    \item The $M_*/M_h$ ratio for central galaxies -- which may be interpreted as the integrated star formation efficiency (SFE) -- strongly depends on halo mass, increasing up to a peak at halo masses of around $2\times10^{12}$ and then decreasing again as the halo mass increases. The SFE shows little-to-no evolution with redshift and remains lower than $20 \%$ at least up to $z\sim3$, indicating a globally inefficient galaxy formation process. The peak levels off with increasing redshift, consistent with a scenario in which AGN feedback in higher mass halos is less important at earlier times.
    
    
    \item The halo mass and stellar mass scale at which the SFE peaks, $M_h^{\rm peak}$ and $M_*^{\rm peak}$, increase continuously with redshift at least to $z\sim4$. This stands in contrast to other works \citep[e.g.,][]{behroozi_universemachine_2019} where the peak halo mass decreases beyond $z \gtrsim 3$. However, given our errors the peak halo mass evolution at these high redshifts remains uncertain.  
    
    \item The total stellar mass content of halos, $M_*^{\rm tot}/M_h$, shows that at $M_h \lesssim 10^{13} \, M_{\odot}$ central galaxies completely dominate the stellar mass budget of the halo at all redshifts. The peak of the $M_*^{\rm tot}/M_h$ ratio is set by the central galaxy, indicating that the physical processes that shape the peak efficiency are related to the quenching of central galaxies.
    
    \item Satellite galaxies start to dominate the total stellar mass budget at $M_h > 2 \times 10^{13} \, M_{\odot}$ and the $M_*^{\rm tot}/M_h$ ratio flattens out as the halo mass increases. The fact that in the satellite-dominated regime, the $M_*^{\rm tot}/M_h$ ratio is lower than the peak means that strong quenching mechanisms must be present in massive halos that quench the mass assembly of satellites.
   
     \item For all samples, the satellite fraction decreases at higher redshifts. Moreover, there are always more low-mass satellites than high-mass ones. The satellite fraction increases more steeply for lower-mass satellites and reaches $\sim 25 \%$ at low redshifts, whereas the most massive galaxies reach up to $\sim 15 \%$.

\end{itemize}

We compared our SHMR measurements with three state-of-the-art hydrodynamical simulations {\sc Horizon-AGN}, {\sc TNG100,} and {\sc EAGLE}. For low halo masses, our results are in general agreement with {\sc TNG100} and {\sc EAGLE} in terms of slope and peak of the $M_*^{\rm tot}/M_h$ ratio. However, there is a significant discrepancy with {\sc Horizon-AGN} that can be explained by insufficient stellar feedback.
The most significant discrepancies with the simulations are for masses above the peak. The $M_*^{\rm tot}/M_h$ ratio in the simulations flattens out at the peak and has a larger value at higher masses compared to our results, which is mainly driven by a higher satellite contribution at all redshifts in the simulations -- this excess in the satellite contribution relative to the central at the high-end is higher in {\sc TNG100} and {\sc EAGLE} than in {\sc Horizon-AGN}. Furthermore, the simulations show higher fractions of satellite galaxies at all redshifts and all masses by about a factor of two. Both these findings at the high-mass end -- excess of $M_*^{tot}/M_h$ and high satellite fractions -- suggest that the environmental quenching of satellites is less efficient in the simulations or that quenching occurs much later. This can be due to either an inefficient hot halo quenching mode, or from other environmental effects, such as ram-pressure or tidal stripping or harassment from other satellites, which are not well captured in the simulations -- probably due to resolution effects. Lack of pre-processing by stellar radiative feedback could also have an impact. 

To date, the COSMOS photometric redshift catalogs have provided the only homogeneous dataset to consistently study the evolution of the SHMR over a large redshift range (since $z\sim5$) for a statistically representative area of the sky. However, the $2 \, \rm deg^2$ of the survey does not eliminate the effects of cosmic variance; at $z>2.5$, the uncertainty in our results makes it difficult to provide a convincing interpretation and comparison with simulations. 

Our analysis is based on a phenomenological model that cannot provide any further insight into the physical processes governing the shape of the SHMR, especially the relative contribution of different feedback modes. But comparisons with hydrodynamical simulations, where the effects of  feedback on the SHMR can be traced back, could provide insights into the relative importance of the different feedback mechanisms acting at different mass scales and environments and, ultimately, can help bridge the gap between observations and simulations. 

One of the main issues raised by our comparisons with hydrodynamical simulations concerns satellite quenching. To investigate how satellites are quenched with respect to their DM halo throughout cosmic time, we could further separate the sample into star-forming and quenched galaxies. Upcoming works and surveys will make this possible. In the imminent future, COSMOS-Web will provide JWST observations in four NIR bands down to ${\rm AB}\sim 27$ over $0.6 \, \deg^2$ in COSMOS. This unprecedented resolution and depth in NIR will enable lensing measurements (such as galaxy-galaxy lensing) to $z\sim 2.5$ and make it possible to measure the SHMR dependence on star-formation activity and even color gradients. At a slightly longer timescale, the Cosmic Dawn Survey will carry out deep NUV to MIR observations in $\sim 50 \, \deg^2$ in the Euclid Deep Fields. Accurate photometric redshift and stellar mass measurements from this survey will probe the most massive end of the SHMR while greatly reducing cosmic variance.

\begin{acknowledgements}

This research is partly supported by the Centre National d'\'Etudes Spatiales (CNES).
MS acknowledges Elena Sarpa, Lukas Furtak and Louis Legrand  for useful discussions. Part of this study benefitted from earlier, unpublished work by J. Coupon.
MS acknowledges the partial thesis funding by \textit{Euclid}-CNES.
HJMcC and CL acknowledge support from the \textit{Programme National Cosmology et Galaxies} (PNCG) of CNRS/INSU with INP and IN2P3, co-funded by CEA and CNES.
I.D. has received funding from the European Union's Horizon 2020
research and innovation program under the Marie Sk\l{}odowska-
Curie grant agreement No. 896225.
This work used the CANDIDE computer system at the IAP supported by grants from the PNCG, DIM-ACAV and CNES and maintained by S. Rouberol. 
This work is based on data products from observations made with ESO Telescopes at the La Silla Paranal Observatory under ESO program ID 179.A-2005 and on data products produced by CALET and the Cambridge Astronomy Survey Unit on behalf of the UltraVISTA consortium. This work is based in part on observations made with the NASA/ESA \textit{Hubble} Space Telescope, obtained from the Data Archive at the Space Telescope Science Institute, which is operated by the Association of Universities for Research in Astronomy, Inc., under NASA contract NAS 5-26555.

\end{acknowledgements}
\bibliographystyle{aa} 
\bibliography{references.bib}

\begin{appendix}
\section{Impact of $N(z)$ on $w(\theta)$} \label{apdx:nofz-on-w}
\begin{figure}[h]
    \centering
    \includegraphics[width=0.99\columnwidth]{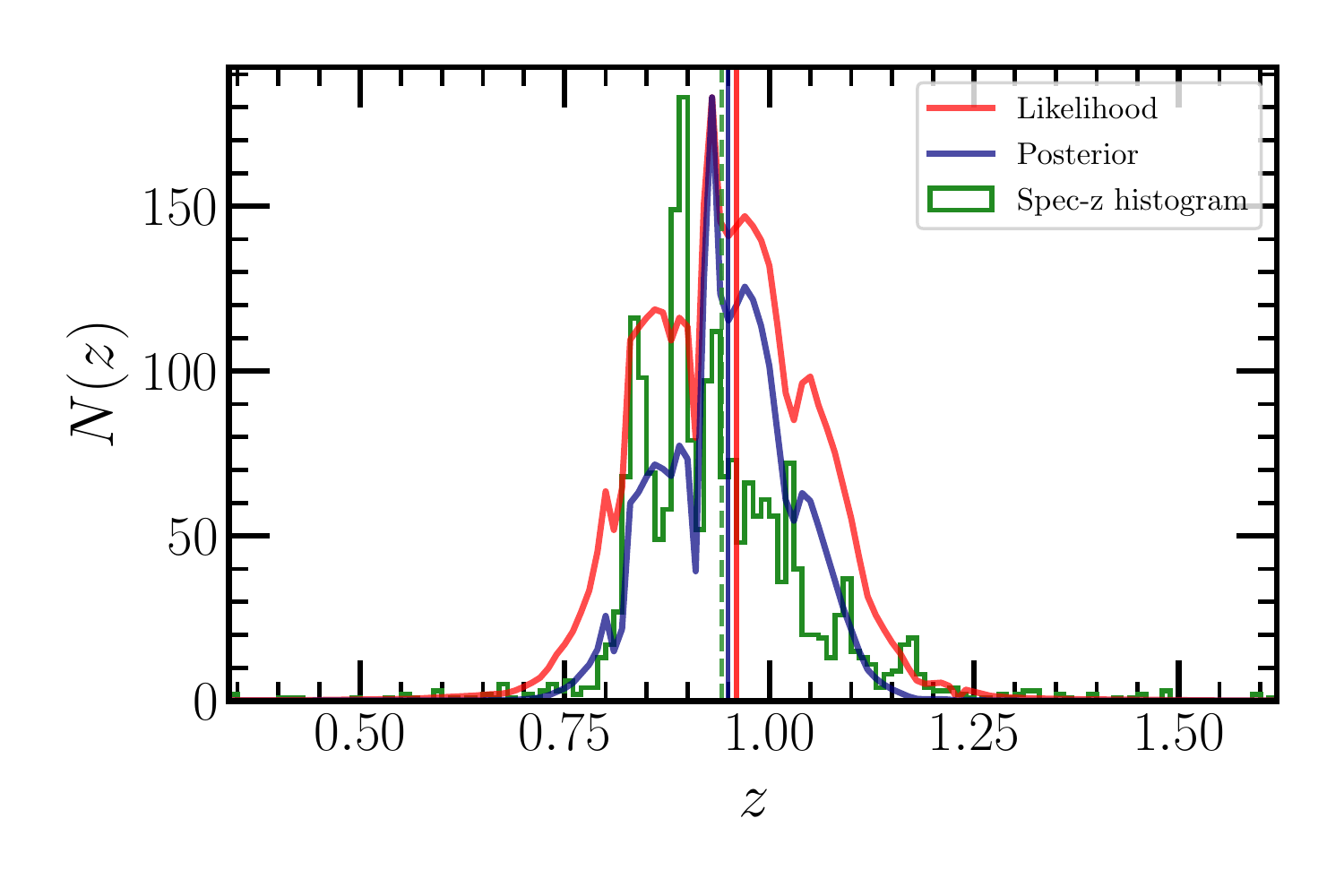}
    \caption{ Redshift distributions based on spec-$z$ histogram and stacking of photo-$z$ likelihood and posterior distributions. All distributions are normalized to the maximum of the spec-$z$ histogram.}
    \label{fig:appendix-nofz}
\end{figure}

\begin{figure}[h]
    \centering
    \includegraphics[width=0.99\columnwidth]{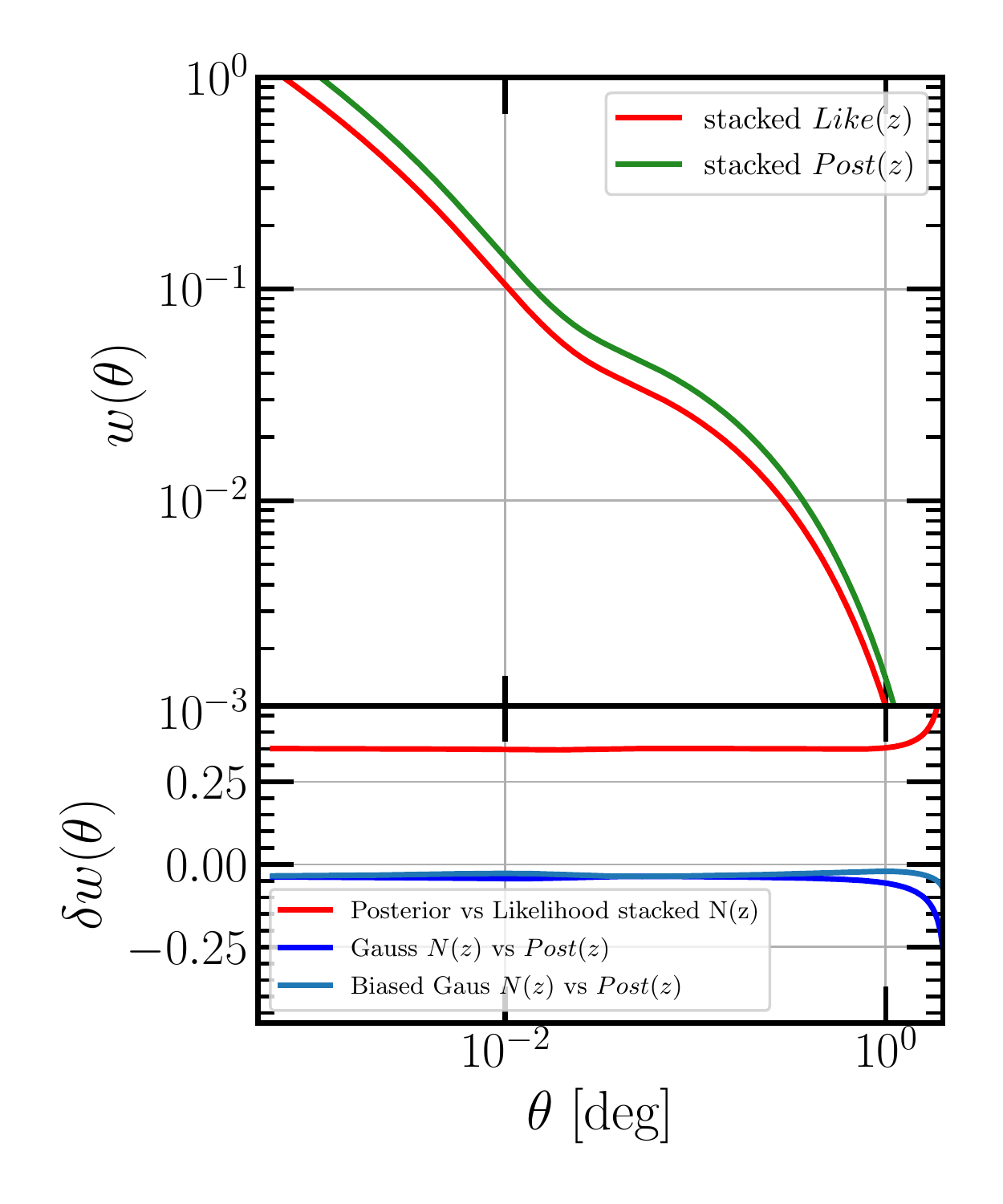}
    \caption{Effect of the redshift distribution on the model correlation function}
    \label{fig:appendix-wofth}
\end{figure}

As discussed in Section \ref{sec:sample-selection}, the assumed redshift distribution of the sample can lead to differences in the modeled $w(\theta)$ that are considerably larger than the measurement uncertainties that may bias the inferred model parameters. We acknowledge, following the discussion in \cite{ilbert_euclid_2021}, that the $N(z)$ constructed by stacking the posteriors $\mathcal{P}(z)$ for each individual galaxy is more representative of the true underlying distribution of the sample, compared to simply stacking the likelihoods $\mathcal{L}(z)$ as outputted from the template fitting code \texttt{LePhare}. In Fig. \ref{fig:appendix-nofz}, we show the $N(z)$ for the sample in the bin $0.8 < \zphot < 1.1$ obtained using $\mathcal{L}(z)$ in red, $\mathcal{P}(z)$ in blue, and the histogram of the sources that have a spectroscopic redshift available in green. The figure shows that indeed stacking the individual $\mathcal{P}(z)$ agrees better with the spec-$z$ histogram, especially in the tails which contribute the most in the $w(\theta)$ amplitude. Fig \ref{fig:appendix-wofth} shows the model $w(\theta)$ of the two $N(z)$ (upper panel). The relative difference between the two is about $35\%$ as shown in the bottom panel. Furthermore, we explore the effect of a bias in the mean redshift as inferred from the N(z) on the $w(\theta)$. For this we use a Gaussian distribution centered at the mean redshift of the bin and then move the mean by $-0.02$. This results in relative difference in less than $3\%$ which is safely within the error bars of the measurements (which are about $10\%$).

\section{Details of the HOD-model derivation of the correlation functions} \label{apdx:woftheta-model}

\subsection{Clustering correlation function}
In brief, one central component of the model is the galaxy-galaxy power spectrum, which can be separated in contributions from clustering of galaxies within the same halo (one-halo term) and between different halos (two-halo term):
\begin{equation}
    P_{\rm gg}(k,z) = P_{\rm gg}^{1\rm h}(k,z) + P_{\rm gg}^{2\rm h}(k,z).
\end{equation}
These two components of the power spectrum of galaxies can be modeled under the HOD framework \citep{cooray_halo_2002, berlind_halo_2002} where the one-halo and two-halo terms are given by:
\begin{equation} \label{eq:power-spec-1h-2h}
    \begin{split}
        & P_{\rm gg}^{1\rm h}(k,z) = \dfrac{1}{\bar{n}_g^2}\displaystyle\int \diff M_h \dfrac{\diff n}{\diff M_h} \left[ \langle N_s \rangle^2 u_s^2(k) + 2 \langle N_s \rangle \, u_s(k) \right], \\
        & P_{\rm gg}^{2\rm h}(k,z) = \\
        & \dfrac{1}{\bar{n}_g^2} \left[  \displaystyle\int \diff M_h \dfrac{\diff n}{\diff M_h} b_h(M_h,z) \left[ \langle N_c \rangle + \langle N_s \rangle \, u_s(k) \right] \right]^2 P_{\rm lin}(k,z).
    \end{split}
\end{equation}
In these equations, $\bar{n}_g = \int \diff M_h \frac{\diff n}{\diff M_h} (\langle N_c \rangle + \langle N_s \rangle)$ is the mean number density of galaxies, $b_h(M_h,z)$ is the large-scale halo bias which we chose the one given by \cite{tinker_large_2010}. $u_s(k)$ is the Fourier transform of the over-density profile of satellite galaxies, for which we assume that it follows the NFW profile \cite{navarro_universal_1997} with a mass-concentration relation as calibrated by \cite{duffy_dark_2008}. The important thing to note here is that the power spectrum is defined in terms of the occupation distributions of centrals and satellites $\langle N_c \rangle$ and $\langle N_s \rangle$ specified by Eqs. \ref{eq:Ncent} and \ref{eq:Nsat} and is where our parametrization of the SHMR enters the model. The linear power spectrum $P_{\rm lin} (k,z)$ enters the two-halo term and dominates large scales. The halo model-based power spectrum is known to be inaccurate on quasi-linear scales at about $k\sim0.5 \, \rm Mpc^{-1}$ at $z=0$ which falls at the transition between the one- and two-halo terms \citep{mead_accurate_2015, mead_accurate_2016, chisari_core_2019}. We correct this by following Eq. 23 of \cite{mead_hmcode-2020_2021}, where a time-dependent function $\alpha(z)$ is applied to the one- and two-halo power spectrum terms.

Having the galaxy-galaxy power spectrum, one can then compute the angular power spectrum via the Limber equation \citep{limber_analysis_1953, kaiser_weak_1992, joachimi_simultaneous_2010}
\begin{equation} \label{eq:angular-powerspectrum}
    C_{\ell}^{i,j} = \displaystyle\int \diff z \; \dfrac{H(z)}{c} \dfrac{N^i(z)\;N^j(z)}{\chi^2(z)} P_{\rm gg}\left(z, k = \dfrac{\ell+1/2}{\chi(z)}\right),
\end{equation}
where $H(z)$ is the expansion rate at redshift $z$, $\chi(z)$ is the radial comoving distance at $z$, and $N^i(z)$ and $N^j(z)$ are the redshift distribution of the galaxies in the samples $i$ and $j$. Since we're interested in the angular auto-correlation function, both $i$ and $j$ are  the same sample. Finally, to arrive at the angular correlation function $w(\theta)$, we perform an inverse Fourier transform of $C_{\ell}$ numerically through the Hankel transforms under the flat sky approximation:
\begin{equation} \label{eq:hankel-transform}
    w(\theta) = \displaystyle\int \diff \ell \dfrac{\ell}{2\pi} C_{\ell} \, J_0(\theta \, \ell),
\end{equation}
where $J_0(\theta \, \ell)$ is the 0-th order Bessel function \citep[for more details see][]{chisari_core_2019}.

\section{Description of the hydrodynamical simulations} \label{apdx:sims-described}

\subsection{Main characteristics and catalog extraction} 
 \paragraph*{{\sc Horizon-AGN}} has been produced  with the adaptive mesh refinement code RAMSES \citep{teyssier_cosmological_2002}, using WMAP7 cosmological parameters \citep{komatsu_seven-year_2011} in a box of size $100 \,h^{-1}\,{\rm Mpc}$ a side \footnote{A light cone has also been extracted on-the-fly in order to build realistic mocks \citep{laigle_horizon-agn_2019,gouin_weak_2019}. In particular, based on these mocks, \cite{hatfield_comparing_2019} showed that the propagation of statistical and systematic uncertainties inherited from redshift and mass photometric estimates lead to an underestimation of the clustering amplitude by $\sim 0.1$ dex.  In the present work, we use galaxy and halo catalogs extracted from the snapshot data, but we checked that using the light cone data does not significantly change the measured SHMR.} and a DM mass resolution of $M_{\rm  DM, res}=8\times 10^7 \, \rm M_\odot$. The simulation includes gas heating, cooling, star-formation, feedback from stellar winds, type Ia and type II supernovae with mass, energy, and metal release in the interstellar-medium. The simulation also follows the formation of black holes and energy release from AGN in a quasar or radio mode depending on the accretion rate.  Full details on the subgrid implementation are given in \cite{dubois_dancing_2014}.
 
 Galaxies and halos are identified using the {\sc AdaptaHOP} structure finder \citep{2004MNRAS.352..376A, tweed_building_2009} applied to the distribution of star and DM particles respectively, as described in previous works \citep[e.g.,][]{dubois_dancing_2014}. Galaxy masses are obtained by summing the masses of all individual stellar particles within twice the effective radius of the galaxies, while halo masses are obtained by summing all DM particles within the halos. Galaxies and halos are positionally matched. The central galaxies of a halo is defined as the most massive galaxies within 0.1 times the halo virial radius. 
 
 We note that stellar mass losses in {\sc Horizon-AGN} were implemented assuming a Salpeter IMF \citep{salpeter_luminosity_1955}. This can lead to $\sim 0.1$ dex more stellar masses at later time than when assuming a Chabrier IMF \citep{chabrier_galactic_2003}. The simulated masses were therefore rescaled accordingly for a consistent comparison with both the other simulations and the observational data (which assumes a Chabrier IMF when deriving the mass from the photometry). 
 
 The simulation is in relatively good agreement with observations up to $z=4$. Known discrepancies include the overestimation of galaxy masses, especially at low mass \citep{kaviraj_horizon-agn_2017}. These low-mass galaxies are on overall too quenched, an indication that star formation at high redshift was not regulated enough and the gas in the environment of these galaxies was too rapidly consumed. In addition, the bimodality is not as well-marked as in observations due to residual star-formation in massive galaxies, possibly because of a slightly insufficient strength for the AGN feedback.

\paragraph*{ {\sc  TNG100-1}} has been produced with the moving-mesh code AREPO \citep{springel_smoothed_2010} using cosmological parameters from \cite{planck_collaboration_planck_2016} in a box of size $75 ~h^{-1}\,{\rm Mpc}$ a side and a DM mass resolution of $M_{\rm  DM, res}=7.5\times 10^6 \,\rm M_\odot$. It follows magnetic fields in addition to the hydrodynamical processes above-mentioned for the {\sc Horizon-AGN} simulation (with different subgrid implementations, see the discussion below).  Full details on the subgrid implementation are given in \cite{Pillepich2018_presentation,springel_first_2018}. 

Halos, subhalos, and galaxies have been identified using the {\sc FoF} \citep{davis_evolution_1985} and {\sc subfind} (which, within a halo identified with the {\sc FoF} technique, relies on all particle species to identify the galaxy, see \citealp{springel_populating_2001}) algorithms, as described in, for instance, \cite{pillepich_first_2018}. We downloaded the halo and galaxy catalogs from the public website\footnote{\href{https://www.tng-project.org/data/}{https://www.tng-project.org/data/}}. In the following, galaxy masses are estimated by summing all stellar mass particles within twice the stellar effective radius, while halo masses are taken as being the sum of all individual dark matter particles in the identified halos, which matches the definition of halo and galaxy masses in {\sc Horizon-AGN}. Galaxies that do not reside within $R_{200}$ of a larger halo are identified as centrals. Galaxy clustering has been measured in \cite{springel_first_2018} {and the SHMR has been presented in \citealp{pillepich_first_2018} (see also \citealp{engler_distinct_2020}). In particular, these authors investigated how the chosen estimate for stellar mass modifies the measured relation. While we are aware of this discussion, we chose in this work to measure stellar mass within twice the effective radius for consistency between different simulated datasets. Such measurement is also  closer to our observational estimate of stellar masses (which derive from SED-fitting on the total magnitudes -- and not on aperture magnitudes).}

 \paragraph*{ {\sc EAGLE}} has been produced using  a modified version of the N-Body Tree-PM smoothed particle hydrodynamics (SPH) code {\sc GADGET-3} \citep{2005MNRAS.364.1105S}, adopting cosmological parameters from Planck \citep{planck_collaboration_planck_2014}. In this work, we use the reference run called {Ref-L0100N1504},corresponding to a box of $100$ comoving~${\rm Mpc}$ a side with a DM mass resolution of  $M_{\rm  DM, res}=9.7\times 10^6 \, \rm M_\odot$. {\sc EAGLE} follows gas heating and cooling, star-formation, feedback from stellar winds and AGB stars, along with type Ia and type II supernovae and from AGN. Full details on the subgrid implementation are given in \cite{crain_eagle_2015} and \cite{schaye_eagle_2015}.
 
 As for {TNG100-1}, the halos, subhalos, and galaxies have been identified using the {\sc FoF} \citep{davis_evolution_1985} and {\sc subfind} \citep{springel_populating_2001} algorithms, as described in \cite{mcalpine_eagle_2016}. We downloaded halo and galaxy catalogs from the public website\footnote{\href{http://virgodb.dur.ac.uk:8080/Eagle/}{http://virgodb.dur.ac.uk:8080/Eagle/}}. Galaxy and halo masses were obtained by summing the masses of all stellar and dark matter particles, respectively, which are part of the objects. The central galaxy is taken as the one which contains the particle with the lowest value of the gravitational potential. The redshift evolution of the SHMR for central galaxies in {\sc EAGLE} simulation was briefly presented in \cite{matthee_origin_2017}. 

\subsection{Implementation of subgrid recipes} \label{apdx:subgrid}


Our comparisons of the SHMR in observations and simulations has highlighted two main discrepancies: (1) in {\sc Horizon-AGN}, the SHMR is always overestimated with respect to observations and the two other simulations, especially at low masses; (2) in all simulations, the relative contribution of satellites with respect to centrals is too large compared to observations. 

With regard to (1), we note first that in {\sc EAGLE}, the free parameters of the subgrid (stellar and AGN) feedback model were calibrated so that the simulations reproduce the galaxy stellar mass function and SHMR, galaxy sizes, and the empirical relation between black hole mass and stellar mass all at $z=0$ \citep{schaye_eagle_2015}. In {\sc TNG100-1,} the calibration was performed against the star formation rate density as a function of cosmic time and the stellar mass function and SHMR both at $z=0$ \citep{Pillepich2018_presentation}. In {\sc Horizon-AGN} however, the calibration was less constraining since only the efficiency of AGN feedback was tuned so that the black hole--galaxy scaling relation at $z=0$ was reproduced \citep[see][for details]{dubois_self-regulated_2012}. It is therefore not surprising that both {\sc EAGLE} and {\sc TNG100-1} better reproduce, overall, the stellar mass function \citep{schaye_eagle_2015,pillepich_first_2018} and SHMR (Fig.~\ref{fig:SHMR-tot-all-vs-lit}) at low masses, because their calibration specifically constrain stellar feedback, which is not the case  for {\sc Horizon-AGN}. 

Without pretending to  exhaustively discuss the differences in stellar feedback and star-formation implementation between {\sc Horizon-AGN} and the two other simulations, we instead opt to  highlight a few aspects. 
In the {\sc EAGLE} simulation \cite[as explained in][]{crain_eagle_2015}, the amount of injected energy from feedback depends on the local properties of the gas (it decreases with metallicity and increases with gas density), the calibration being adjusted in order to reproduce the local stellar mass function. This tuning contributes to increase the supernova feedback energy at high redshift {beyond the energy available for their adopted Chabrier IMF~\citep{crain_eagle_2015}}. {Furthermore, the stochasticity of energy deposition of supernovae is artificially increased to enhance the impact of their feedback in terms of wind mass loading and quenching of star formation\footnote{While this implementation of stochastic feedback enhance the impact of feedback in those (GADGET) SPH simulation, it makes no difference with standard thermal release when simulated with (RAMSES) AMR~\citep{rosdahl2017}.}~\citep{dellavecchia2012}.}
Finally, the {\sc EAGLE} star-formation law does not follow the standard Kennicutt-Schmidt prescription \citep[][as adopted in {\sc Horizon-AGN}]{kennicutt1998}; rather, it depends on pressure instead of density and includes a metallicity-dependent density threshold (against a simple density threshold in {\sc Horizon-AGN}), which tends to reduce star-formation in metal-rich regions. Those features are likely to contribute to a more efficient quenching of star-formation in {\sc EAGLE} compared to {\sc Horizon-AGN}.

Star-formation  in {\sc TNG100} follows the empirical Kennicutt-Schmidt relation, however, feedback implementation is sensibly different, as described in \cite{Pillepich2018_presentation}. More specifically, the energy transfer from SNe to large-scale galactic winds is very efficient because of the hydrodynamical decoupling of the launched wind gas from the dense star-forming gas, until they recouple hydrodynamically with the circum-galactic gas~\citep{springel_cosmological_2003, vogelsberger_model_2013}. These authors used a wind velocity that is proportional to the local DM velocity dispersion and cosmic time so that the winds are faster in more massive halos and at lower redshift. Similarly to the {\sc EAGLE} model, the given supernovae energy to the gas is higher for energy deposit in lower metallicity gas, with the same scaling with metallicity as in EAGLE. Taken together, these features contribute to make stellar feedback more effective at high redshift in {\sc TNG100} compared to {\sc Horizon-AGN}.

Finally, we note that many missing mechanisms could be naturally added to the {\sc Horizon-AGN} model in order to increase the strength of stellar feedback in low-mass galaxies at high redshift in a physically motivated way (without necessarily requiring empirical tuning of parameters). These processes include radiation from stars (see the discussion below), cosmic ray feedback from supernovae \citep[e.g.,][]{booth2013,salem2014,dashyan2020}, a gravo-turbulent model for star formation (e.g.,~\citealp{nunez2021,dubois2021}), or adopting an IMF varying with redshift, stellar density, or metallicity \citep[e.g.,][]{applebaum2020,progmet2021}.

With regard to (2), however, the efficient stellar feedback implemented in {TNG100} and {\sc EAGLE} do not prevent the too large satellite fraction (Fig.~\ref{fig:fsat-vs-z}) and excessive contribution of satellites to the total SHMR (Fig.~\ref{fig:SHMR-all-vs-lit}). As previously noted in Sec.~\ref{sec:comparison-hydro}, such a discrepancy must be due to a lack of satellite quenching at high redshift, making these satellites to grow too massive (with respect to observations) before being eventually quenched. Galaxy pre-processing by stellar radiation feedback \citep[e.g.,][]{rosdahl2013,hopkins2020} resulting in less tightly-bound galaxies could be a prerequisite to efficiently quench satellites (once they have been accreted) through tidal stripping, as suggested by \cite{Costa2019}. 

Finally, some studies mention the possible role of AGN feedback from the central galaxy in quenching its satellites \citep[e.g.,][]{dashyan2019,martinnavarro2019}: the lack of quenching in high-redshift satellites could therefore also be the consequence of an imperfect AGN feedback implementation at high redshift, {although hydrodynamical simulations show that the activity of SNe in low mass galaxies quench the growth of black holes and their associated AGN feedback~\citep[e.g.,][]{dubois2015,habouzit2017,angles2017}}.

\section{Data}
The data used to make all the plots in this paper -- measurements of the clustering and GSMF, best-fit models, derived quantities, and so on, are shared publicly with the community in tabular form on the following github repository: \href{https://github.com/mShuntov/SHMR_COSMOS2020}{https://github.com/mShuntov/SHMR\_COSMOS2020}

\section{Posterior probabilities of the parameters} \label{apdx:posteriors}

  \begin{figure*}[hp]
   \begin{subfigure}[b]{0.5\textwidth}
            \includegraphics[width=1\hsize]{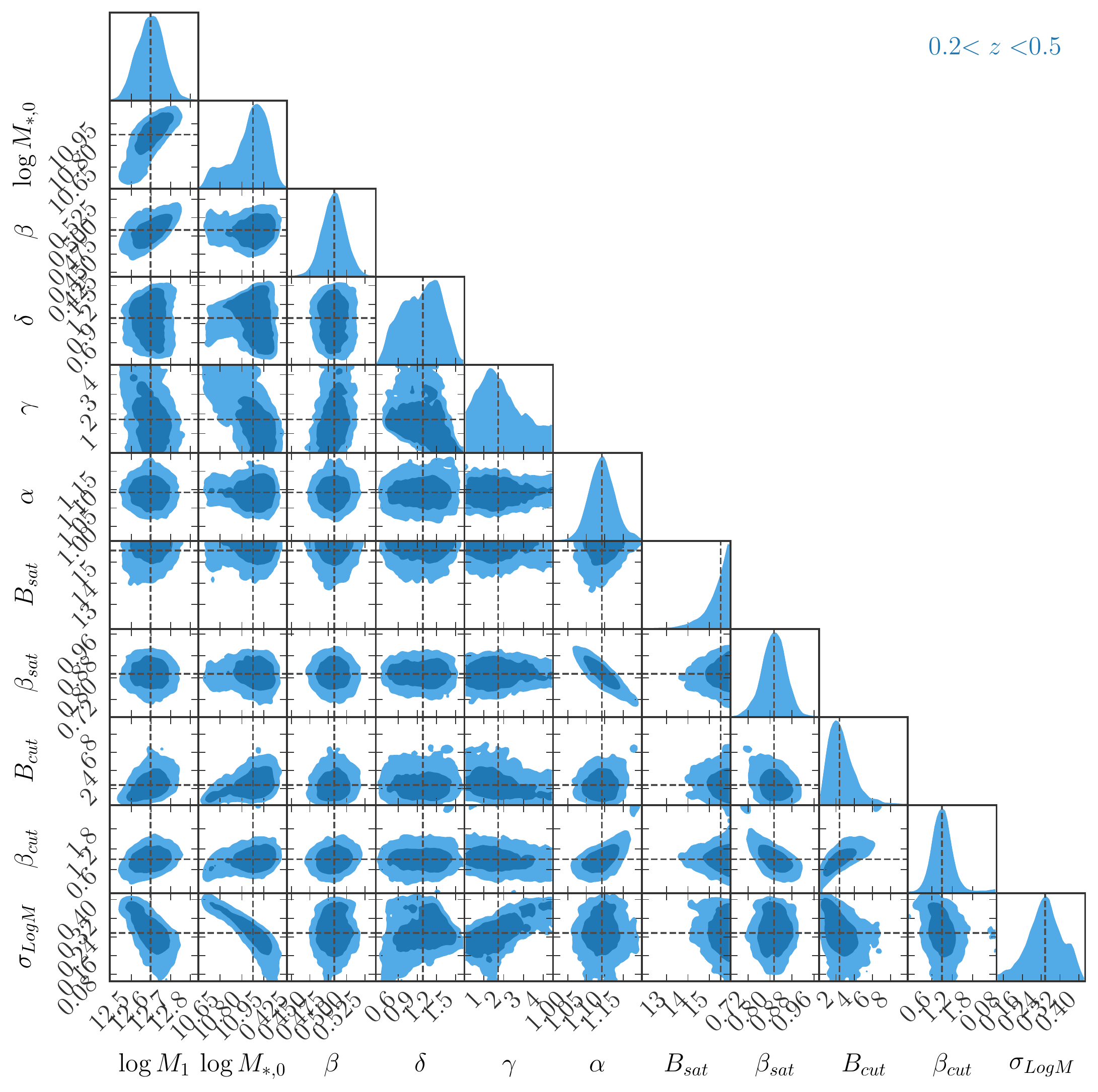}     
   \end{subfigure}
   \hfill
   \begin{subfigure}[b]{0.5\textwidth}
            \includegraphics[width=1\hsize]{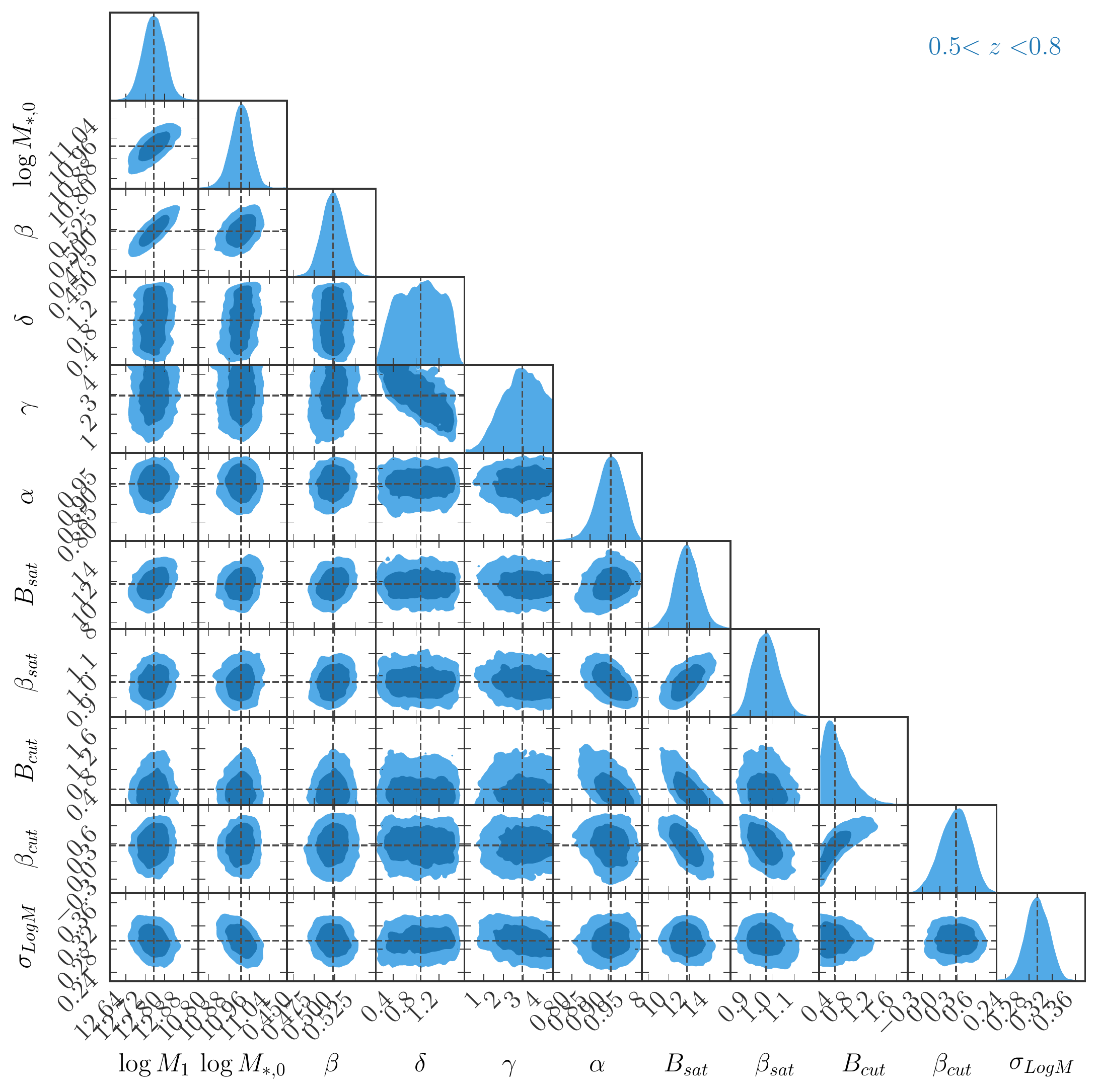}     
   \end{subfigure}
   \hfill
   \begin{subfigure}[b]{0.5\textwidth}
            \includegraphics[width=1\hsize]{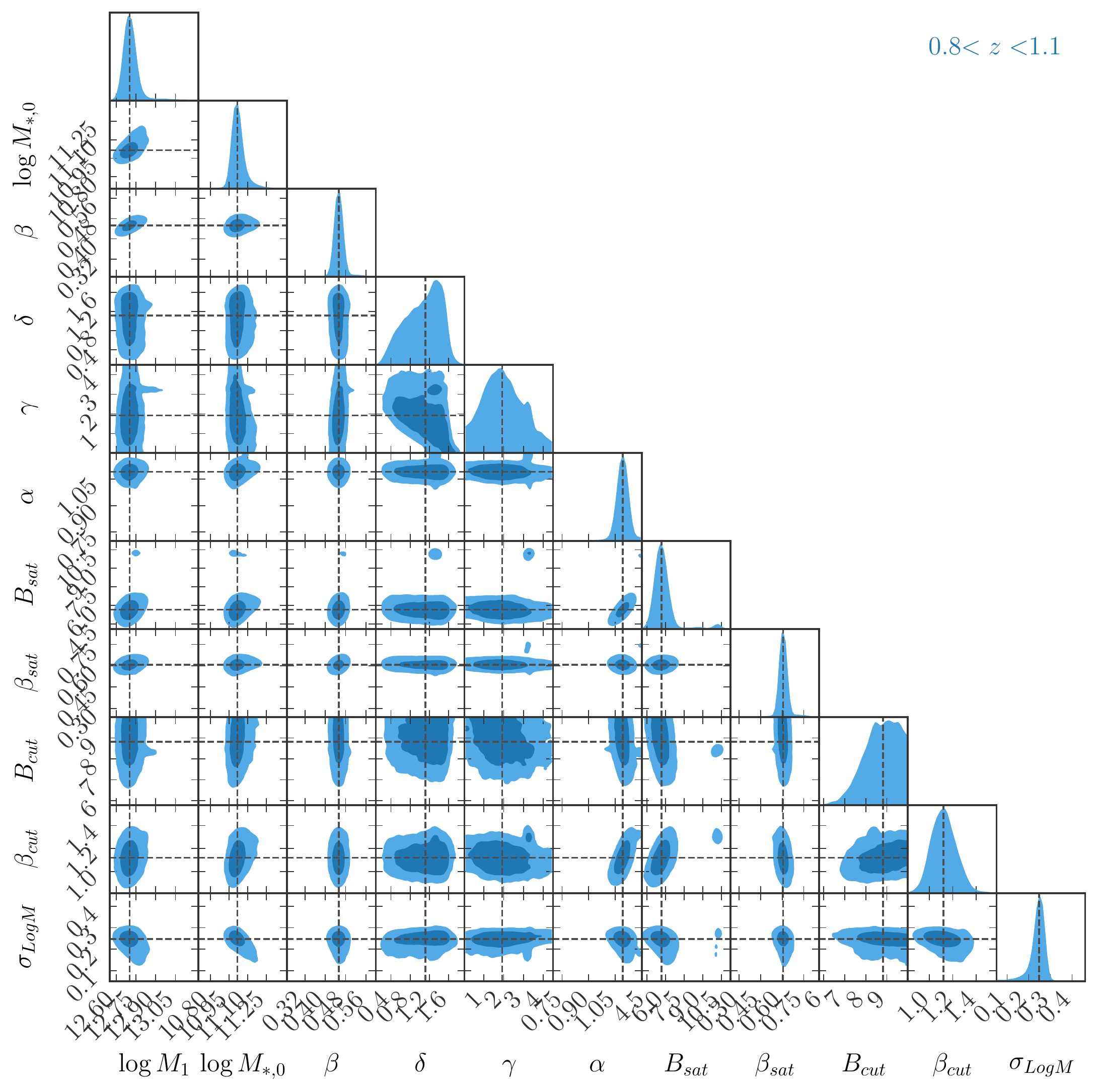}     
   \end{subfigure}
   \hfill
   \begin{subfigure}[b]{0.5\textwidth}
            \includegraphics[width=1\hsize]{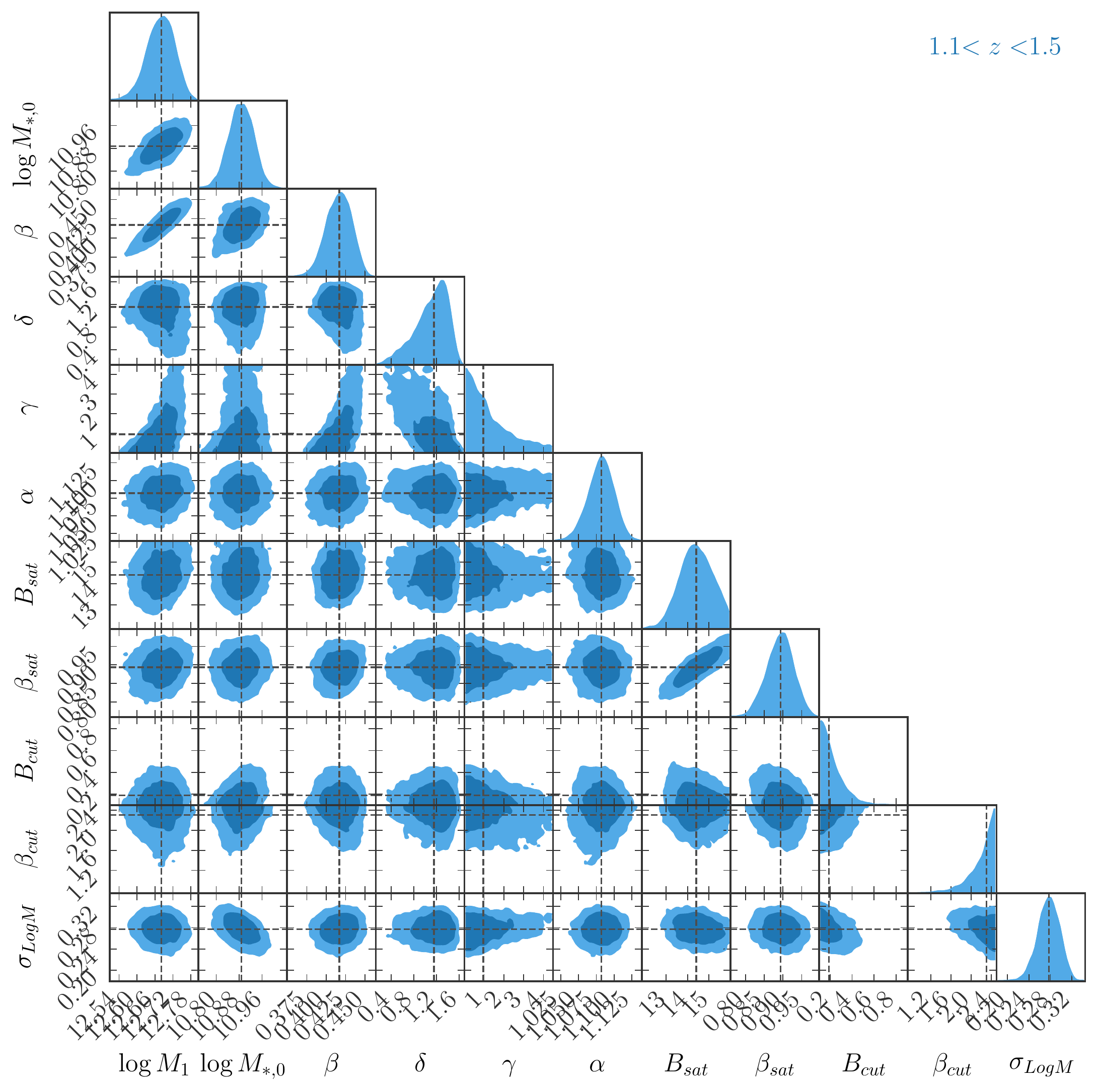}     
    \end{subfigure}
    \hfill

  \caption{Posterior distributions for each redshift fit}
   \label{fig:Contours}
        \end{figure*}

\begin{figure*}[hp]
  
   \begin{subfigure}[b]{0.42\textwidth}
            \includegraphics[width=1\hsize]{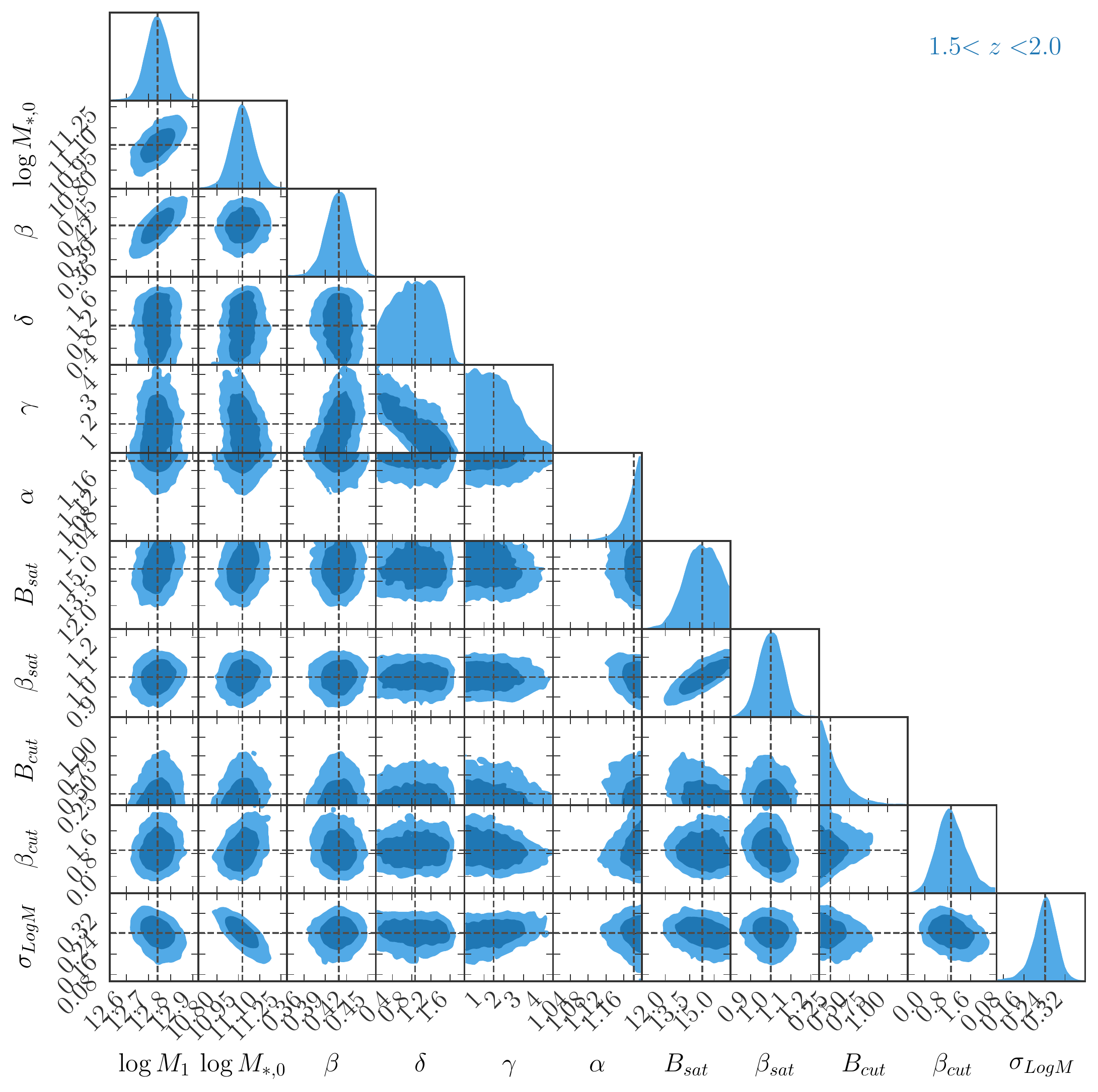}     
   \end{subfigure}
   \hfill
   \begin{subfigure}[b]{0.42\textwidth}
            \includegraphics[width=1\hsize]{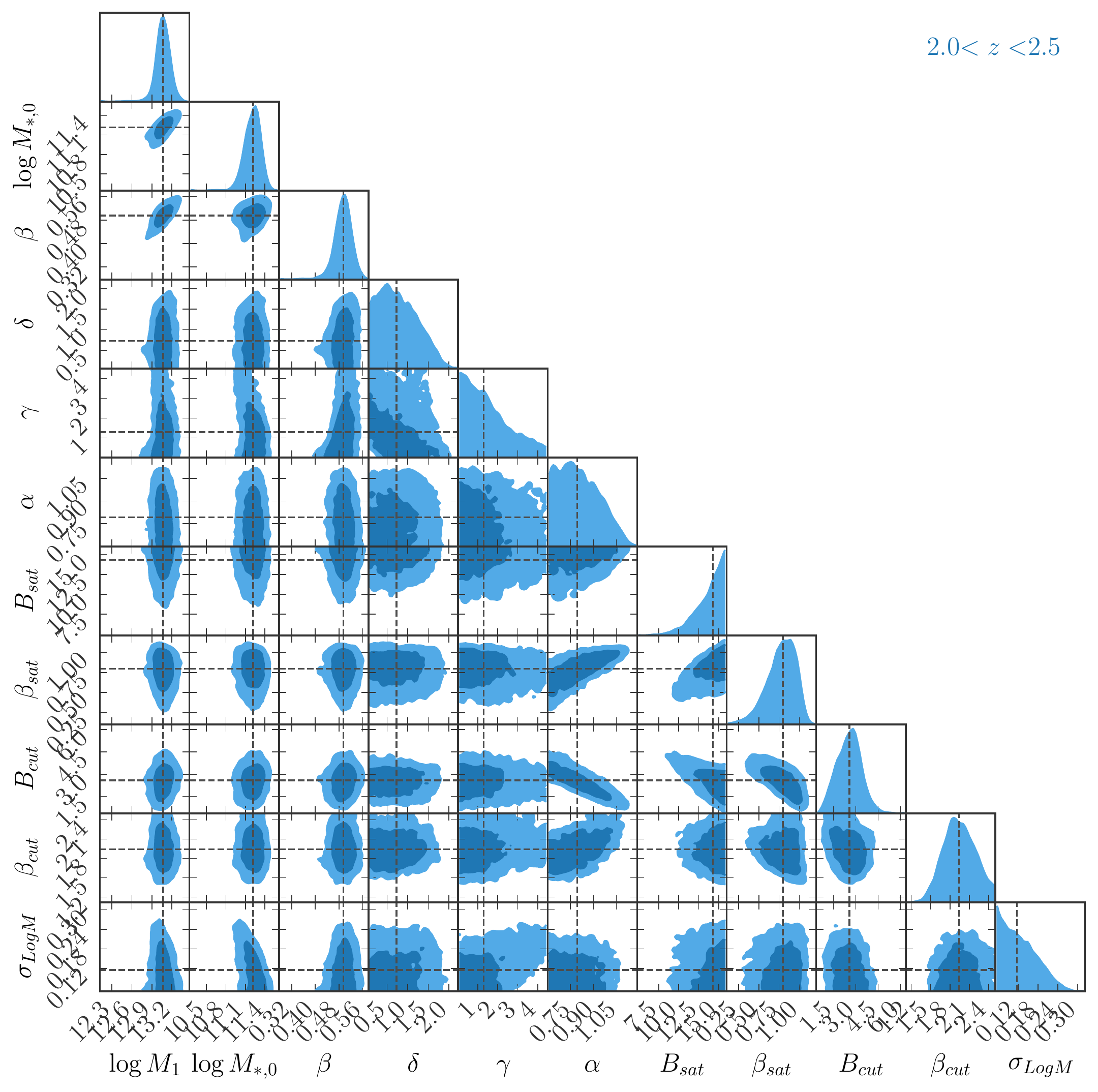}     
   \end{subfigure}
   \hfill
   \begin{subfigure}[b]{0.42\textwidth}
            \includegraphics[width=1\hsize]{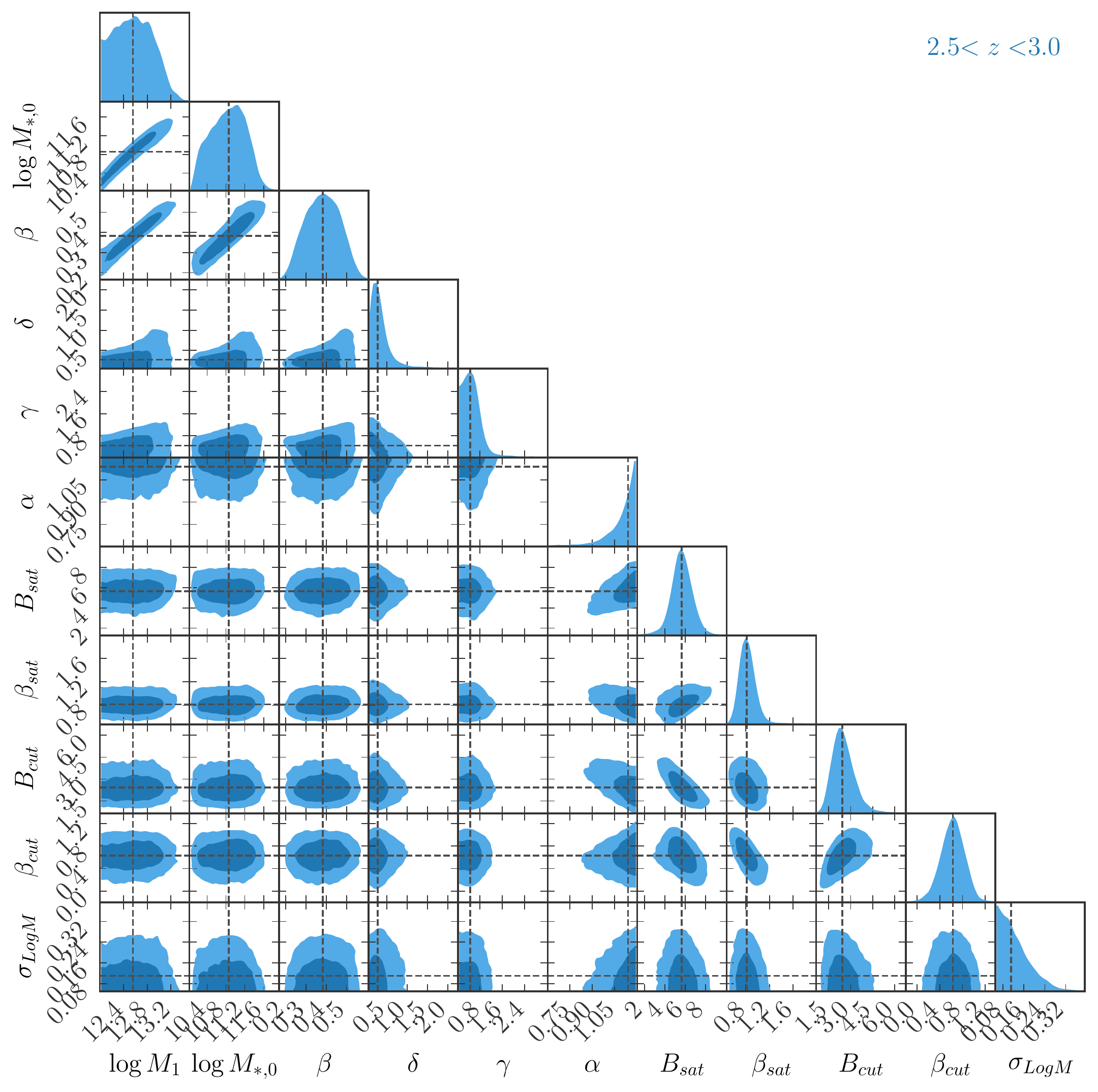}     
   \end{subfigure}
   \hfill
   \begin{subfigure}[b]{0.42\textwidth}
            \includegraphics[width=1\hsize]{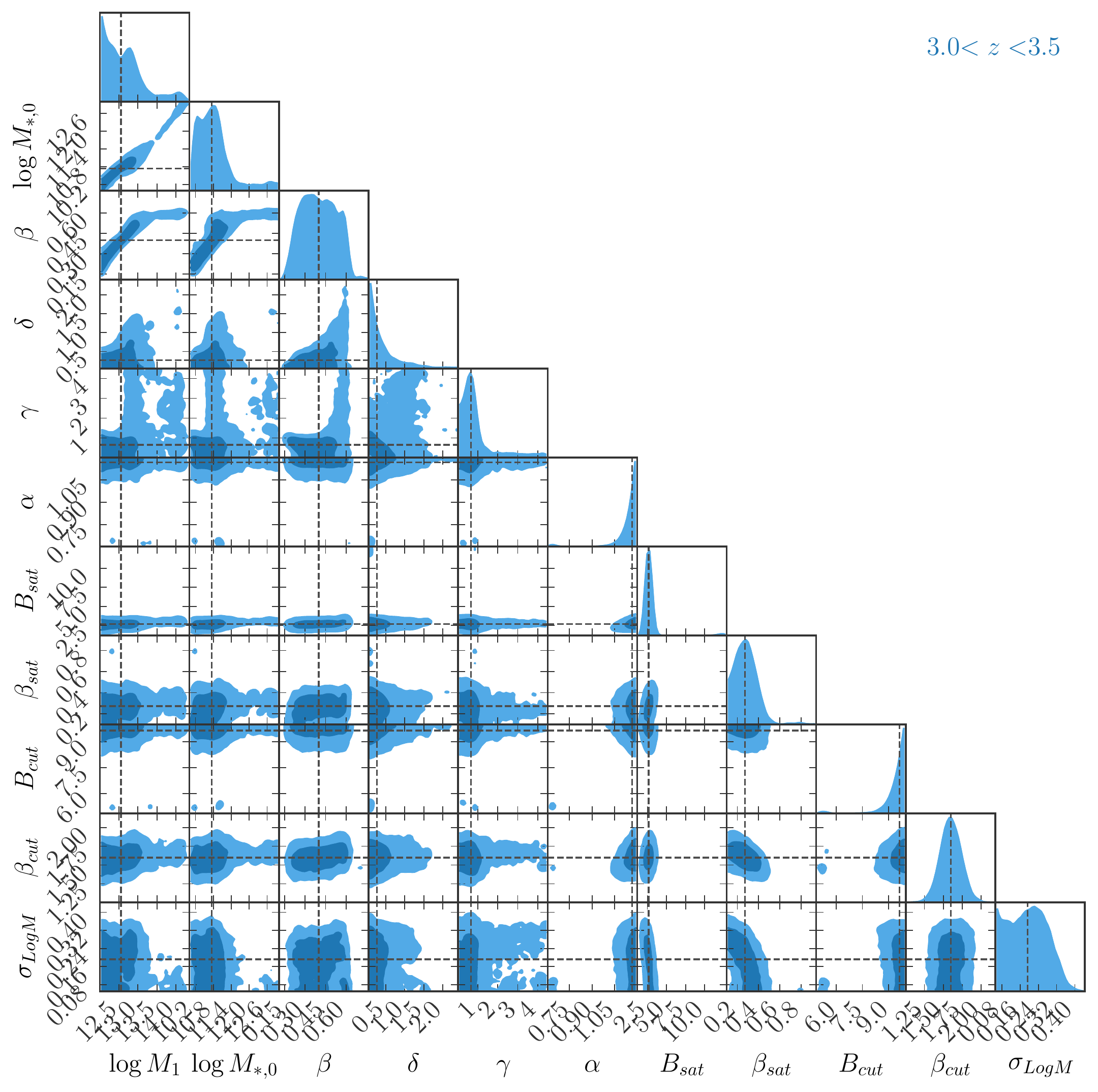}     
   \end{subfigure}
   \hfill
   \begin{subfigure}[b]{0.42\textwidth}
            \includegraphics[width=1\hsize]{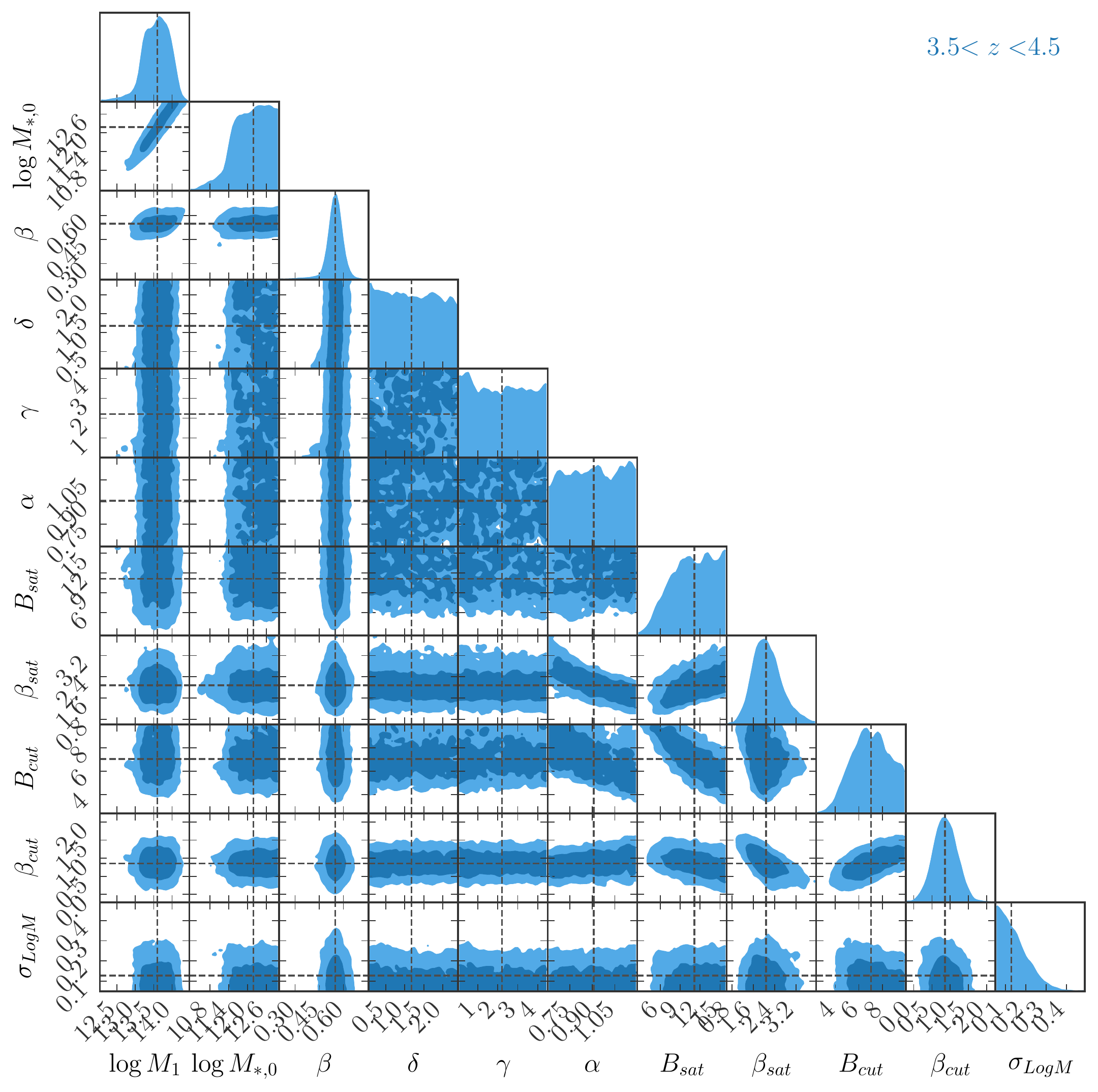}     
   \end{subfigure}
   \hfill
   \begin{subfigure}[b]{0.42\textwidth}
\includegraphics[width=1\hsize]{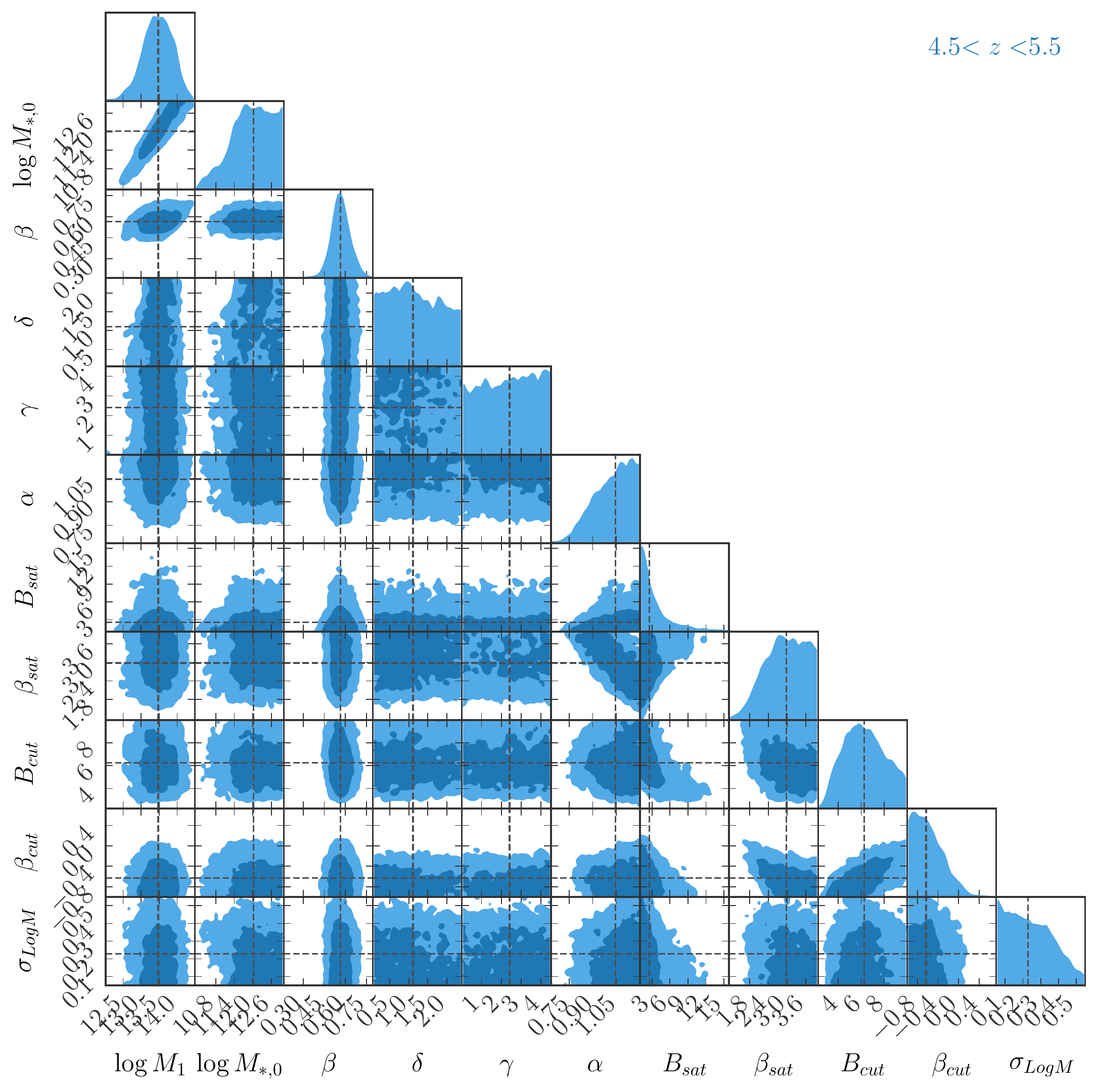}     
   \end{subfigure}
  \caption{Posterior distributions for each redshift fit}
   \label{fig:Contours}
        \end{figure*}

\vfill

\newgeometry{margin=1cm} 
\begin{landscape}
\section{Best-fit values of the model parameters} \label{apdx:bestfit-table}
\vspace{5cm}
\begin{table*}[h]
\footnotesize\centering
\renewcommand{\arraystretch}{1.35}
\setlength{\tabcolsep}{7pt} 
        \begin{center}
                \begin{tabular}{ccccccccccccc}
                \hline
$z$-bin  & log$\, M_{1}$ &  log$\,M_{*,0}$ &  $\beta$ &  $\delta$ &  $\gamma$ &  $\alpha$ &   $B_{\rm{sat}}$ &  $\beta_{\rm{sat}}$ &  $B_{\rm{cut}}$  &  $\beta_{\rm{cut}}$ &  $\sigma_{\rm{Log}M}$ & $\chi^2_{\rm reduced}$ \\
                        \hline
                        \hline
                        $0.2< z < 0.5$ & $ 12.629_{-  0.072}^{+  0.069}$ & $ 10.855_{-  0.216}^{+  0.099}$ & $  0.487_{-  0.016}^{+  0.016}$ & $  0.935_{-  0.335}^{+  0.295}$ & $  1.939_{-  1.058}^{+  2.515}$ & $  1.056_{-  0.030}^{+  0.029}$ & $ 15.467_{-  0.753}^{+  0.394}$ & $  0.845_{-  0.040}^{+  0.040}$ & $  2.045_{-  0.974}^{+  1.288}$ & $  0.745_{-  0.251}^{+  0.231}$ & $  0.268_{-  0.080}^{+  0.009}$ & $ 2.6$ \\
            \hline
            $0.5< z < 0.8$ & $ 12.793_{-  0.043}^{+  0.044}$ & $ 10.927_{-  0.041}^{+  0.038}$ & $  0.502_{-  0.013}^{+  0.013}$ & $  0.802_{-  0.407}^{+  0.446}$ & $  3.132_{-  1.143}^{+  1.488}$ & $  0.905_{-  0.038}^{+  0.033}$ & $ 12.221_{-  1.206}^{+  1.082}$ & $  0.985_{-  0.052}^{+  0.057}$ & $  0.277_{-  0.168}^{+  0.290}$ & $  0.076_{-  0.386}^{+  0.303}$ & $  0.293_{-  0.019}^{+  0.002}$ & $ 2.5$ \\
            \hline
            $0.8< z < 1.1$ & $ 12.730_{-  0.039}^{+  0.044}$ & $ 11.013_{-  0.032}^{+  0.057}$ & $  0.454_{-  0.014}^{+  0.014}$ & $  1.109_{-  0.455}^{+  0.329}$ & $  1.925_{-  1.014}^{+  1.441}$ & $  1.065_{-  0.030}^{+  0.030}$ & $  5.416_{-  0.507}^{+  0.552}$ & $  0.612_{-  0.018}^{+  0.019}$ & $  8.845_{-  1.020}^{+  1.261}$ & $  1.098_{-  0.095}^{+  0.113}$ & $  0.250_{-  0.032}^{+  0.002}$ & $10.2$ \\
            \hline
            $1.1< z < 1.5$ & $ 12.673_{-  0.085}^{+  0.065}$ & $ 10.967_{-  0.076}^{+  0.071}$ & $  0.393_{-  0.026}^{+  0.022}$ & $  0.746_{-  0.189}^{+  0.292}$ & $  0.335_{-  0.235}^{+  0.443}$ & $  1.078_{-  0.017}^{+  0.017}$ & $ 15.015_{-  0.821}^{+  0.639}$ & $  0.906_{-  0.038}^{+  0.031}$ & $  0.101_{-  0.046}^{+  0.065}$ & $  4.197_{-  0.463}^{+  0.222}$ & $  0.167_{-  0.077}^{+  0.008}$ & $ 3.8$ \\
            \hline
            $1.5< z < 2.0$ & $ 12.787_{-  0.065}^{+  0.067}$ & $ 11.040_{-  0.137}^{+  0.163}$ & $  0.410_{-  0.018}^{+  0.015}$ & $  0.716_{-  0.423}^{+  0.480}$ & $  1.312_{-  0.878}^{+  1.431}$ & $  1.213_{-  0.031}^{+  0.027}$ & $ 14.168_{-  0.984}^{+  0.981}$ & $  0.951_{-  0.072}^{+  0.063}$ & $  0.099_{-  0.068}^{+  0.168}$ & $  1.848_{-  1.209}^{+  1.667}$ & $  0.211_{-  0.203}^{+  0.012}$ & $ 6.1$ \\
            \hline
            $2.0< z < 2.5$ & $ 13.097_{-  0.101}^{+  0.087}$ & $ 11.254_{-  0.152}^{+  0.101}$ & $  0.495_{-  0.028}^{+  0.022}$ & $  0.668_{-  0.420}^{+  0.562}$ & $  1.077_{-  0.765}^{+  1.325}$ & $  0.793_{-  0.130}^{+  0.158}$ & $ 14.156_{-  2.419}^{+  1.358}$ & $  0.751_{-  0.223}^{+  0.164}$ & $  2.501_{-  0.860}^{+  0.776}$ & $  1.968_{-  0.236}^{+  0.290}$ & $  0.050_{-  0.041}^{+  0.010}$ & $ 3.8$ \\
            \hline
            $2.5< z < 3.0$ & $ 12.627_{-  0.380}^{+  0.325}$ & $ 10.920_{-  0.440}^{+  0.333}$ & $  0.393_{-  0.089}^{+  0.077}$ & $  0.274_{-  0.130}^{+  0.218}$ & $  0.446_{-  0.295}^{+  0.342}$ & $  1.246_{-  0.098}^{+  0.041}$ & $  6.539_{-  0.862}^{+  0.921}$ & $  0.772_{-  0.093}^{+  0.107}$ & $  1.763_{-  0.528}^{+  0.679}$ & $  0.686_{-  0.248}^{+  0.237}$ & $  0.083_{-  0.057}^{+  0.007}$ & $ 5.7$ \\
            \hline
            $3.0< z < 3.5$ & $ 12.820_{-  0.519}^{+  0.972}$ & $ 11.067_{-  0.526}^{+  1.124}$ & $  0.465_{-  0.154}^{+  0.138}$ & $  0.354_{-  0.260}^{+  0.988}$ & $  0.741_{-  0.443}^{+  2.066}$ & $  1.251_{-  0.092}^{+  0.037}$ & $  2.592_{-  0.293}^{+  0.425}$ & $  0.067_{-  0.043}^{+  0.109}$ & $ 13.122_{-  1.955}^{+  1.195}$ & $  2.075_{-  0.224}^{+  0.181}$ & $  0.013_{-  0.009}^{+  0.001}$ & $ 4.1$ \\
            \hline
            $3.5< z < 4.5$ & $ 13.638_{-  0.331}^{+  0.303}$ & $ 12.222_{-  0.558}^{+  0.533}$ & $  0.551_{-  0.034}^{+  0.038}$ & $  1.557_{-  1.075}^{+  1.187}$ & $  3.149_{-  2.236}^{+  2.279}$ & $  0.930_{-  0.231}^{+  0.253}$ & $  9.930_{-  4.205}^{+  4.068}$ & $  1.838_{-  0.579}^{+  0.694}$ & $  7.588_{-  2.038}^{+  2.985}$ & $  1.017_{-  0.368}^{+  0.415}$ & $  0.085_{-  0.061}^{+  0.008}$ & $ 3.5$ \\
            \hline
            $4.5< z < 5.5$ & $ 13.547_{-  0.392}^{+  0.378}$ & $ 12.105_{-  0.676}^{+  0.605}$ & $  0.567_{-  0.062}^{+  0.066}$ & $  1.427_{-  0.971}^{+  1.262}$ & $  3.225_{-  2.167}^{+  2.248}$ & $  1.031_{-  0.264}^{+  0.194}$ & $  2.630_{-  1.282}^{+  4.383}$ & $  3.108_{-  1.062}^{+  1.816}$ & $  7.020_{-  2.105}^{+  3.104}$ & $ -0.621_{-  0.274}^{+  0.446}$ & $  0.226_{-  0.151}^{+  0.017}$ & $ 1.9$ \\
                        \hline
                \end{tabular}
        \end{center}
        \caption{Best-fit values of the model parameters in the ten redshift bins}
        \label{tab:best-fit_values}
\end{table*}
\end{landscape}

\end{appendix}

\end{document}